\documentclass{article}
\usepackage{arxiv}

\usepackage[utf8]{inputenc}
\usepackage{graphicx}
\usepackage{amsmath}
\usepackage{amsfonts}
\usepackage{hyperref}
\usepackage{booktabs}
\usepackage{caption}
\usepackage{subcaption}
\usepackage{multirow}

\title{Latent Space Projections and Atlases, A Cautionary Tale in Deep Neuroimaging using Autoencoders}

\author{
  J.M. Gorriz\thanks{Corresponding author} , F. Segovia, C. Jimenez, J.E. Arco, F.J. Martinez, J Ramirez\\
  Data Science and Computational Intelligence Institute\\
  University of Granada\\
  Granada, Spain \\
  \texttt{jg825@cam.ac.uk}\
     \And
 S. Abulikemu, J. Suckling \\
  Department of Psychiatry\\
  University of Cambridge\\
  Cambridge, UK\\
  \texttt{js369@cam.ac.uk} \\
       \And
       International Initiatives\\
for the Alzheimer's Disease Neuroimaging Initiative (ADNI)\\
}

\begin{document}
\maketitle

\begin{abstract}
This study introduces a deep learning framework for the inferential exploration of latent representations in 3D brain MRI, leveraging a simple convolutional autoencoder with a hierarchical encoder and a compact latent space. Trained on segmented gray matter images from the Alzheimer’s Disease Neuroimaging Initiative (ADNI) dataset, the model learns latent representations that preserve neuroanatomical structure and reflect clinical variability across cognitive status. Dimensionality reduction techniques (PCA, t-SNE, PLS, UMAP) were applied to visualize and interpret the latent space, correlating it with anatomical regions defined by the AAL atlas. As a novel contribution, the Latent–Regional Correlation Profiling (LRCP) framework, which combines statistical association and supervised discriminability to identify brain regions that encode clinically relevant latent information is proposed. Our results show that even minimal architectures capture meaningful patterns associated with progression to Alzheimer’s disease. Interpretability is assessed by applying SHAP-based regression to a post-hoc model that predicts reconstruction error from atlas-based regional gray matter intensities, thereby identifying anatomically meaningful regions involved in class-specific reconstruction strategies. These findings are further validated using statistical agnostic methods, highlighting the importance of rigorous evaluation in neuroimaging. This work demonstrates the potential of autoencoders as exploratory tools for biomarker discovery and hypothesis generation in clinical neuroscience.
\end{abstract}


\section{Introduction}

Open access to large-scale neuroimaging data has motivated and encouraged the development of machine learning (ML) methods to extract clinically and biologically meaningful features \cite{Hofmann,42,Zeineldin,Adeli2005}. Among these, autoencoders are unsupervised neural networks designed to learn input data through compressed representations \cite{Martinez}. While the use of autoencoders and related models for unsupervised representation learning is promising, differential analyses based on latent embeddings should be approached with caution \cite{Gorrizarxiv,Eklund16,Noble20,Varoquaux18,Varma2006}. 

Notwithstanding this good advice, the use of autoencoders (AE) and other neural architectures to analyze such representations, can facilitate the discovery of patterns associated with neurodegenerative diseases, including Mild Cognitive Impairment (MCI), Alzheimer’s Disease (AD), and the neurological  progression of the former to the latter \cite{Bass2023}. This objective is central to so-called feature attribution (FA), which is applied after training a classification model and involves importance or saliency mapping—typically using gradients or activations with respect to the input—such as Grad-CAM \cite{Selvaraju17}, Shapley additive explanations (SHAP) \cite{BhattaraideepSHAP}, and  guided backpropagation \cite{eitel2019guidedbackprop}, among others. Most existing latent-space interpretation approaches in neuroimaging are primarily descriptive, relying on correlation analyses or visual inspection of latent features. While these methods can provide useful insights, they do not offer statistical guarantees regarding the reliability or generalizability of the observed relationships.

Many studies employing these methods concentrate primarily on classification performance \cite{Martinez,Martinez2,Martinez3, Feng2020,Bass2022} or rely on supervised learning to identify patterns associated with clinical labels (i.e. ground truth) particularly in datasets such as Alzheimer's Disease Neuroimaging Initiative (ADNI) where the disease of individuals with MCI can progress to AD  \cite{Biffi,Gorriz}. However, such approaches often limit the interpretability of learned representations and may overlook the exploratory potential of unsupervised models. 

In ML with neuroimaging, robust validation demands careful cross-validation (CV) strategies—ideally nested—to prevent data leakage and promote generalization \cite{Bates2023}. Estimates of statistical uncertainty, such as confidence intervals and variability across folds, are essential when assessing the robustness of both model performance and interpretability outputs \cite{Gorriz2025}. Attribution methods \cite{42,Bass2022}, like feature maps, should be quantitatively evaluated against ground truth or independent biological markers rather than relying solely on visual plausibility. Finally, disentanglement or attribution processes must avoid circular analysis by strictly separating training data from evaluation pipelines. Together, these practices strengthen the credibility and reproducibility of findings in high-dimensional (neuroimaging) contexts.

Most existing latent-space interpretation approaches in neuroimaging are primarily descriptive, relying on correlation analyses or visual inspection of latent features. While these methods can provide useful insights, they do not offer statistical guarantees regarding the reliability or generalizability of the observed relationships. As a result, strong neurobiological conclusions may be drawn from patterns that are not robust or may be driven by high-dimensional effects, noise, or spatial dependencies. This highlights a critical gap in current methodologies: the lack of validation frameworks capable of assessing whether latent representations encode statistically meaningful information.

In this context, there is a growing interest in human-centered machine learning approaches, where model outputs are not used in isolation but are integrated into expert-driven decision-making processes. In neuroimaging, this often takes the form of human-in-the-loop frameworks, where clinicians or imaging specialists review and interpret model predictions, particularly in borderline or uncertain cases. Such approaches require models to provide not only accurate predictions but also interpretable and statistically reliable information that can support expert judgment. In this sense, improving the interpretability and validation of latent representations is a key step toward enabling the integration of deep learning models into clinically meaningful workflows.

In this work, we address this limitation by proposing a validation-oriented framework that moves beyond descriptive analysis toward statistically grounded interpretation of latent representations. This work proposes a simple and interpretable 3D convolutional autoencoder applied to segmented T1-weighted MRI scans to model gray and white matter distributions. Rather than prioritizing architectural complexity or purely predictive performance, the primary contribution of this study lies in the systematic validation and interpretability of the learned representations. The model is trained to reconstruct input volumes while learning a compact latent space that captures meaningful neuroanatomical structure. We analyze this latent space using multiple dimensionality reduction (DR) techniques \cite{Hinton}, which are widely employed in neuroimaging studies \cite{Zubasti2025}to explore high-dimensional representations and facilitate interpretation. In this work, we place particular emphasis on validating the resulting embeddings beyond qualitative visualization. Specifically, we rigorously assess whether the learned latent dimensions and associated relevance maps correspond to anatomically and clinically meaningful brain regions, as defined by a standardized brain atlas \cite{atlas}. This validation is performed using an upper-bounding strategy \cite{Gorrizarxiv}, enabling a quantitative comparison between model-derived regions of importance and established neuroanatomical knowledge. By explicitly correlating model outputs with findings reported in the medical literature, this work moves beyond opaque, highly complex deep learning architectures and instead demonstrates how simpler models can yield explainable and clinically grounded insights, thereby enhancing trust and interpretability in neuroimaging-based learning frameworks.

\subsection{Related work}

Deep generative models for image-to-image translation have advanced significantly offering powerful tools for learning latent representations that disentangle meaningful variations across domains \cite{41,45,49,50,51}. These approaches have been applied across a range of tasks in computer vision and medical imaging \cite{40,52,53,54}, including domain adaptation, unsupervised feature learning \cite{Madni2025}, and modality transfer. One notable example is the architecture proposed in \cite{42}, which introduces a dual-latent space formulation separating content (shared information) from attributes (domain-specific information) along with adversarial mechanisms to enforce this separation during training. Moreover, recent developments in interpretable machine learning have enabled more targeted analysis of neuroimaging data, as illustrated by the framework introduced in \cite{Bass2022}, which combined regression and classification to reveal anatomical patterns associated with neurological phenotypes. 

While such methods hold great promise for individualized brain mapping, they also raise concerns about interpretability and statistical rigor. Specifically, interpreting variation in latent representations as biologically or clinically significant without proper safeguards can lead to biased analyses \cite{Poldrack}. This is especially problematic in the context of high-dimensional neuroimaging data \cite{Snoek}, where minor but widespread structural brain differences may be exaggerated due to overfitting, insufficient correction for multiple testing, or limited CV. Robust statistical validation is essential when using latent representations to ensure that findings reflect genuine, reproducible effects rather than noise or methodological artifacts \cite{Gorgen}. 

As an illustration, the use of Pearson correlation to assess the similarity between feature attribution maps and population-level statistical maps is common practice in neuroimaging model validation. However, this approach has important limitations. First, statistical maps derived from population-level analyses—typically obtained through voxel-wise group comparisons—may not accurately reflect the individual-level anatomical variability that data-driven models are designed to capture. Second, Pearson correlation coefficients in the range of 0.5 are often considered moderate, but in the context of high-dimensional and spatially autocorrelated brain data such values can arise even when the actual anatomical overlap is limited. Consequently, statistically significant correlations may not translate into practically meaningful or interpretable spatial alignment. To ensure more robust validation, it is advisable to complement correlation-based metrics with spatial overlap measures (e.g., the Dice coefficient), assess consistency across CV folds, and consider alignment with external clinical or anatomical references that offer more contextually grounded validation.

\begin{figure}[t]
    \centering
    \begin{subfigure}{\textwidth}
        \centering
        \includegraphics[width=0.75\textwidth]{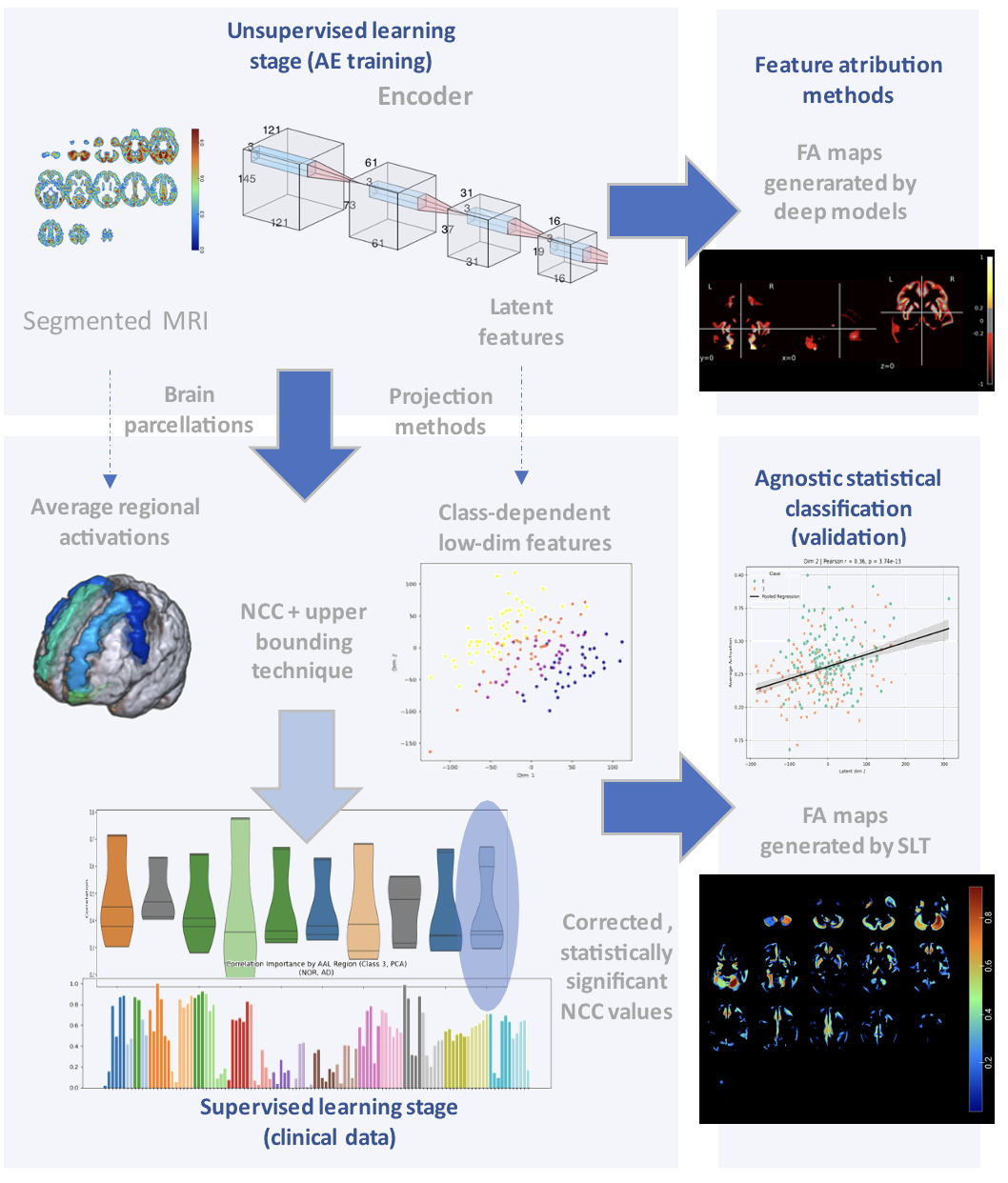}
        \label{fig:subfig1}
    \end{subfigure}
    \caption{Overview of the analysis methods used to provide interpretability of the latent space. FA: feature attribution; NCC: normalized correlation coefficient; SLT: Statistical Learning Theory. Both approaches for estimating FA maps share an unsupervised learning stage in which the autoencoder is trained to efficiently reconstruct the whole brain volume. Our approach employs clinically defined groups to identify region-wise relevance maps by computing correlations and selecting those validated by SLT.}
    \label{fig:overview}
\end{figure}

Our work aligns with a growing perspective in cognitive computational neuroscience that emphasizes exploration as a fundamental and underappreciated function of deep neural networks \cite{Cichy,Cichy2016}. We directly utilize the latent features, used by the generative model (decoder) for classification and group comparisons, without requiring reconstruction or explicit attribution (see figure \ref{fig:overview}). This enables a more streamlined analysis pipeline while capitalizing on the representational efficiency of the latent space. Specifically, we adopt a data-driven framework to interrogate the latent space of a 3D convolutional AE trained on tissue-segmented structural brain images, not only to reconstruct inputs but also to discover meaningful anatomical and diagnostic relationships through unsupervised embedding. By leveraging interpretable techniques such as region-wise error attribution and SHAP-based regression analysis \cite{Chatterjee} as a baseline, we aim to uncover latent representations that are scientifically informative rather than solely predictive, contributing to the broader agenda of understanding how deep learning models can function as tools for model-based hypothesis generation and neuroscientific discovery. 

\section{Materials and Methods}

\subsection{Dataset and preprocessing}
Data used in preparation of this paper were obtained from the ADNI database (adni.loni.usc.edu). The database contains T1-weighted MRI scans acquired on 1.5 T and 3.0 T MRI scanners from individuals with a diagnosis of AD or MCI, or were enrolled as cognitively normal controls (NOR). Images were acquired longitudinally at multiple time points. For this study, we only included 1.5 T structural MRI (sMRI) scans. The original database contained over 12,000 T1-weighted MRI images, including 229 NOR (Class 0), 188 AD (Class 3), 252 MCI (Class 1), 149 participants whose disease progressed from MCI to AD: known as 'converters' (MCIc) (Class 2). 
All images obtained from the ADNI database were processed using a standardized and widely validated preprocessing pipeline consistent with prior ADNI-based studies. Specifically, images underwent correction for intensity non-uniformity (bias field correction), followed by tissue segmentation into gray matter (GM), white matter (WM), and cerebrospinal fluid (CSF) using the CAT12 toolbox implemented in SPM12. The resulting GM maps were spatially normalized to the MNI standard space using DARTEL, modulated to preserve regional volumetric information, and smoothed with an isotropic Gaussian kernel of 8 mm full width at half maximum (FWHM) to account for inter-subject anatomical variability and improve signal-to-noise ratio. Automated and visual quality control procedures provided by CAT12 were applied to identify and exclude images with segmentation errors, motion artifacts, or extreme total intracranial volume values, ensuring a homogeneous and reliable dataset. This preprocessing workflow, which follows established conventions for ADNI 1.5 T MRI data, enhances reproducibility and strengthens the internal validity of the subsequent analyses and representation learning. For this study, only the first medical examination of each participant was considered, resulting in a total of $N = 818$ segmented gray matter (GM) images after standard preprocessing using CAT12 \cite{CAT12} and SPM12 \cite{SPM12}. Demographic data are summarized in table \ref{tab:demog}. 

\begin{table}[ht]
\centering
\caption{Demographics details of the ADNI dataset with group means and their standard deviation}
\label{tab:demog}
\begin{tabular}{@{}lcccr@{}}
\toprule
Status & Number & Age & Gender (M/F) & MMSE \\ 
\midrule
NOR  & 229 & 75.97$\pm$5.0  & 119/110 & 29.00$\pm$1.0 \\
MCI  & 252 & 75.27$\pm$7.25 & 157/95  & 26.85$\pm$2.39 \\
MCIc & 149 & 74.01$\pm$7.03 & 97/52   & 26.97$\pm$1.77 \\
AD   & 188 & 75.36$\pm$7.5  & 99/89   & 23.28$\pm$2.0 \\
\bottomrule
\end{tabular}
\end{table}

\subsection{Model architecture}

\begin{figure}[t]
    \centering
    \begin{subfigure}{\textwidth}
        \centering
        \includegraphics[width=0.75\textwidth]{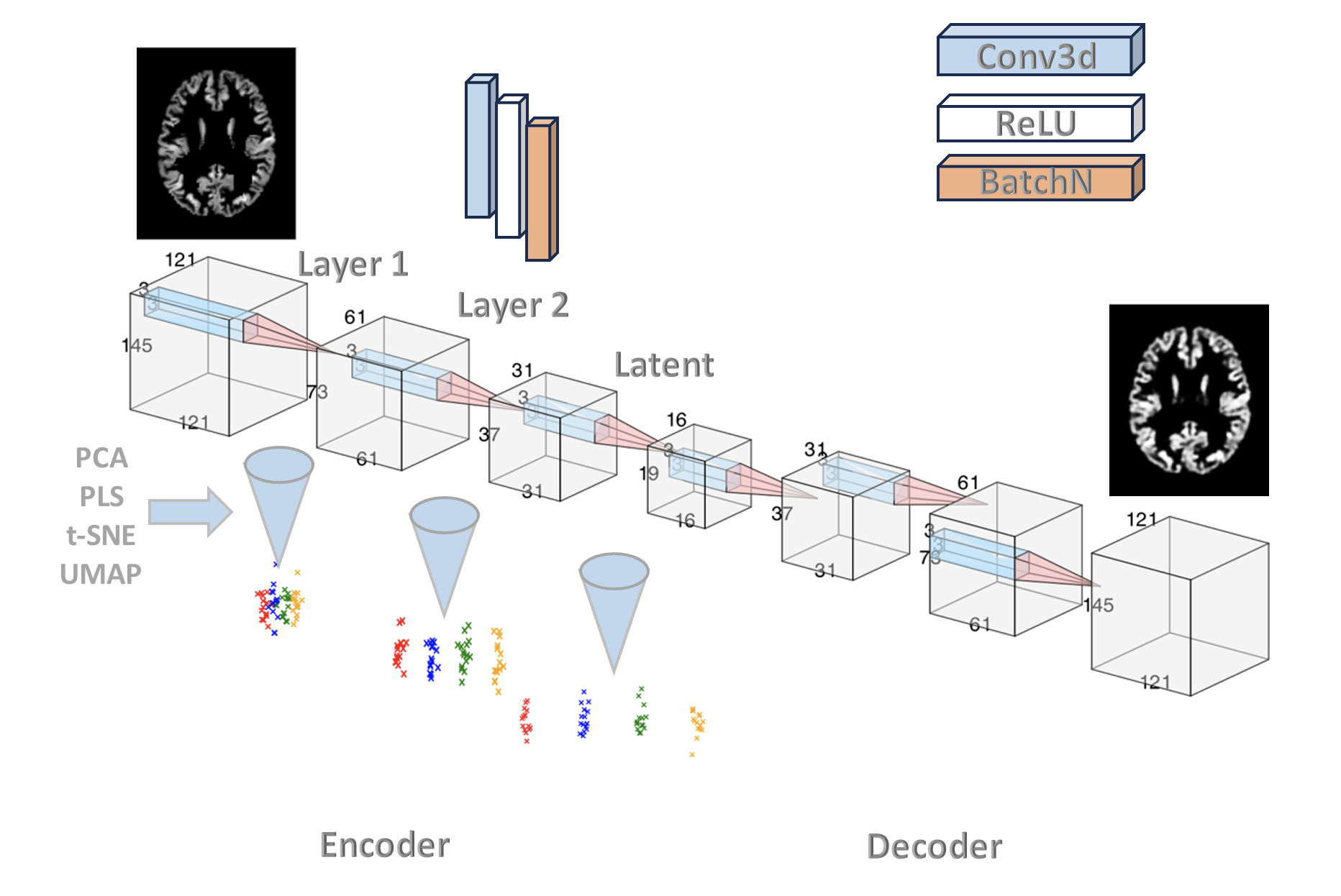}
        \label{fig:subfig2}
    \end{subfigure}
    \caption{Overview of the neural architecture based on Autoencoder (AE). The model compresses an input volume of size $1\times 121\times 145 \times 121$ through an encoder consisting of 3D convolutional layers with ReLU activations and batch normalization. The feature map dimensions progressively reduce as follows: 
    $x_1: [16, 61, 73, 61], x_2: [32, 31, 37, 31], x_3: [64, 16, 19, 16]$, forming the latent representation $z$. The decoder reconstructs the original volume using 3D transposed convolutions, gradually restoring spatial dimensions until the output matches the input size, with a final sigmoid activation to preserve MRI intensity. Intermediate tensors $(x_1, x_2, x_3)$ illustrate feature extraction at multiple levels of abstraction.}
    \label{fig:architecture}
\end{figure}

We implemented a three-dimensional convolutional AE (see figure \ref{fig:architecture}) using PyTorch to learn compact representations of volumetric brain MRI data. The encoder consisted of three sequential 3D convolutional layers with kernel size 3, stride 2, and padding 1, increasing the number of channels from 1 to 16, 32, and 64, respectively. Each convolutional layer was followed by a ReLU activation and batch normalization to promote stable and efficient training. The decoder mirrored this architecture employing three 3D transposed convolutional layers (ConvTranspose3d) with the same kernel size and stride sequentially reducing the number of channels from 64 to 32, 16, and finally 1. Each transposed convolution was followed by a ReLU activation and batch normalization, except for the final layer which used a sigmoid activation to constrain the output values between 0 and 1. The model was trained end-to-end to minimize the mean squared error (MSE) between the input and reconstructed images. Given that GM probability maps are inherently smooth and lack sharp structural boundaries, MSE proved to be an adequate loss function as it did not significantly distort the underlying anatomical information. This architecture enabled the extraction of hierarchical and spatially meaningful latent features from 3D neuroimaging data, facilitating downstream analyses such as clustering or classification in the learned latent space.

\subsection{Low-dimensional projection methods}

To better explore and interpret the latent representations learned from MRI data, we employed several DR techniques—both linear and non-linear, namely Principal Component Analysis (PCA), Partial Least Squares (PLS), t-distributed Stochastic Neighbor Embedding (t-SNE) \cite{Maaten}, and Uniform Manifold Approximation and Projection (UMAP) \cite{McInnes}. These methods facilitate both visualization and quantitative analysis of how latent features relate to anatomical or clinical patterns (see appendix \ref{app:projection}). These DR techniques served not only to enable visual inspection of the latent space, but also to identify class-separable structures, assess correlations with region-wise anatomical signals (e.g., via the AAL atlas), and gain insights into how neurodegenerative conditions manifest within the learned feature spaces. For the downstream analyses, we adopted an upper bounding strategy (as shown in section \ref{sec:agnostic}) to ensure that the evaluation remained extrapolable and to correct for distortions introduced by the low-dimensional projections (e.g., t-SNE, UMAP). This approach allowed us to assess the representational structure of the latent space while accounting for the intrinsic variability and possible errors arising from DR.

\subsection{Model-agnostic statistical validation}\label{sec:agnostic}

In neuroimaging studies, ML models are commonly evaluated using standard K-fold CV. However, when dealing with limited, imbalanced, or heterogeneous samples—as is often the case in clinical applications such as Alzheimer’s disease (AD) research—predictive validation strategies based on CV may yield unstable or overly optimistic estimates of model performance. These limitations stem from the sensitivity of CV to the choice of the number of folds  $K$, which entails a trade-off between increased variance for large $K$ (small test sets) and increased bias for small $K$ (large test sets), as well as from the lack of explicit guarantees on the true generalization error. Such issues become particularly relevant when interpreting learned latent representations or when drawing conclusions from region-level or feature-level analyses, where statistical significance alone does not necessarily imply robustness or practical relevance. To address these challenges, we adopt a conservative, theoretically grounded validation strategy based on the CV Upper Bound on the actual risk (CUBV) \cite{Gorrizarxiv}. This approach provides an explicit upper bound on the generalization error, thereby offering a principled way to assess the robustness of detected patterns beyond purely empirical performance estimates.

To investigate population-level group differences across the clinical spectrum, we adopt an inferential rather than a predictive framework. Features are first extracted using a deep model trained in an unsupervised manner, ensuring that no diagnostic labels are used during representation learning. Linear classifiers are then fitted on the full dataset to quantify group separability in the learned latent space for several contrasts, including NOR–AD, NOR–MCI, NOR–MCIc, and the multi-class setting NOR–MCI–MCIc–AD. Rather than estimating out-of-sample predictive performance, we focus on a conservative theoretical characterization of group separation by computing worst-case upper bounds on the classification error. These bounds provide probabilistic guarantees that, with high confidence, the true population error does not exceed the observed empirical error plus a complexity-dependent term. Under the null hypothesis of no group effect, the expected population error corresponds to chance level for the corresponding classification task. Consequently, demonstrating that the worst-case error bound remains below chance supports the presence of genuine population-level differences among groups, even under highly conservative assumptions. While this approach does not rely on empirical estimates of out-of-sample generalization performance, permutation testing can be used to obtain classical p-values, and the combination of both analyses provides robust inferential evidence that the observed separations across diagnostic groups are unlikely to arise from random variability alone. A detailed theoretical formulation based on SLT is provided in appendix and previous works \cite{Gorrizarxiv}.

\section{MRI analysis for AD progression}

\subsection{Reconstruction analysis}
\begin{figure}[t]
    \centering
    \begin{subfigure}{\textwidth}
        \centering
        \includegraphics[width=0.7\textwidth]{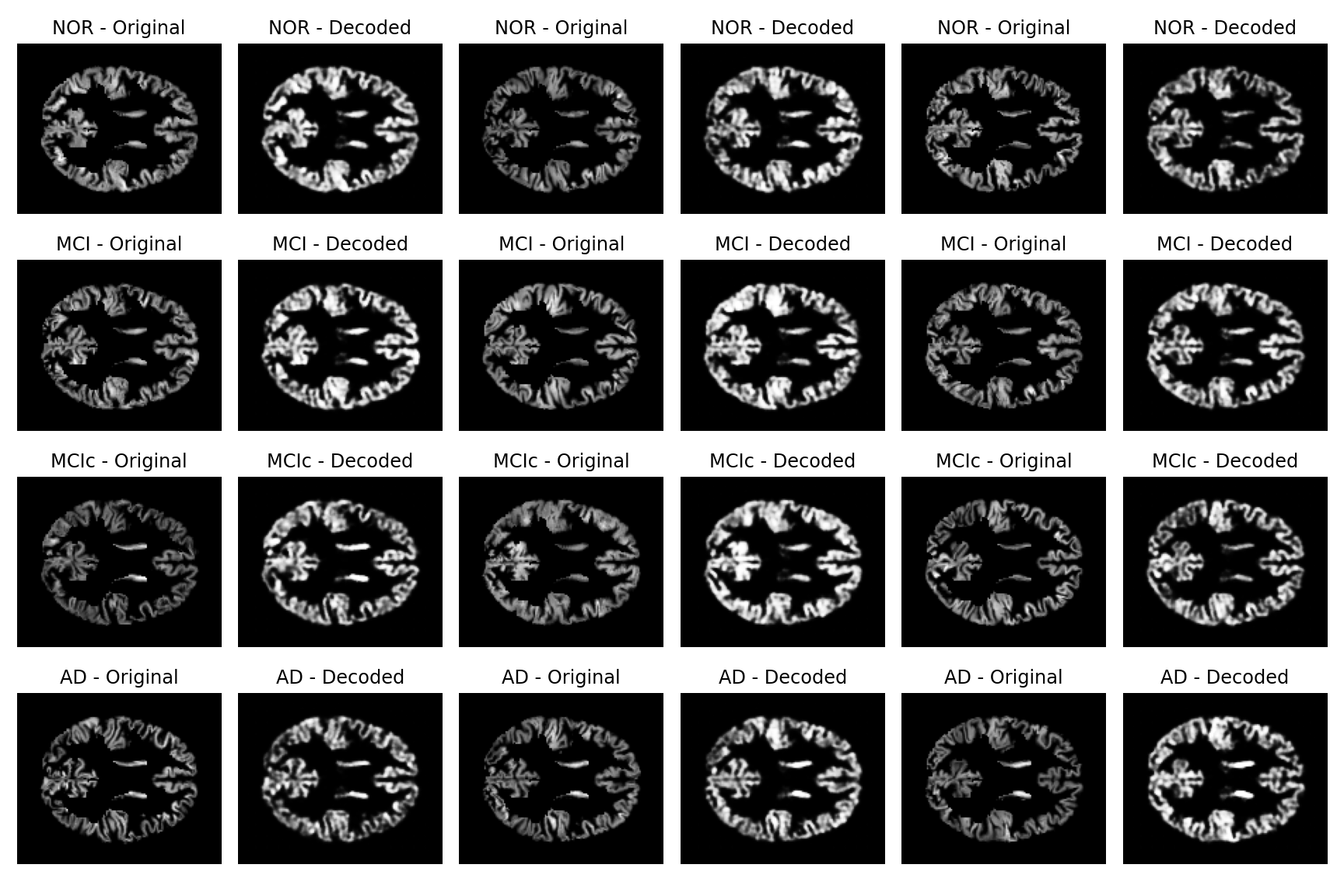}
        \caption{MRI GM input and MSE latent reconstructions.}
        \label{fig:subfig4}
    \end{subfigure}
        \begin{subfigure}{0.7\textwidth}
        \centering
        \includegraphics[width=\textwidth]{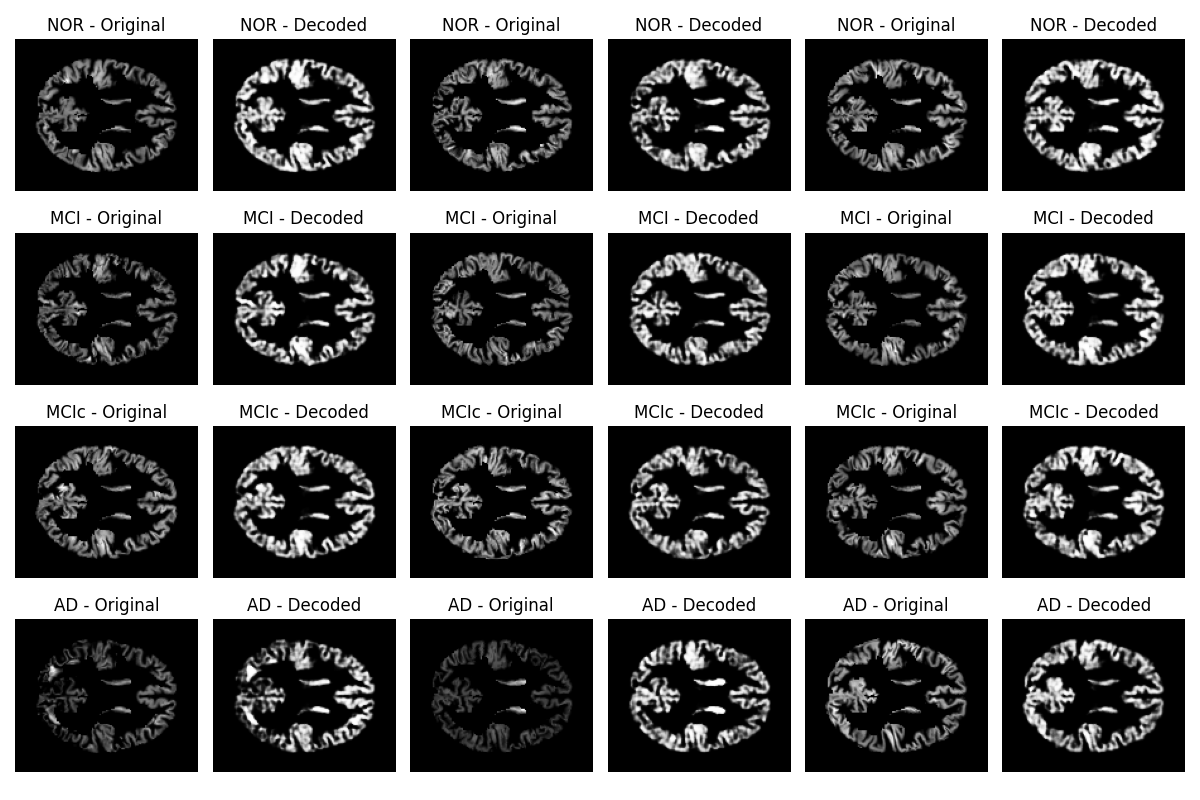}
        \caption{MRI GM input and SSIM+MSE latent reconstructions.}
        \label{fig:subfig5}
    \end{subfigure}
    \caption{Reconstruction quality comparison between MSE (10 epochs) and the combined loss (20 epochs). Each row presents three representative examples per class, displaying the original image alongside its reconstruction, visualized at the $Z=60$ slice.}
    \label{fig:reconstruction}
\end{figure}

Input images were segmented GM maps derived using the CAT12 \cite{CAT12} toolbox for SPM12 \cite{SPM12}. Unlike full volumetric brain scans, these images represented only the distribution of GM tissue constituting a subset of the full voxel space. As a result, the AE was trained and evaluated on anatomically constrained data, focusing specifically on regions relevant for morphological analysis. Although the original T1-weighted MRI scans had a resolution of 121 × 145 × 121 voxels, the effective number of voxels involved in the reconstruction loss was significantly smaller, limited to those classified as GM by the CAT12 segmentation pipeline. This spatial sparsity increased the interpretability of the reconstruction loss, as the model was optimized to preserve the structure of clinically meaningful brain tissue (see figure \ref{fig:reconstruction}).

Given that CAT12 GM images represented GM concentrations or volumes that were not normalized, the MSE values should be interpreted accordingly. In this context, achieving an MSE consistently below 0.01 after 10 training epochs indicates a very low voxel-wise squared difference between original and reconstructed GM maps reflecting a high similarity. A lower MSE thus corresponds to more accurate reconstructions; that is, the AE effectively preserved the anatomical detail of the GM tissue.

The consistently low MSE indicated that the encoder had successfully learned a compact latent representation capable of reconstructing the essential structural features of the GM distribution. This latent space could therefore be meaningfully analyzed in relation to standardized brain regions, such as those defined by the AAL atlas. As the reconstruction process was restricted to GM voxels—where neurodegenerative effects are often most pronounced—the learning signal was more focused and anatomically specific. Consequently, achieving MSE values consistently below 0.01 supports the model’s capacity to encode diagnostically relevant morphological information in a class-sensitive latent space, reinforcing its suitability for subsequent anatomical correlation analyses and group-level comparisons.

\subsection{Atlas-based SHAP/correlation analysis per class}

To assess whether an autoencoder-based neural network is able to capture differences between subjects with distinct clinical conditions through the reconstruction of segmented GM images, we employed SHAP-based feature attribution analysis. If such differences are effectively encoded by the model, the relevance assigned by SHAP should vary across anatomical regions \cite{atlas} and preferentially highlight those brain areas that have been consistently reported as targets of neurodegeneration in the medical literature. Under this hypothesis, region-wise SHAP values are expected to differ between clinical groups, reflecting group-specific reconstruction strategies learned by the network. To test this, we relied on the previously defined clinical groupings and atlas-based regional parcellations, using them to identify which regions contribute most to reconstruction performance in each condition. This approach allows us to go beyond trivial effects driven by large regions or global reconstruction error, and instead focus on anatomically meaningful areas where the network either allocates greater explanatory relevance or struggles to accurately model disease-related structural alterations. Once this behavior is demonstrated, it becomes meaningful to further study the relationship between the low-dimensional latent space and the mean intensity values of each anatomical region, in order to analyze how such relationships are encoded across different clinical groups. To ensure the robustness of this analysis, we incorporate an upper-bounding strategy that protects against spurious associations and false positives that may arise when relying solely on classical model validation or correlation-based approaches.

We sought to further explore how the learned latent representations relate to standardized anatomical structures, such as those defined by the AAL brain atlas \cite{atlas}. For each trained model, we extracted the latent vectors associated with the input images and computed Pearson correlation coefficients between these features and the corresponding gray matter intensities averaged across each AAL region. This approach allowed us to assess the extent to which the latent space retained anatomically meaningful information. Importantly, discrepancies in the correlation patterns between groups would suggest that the encoder adapted differently to the spatial features in each scenario. In other words, if the latent-regional GM correlations were to vary systematically across groups, it would indicate that the encoder emphasizes different brain regions depending on the group to ensure sufficiently accurate reconstructions. Such group-specific adaptations in the latent space reflect how the model implicitly learned to prioritize certain anatomical areas that were more informative or discriminative for reconstructing the brain images of each clinical comparison. Pearson correlation was chosen as it is the primary statistical measure in the extant literature for detecting and quantifying these latent-region associations and their potential divergence across groups.


\subsection{Corrected correlation analysis with statistical agnostic regression}

\begin{figure}[]
  \centering
\includegraphics[width=\textwidth]{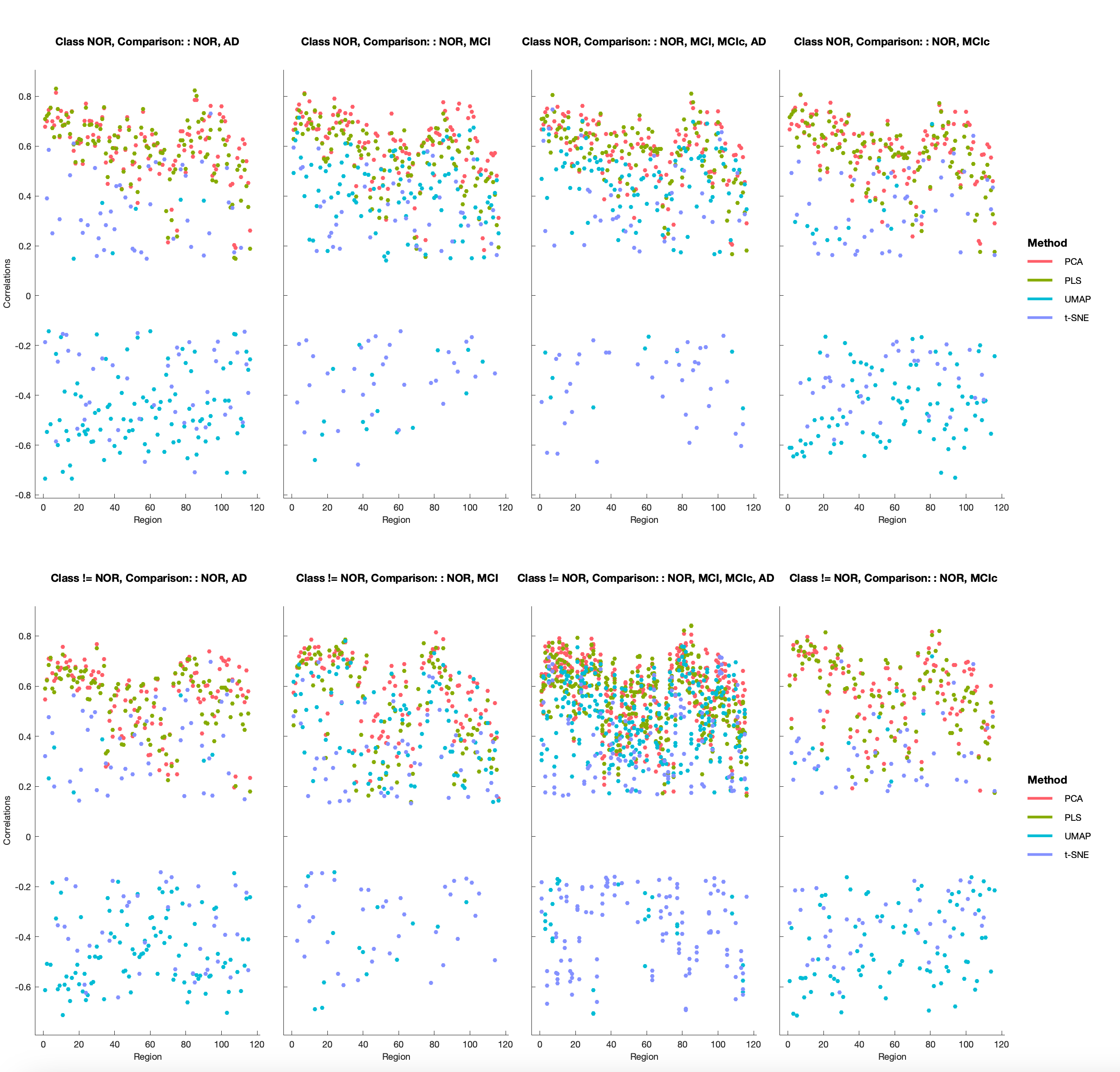}
\caption{Correlation analysis including all comparisons and anatomical AAL regions is shown for the normal class (top row) and the remaining classes (bottom row). Even small correlations are found to be significant across different methods in component 1 and the latent layer. }
\label{fig:corranal}
\end{figure}

\begin{figure}[]
  \centering
\includegraphics[width=\textwidth]{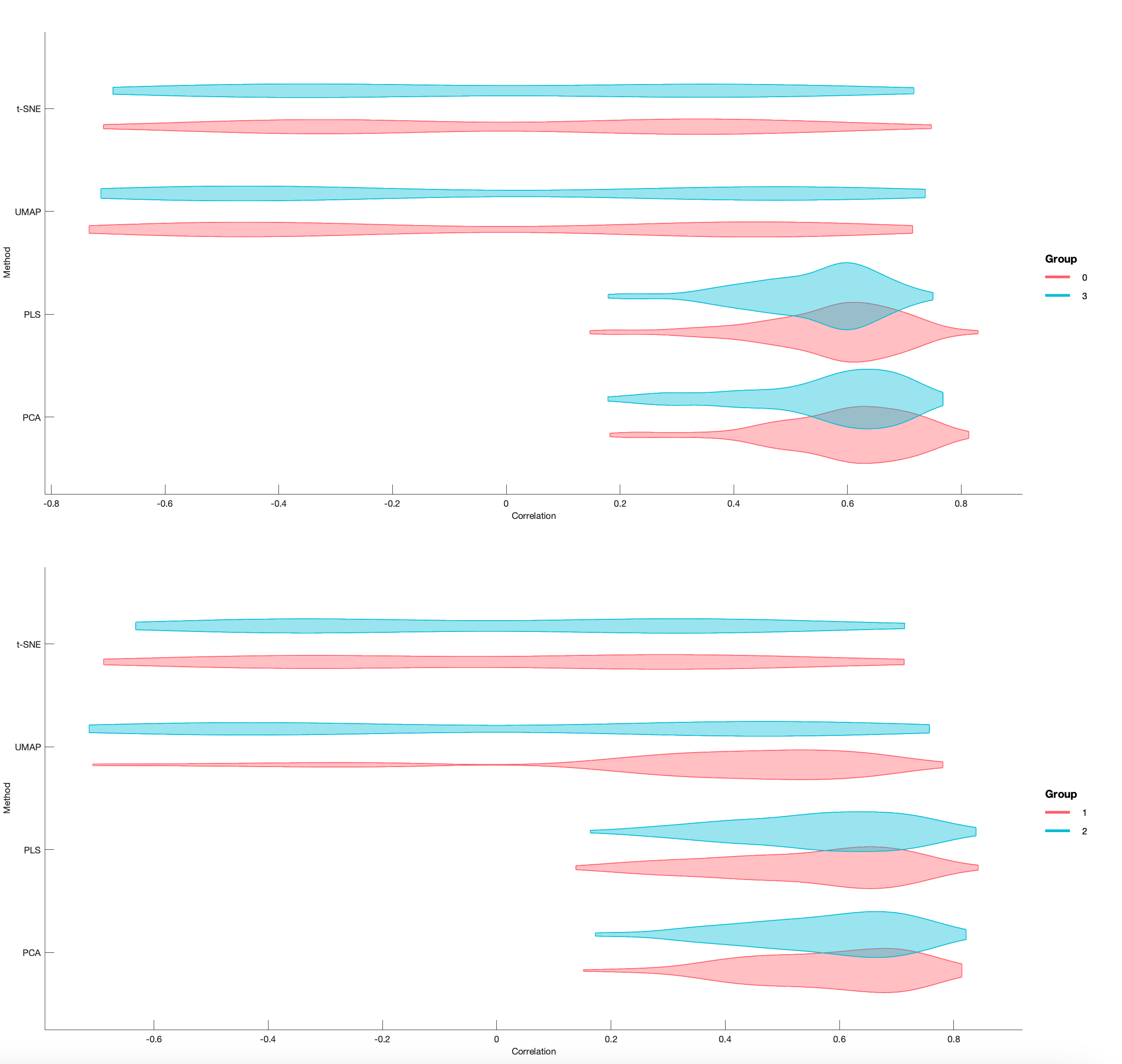}
\caption{Correlation analysis including two comparisons (AD-NOR and MCI-MCIc) and anatomical AAL regions (continuation). The distribution shapes by method and class are displayed.}
\label{fig:corranal2}
\end{figure}

With $300$ samples, even relatively small correlations (e.g., $|r| > 0.11$) can reach statistical significance using Pearson correlation. In terms of $R^2$, this means that only about 1.2\% of the variance in one variable is explained ($R^2 \approx 0.012$), indicating a very low explanatory power despite statistical significance (see figures \ref{fig:corranal} and \ref{fig:corranal2}). This explains why significant results often appear in large datasets even when the actual effect size is small. Such sensitivity to sample size underscores the limitations of using p-values alone for inference. This motivates the use of more robust statistical approaches such as statistical agnostic regression (SAR) \cite{Gorriz2025} that corrects statistical significance by integrating both effect size and multiple comparisons, offering a more reliable assessment of relevance.

Again, for each subject, the latent vectors were obtained from different layers of the model, and correlations were computed separately for each class label. Significant correlations were identified based on associated p-values and optionally corrected using SAR. To aid interpretation, we generated two types of visualizations per class and method (see appendix \ref{app:shap-corr}): bar plots summarizing the normalized mean absolute correlation values across regions, and violin plots showing the distribution of raw correlation values for the top-N most relevant regions. These plots highlight which anatomical regions most consistently relate to specific latent features across different diagnostic groups, providing biologically meaningful insights into the model’s internal representations.

\subsection{Latent–regional correlation profiling analysis}

To further characterize how specific latent dimensions related to regional brain features, we performed a Latent–Regional Correlation Profiling (LRCP) analysis. Pearson correlations were calculated between each latent component and the regional signal across all participants, visualized via scatter plots with pooled regression lines. Additionally, we assessed the discriminative power of each latent–region pair using a statistical upper-bound significance test \cite{Gorrizarxiv}, optionally corrected using SAR \cite{Gorriz2025}, to evaluate whether the observed associations  meaningfully separated diagnostic classes. Plots were represented for each component–region pair, categorized as significant or non-significant based on the upper bounding approach. This process enabled the identification of latent dimensions that consistently encode biologically or diagnostically relevant regional variations.

\section{Results}

Experiments leveraged a Dell server equipped with 4 NVIDIA H100 Tensor Core GPUs connected via NVLink, enabling high-throughput deep learning on large neuroimaging datasets. Given the simplicity of the proposed 3D autoencoder architecture, training times were moderate, with per-epoch durations on the order of minutes depending on dataset size, enabling efficient and reproducible experimentation.

\subsection{Relevant maps using correlation values}

As a critical baseline, we reported classification results based on latent features in \cite{Martinez}. We further provide visualizations of the maximum Pearson correlation values between latent-space activations—projected onto low-dimensional spaces—and region-wise average intensities, stratified by class and fused with gray matter (GM) MRI (figures~\ref{fig:PCAfus} and \ref{fig:tsnefus}). Additional results using SHAP are presented in appendix~\ref{app:shap}. These results explore the correspondence between network activations and anatomical signal distributions and whether it differs across clinical conditions (e.g., AD vs. NOR). This offers a transparent alternative to assess model interpretability and underscores the need for more rigorous and nuanced validation practices in the field. An inspection of the groups and regions with the highest correlations revealed overlapping areas across clinically relevant comparisons in image reconstruction, as summarized for the t-SNE–based projections in table~\ref{tab:region_overlap}. These regions correspond closely to those identified through the SHAP analysis presented in the appendix (table~\ref{tab:shap_regions}).

\begin{figure}[]
 \centering
\includegraphics[width=0.49\textwidth]{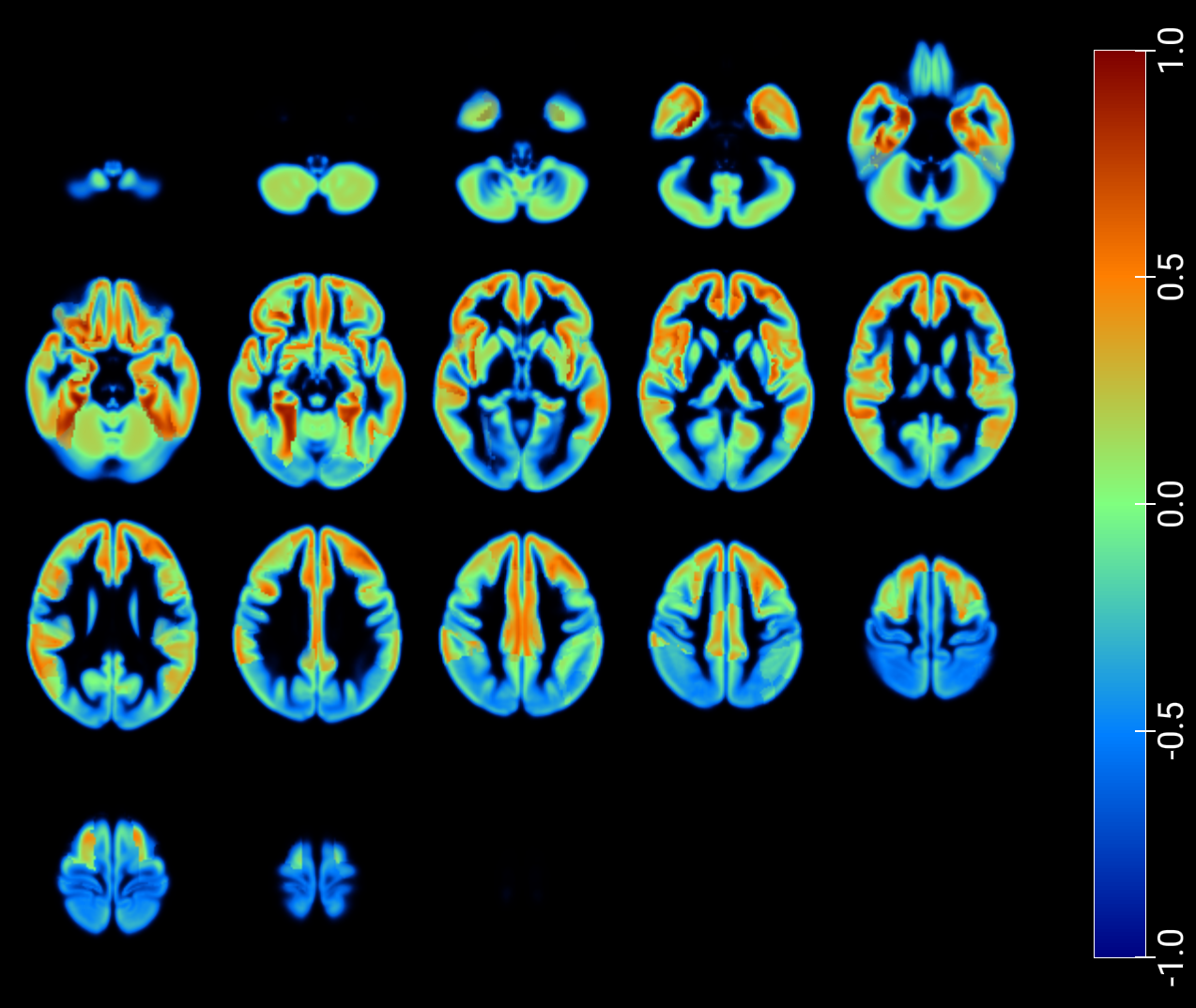}
\includegraphics[width=0.49\textwidth]{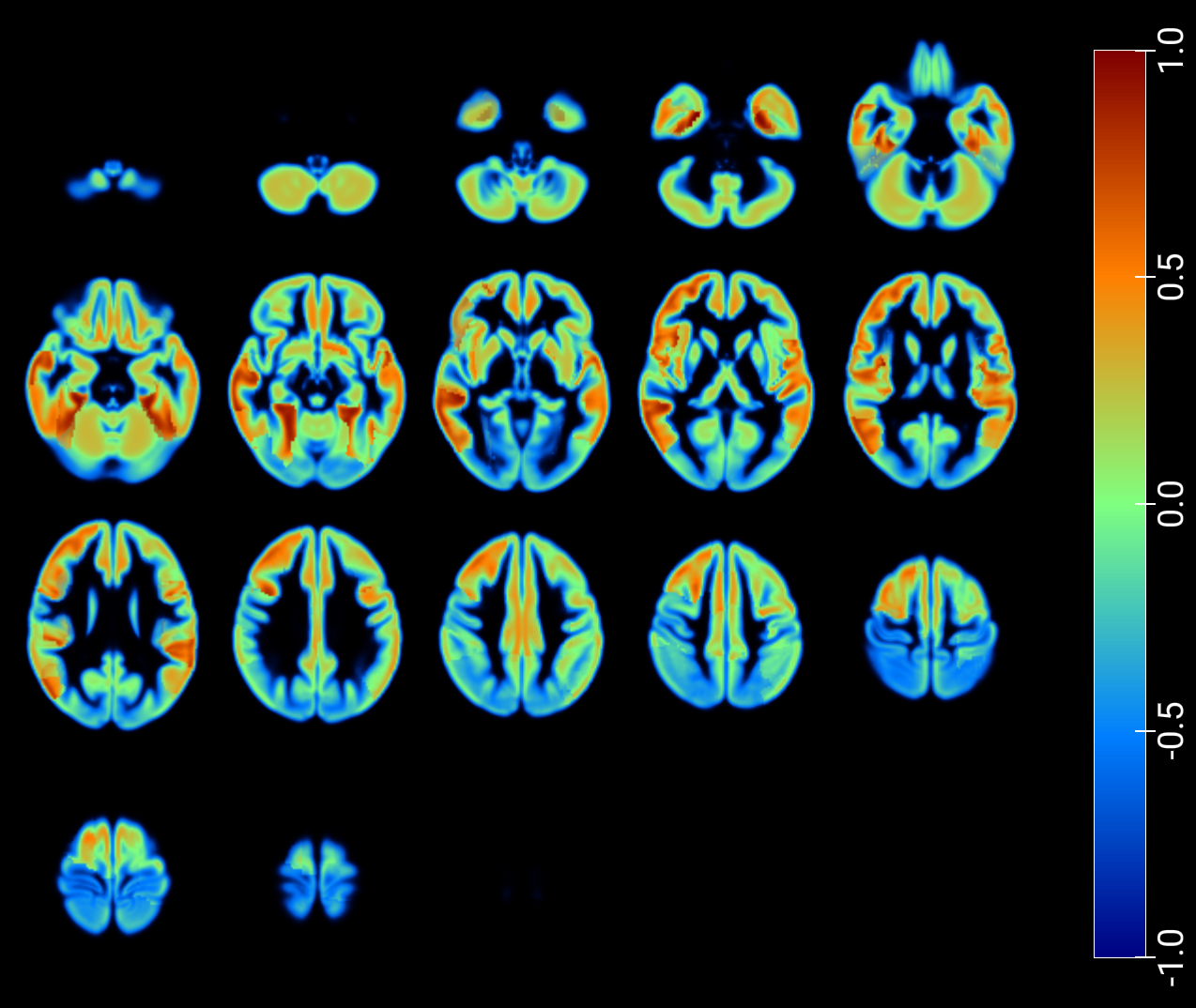}
\caption{Fused neuroanatomical visualization of significant latent-to-anatomy correlations (PCA method, component 1, latent layer). The left panel corresponds to the NOR (cognitively normal) group, while the right panel shows results for the AD group. Each map displays the overlay of significant Pearson correlation values ($p < 0.05$) between latent-space activations and region-wise AAL intensities fused with a high-resolution anatomical MRI image. This fusion enhances interpretability by localizing deep feature correlations within brain structures, stratified by class. The results illustrate how distinct clinical conditions may involve different anatomical substrates reflected in the latent representations learned by the model.}
  \label{fig:PCAfus}
\end{figure}

\begin{figure}[t]
 \centering
\includegraphics[width=0.49\textwidth]{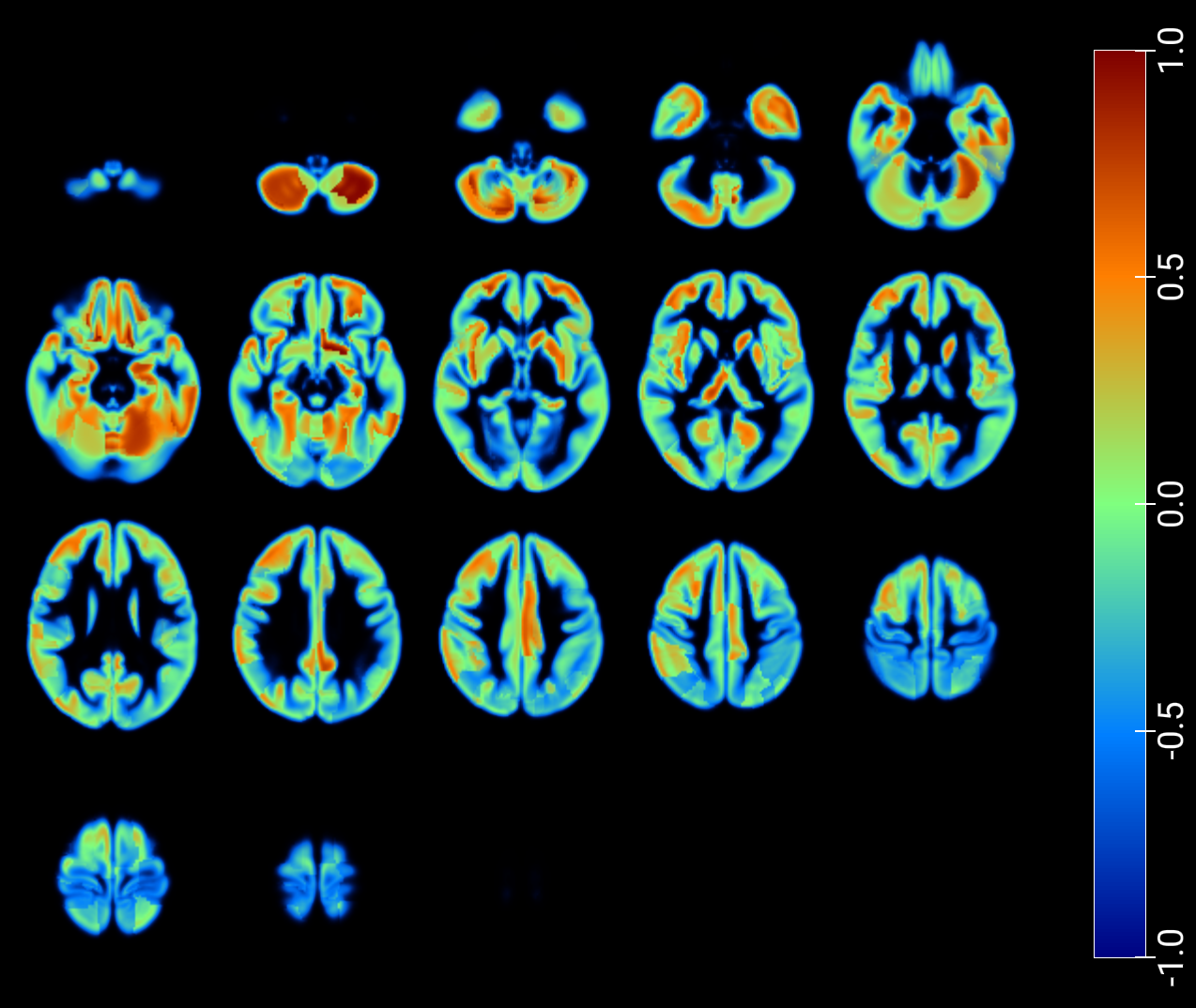}
\includegraphics[width=0.49\textwidth]{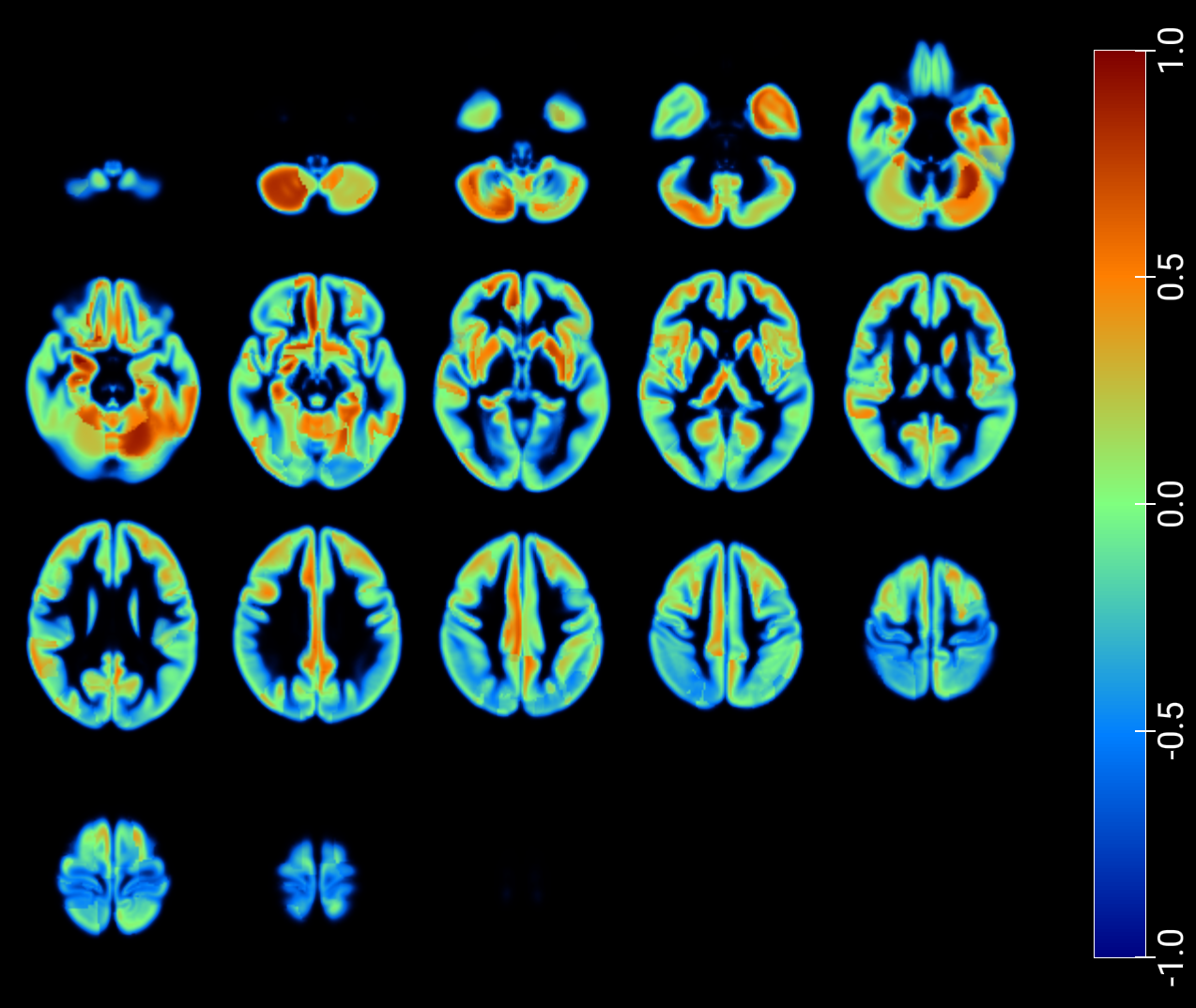}
\caption{Fused neuroanatomical visualization of significant latent-to-anatomy correlations (t-SNE method, component 1 for NOR and AD classes at the latent layer.)}
\label{fig:tsnefus}
\end{figure}

\begin{table*}[]
\centering
\renewcommand{\arraystretch}{1.2}
\begin{tabular}{ccl}
\toprule
\textbf{Group 1} & \textbf{Group 2} & \textbf{Common Regions (Top 10 Correlation)} \\
\midrule
NOR, AD         & NOR, MCI         & Cingulum\_Mid\_R, \textbf{Frontal\_Mid\_L}, \textbf{Insula\_R} \\
         & NOR, MCI, MCIc, AD & Cerebelum\_Crus2\_R, \textbf{Frontal\_Mid\_L}, \textbf{Insula\_R} \\
         & NOR, MCIc        & \textbf{Frontal\_Mid\_L}, Frontal\_Mid\_R, Frontal\_Sup\_L \\\midrule
NOR, MCI        & NOR, MCI, MCIc, AD & Frontal\_Inf\_Oper\_L, \textbf{Frontal\_Mid\_L}, \\
       &  &  \textbf{Insula\_R}, \textbf{Temporal\_Mid\_L} \\
        & NOR, MCIc        & \textbf{Frontal\_Mid\_L}, \textbf{Temporal\_Mid\_L} \\\midrule
NOR, MCI, MCIc, AD & NOR, MCIc     & \textbf{Frontal\_Mid\_L}, Heschl\_L, \textbf{Temporal\_Mid\_L} \\
\bottomrule
\end{tabular}
\caption{Overlap of the top 10 regions with highest correlation across pairwise clinical group comparisons. Note: Regions highlighted in \textbf{bold} appear repeatedly across different comparisons.}
\label{tab:region_overlap}
\end{table*}

\subsection{Correction of correlation-based analyses}

\begin{figure}[t]
\centering
\includegraphics[width=0.75\textwidth]{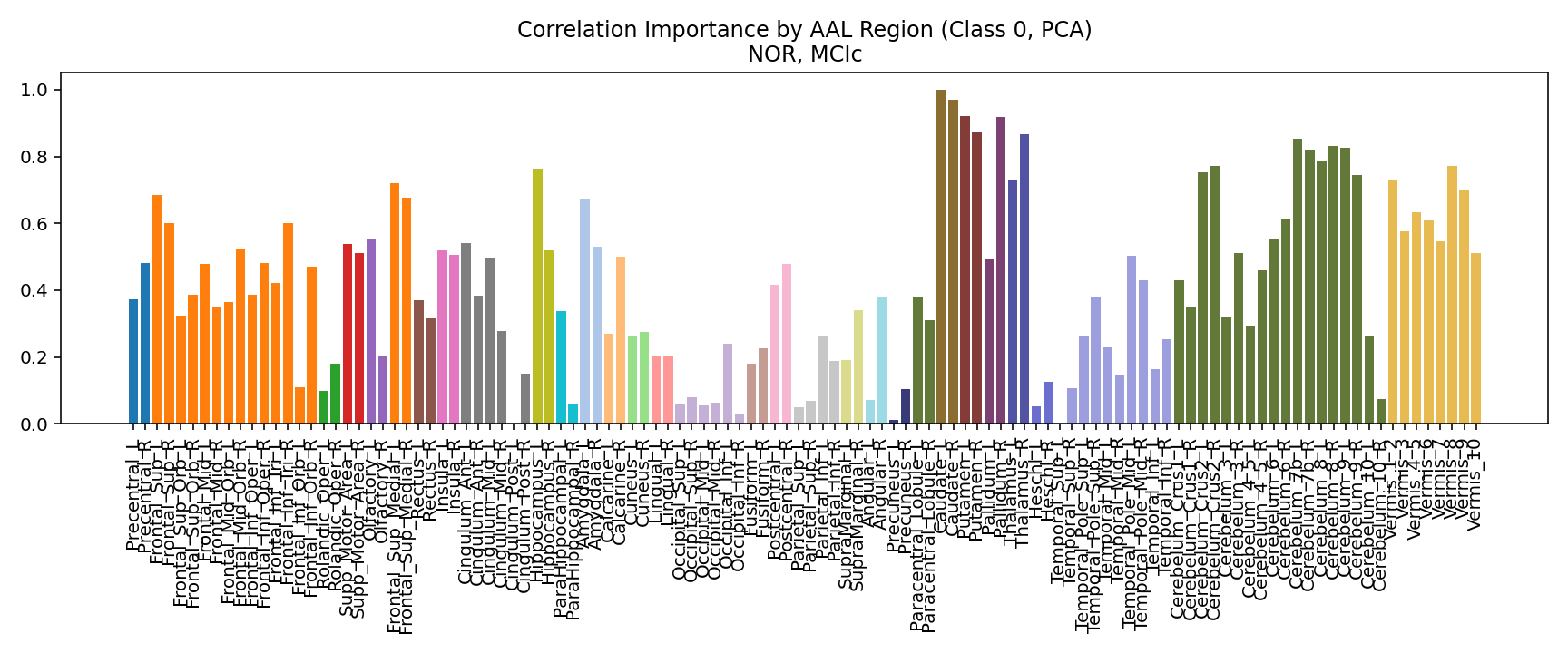}
\includegraphics[width=0.75\textwidth]{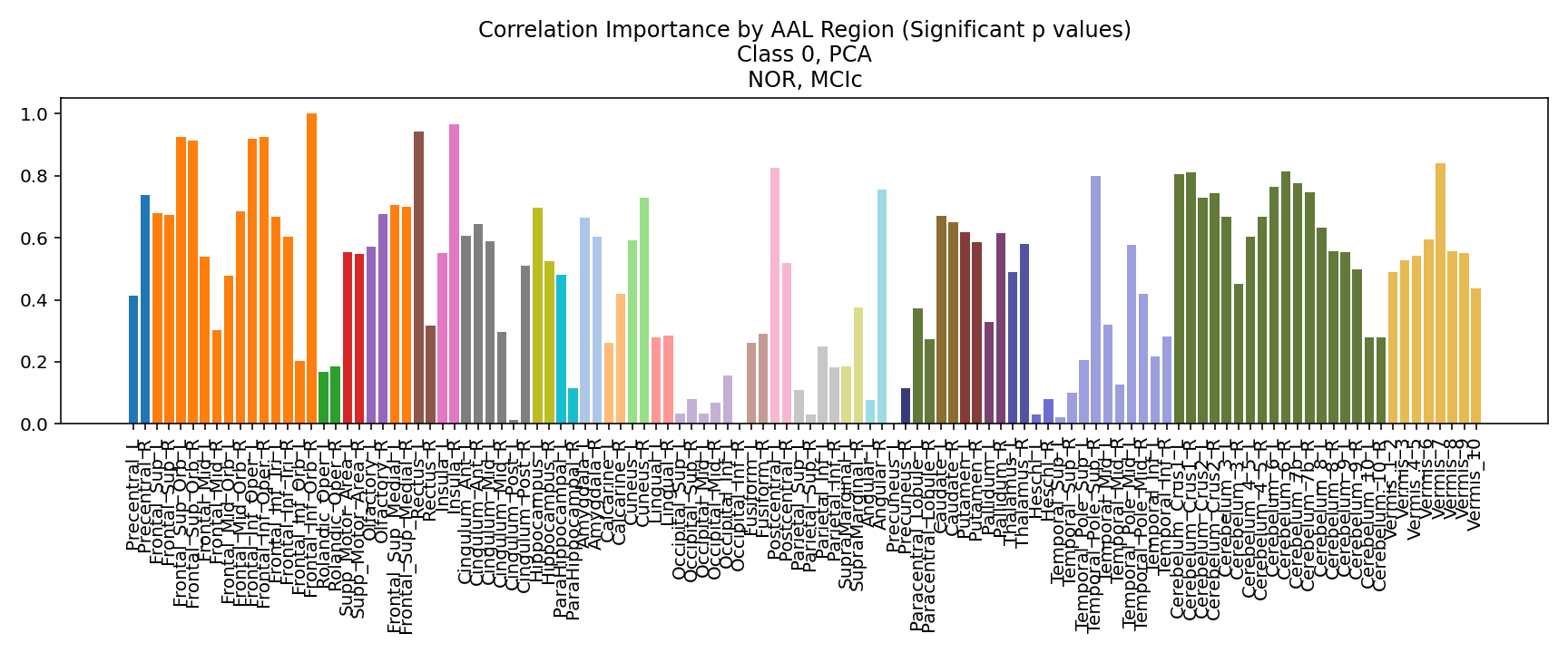}
\includegraphics[width=0.75\textwidth]{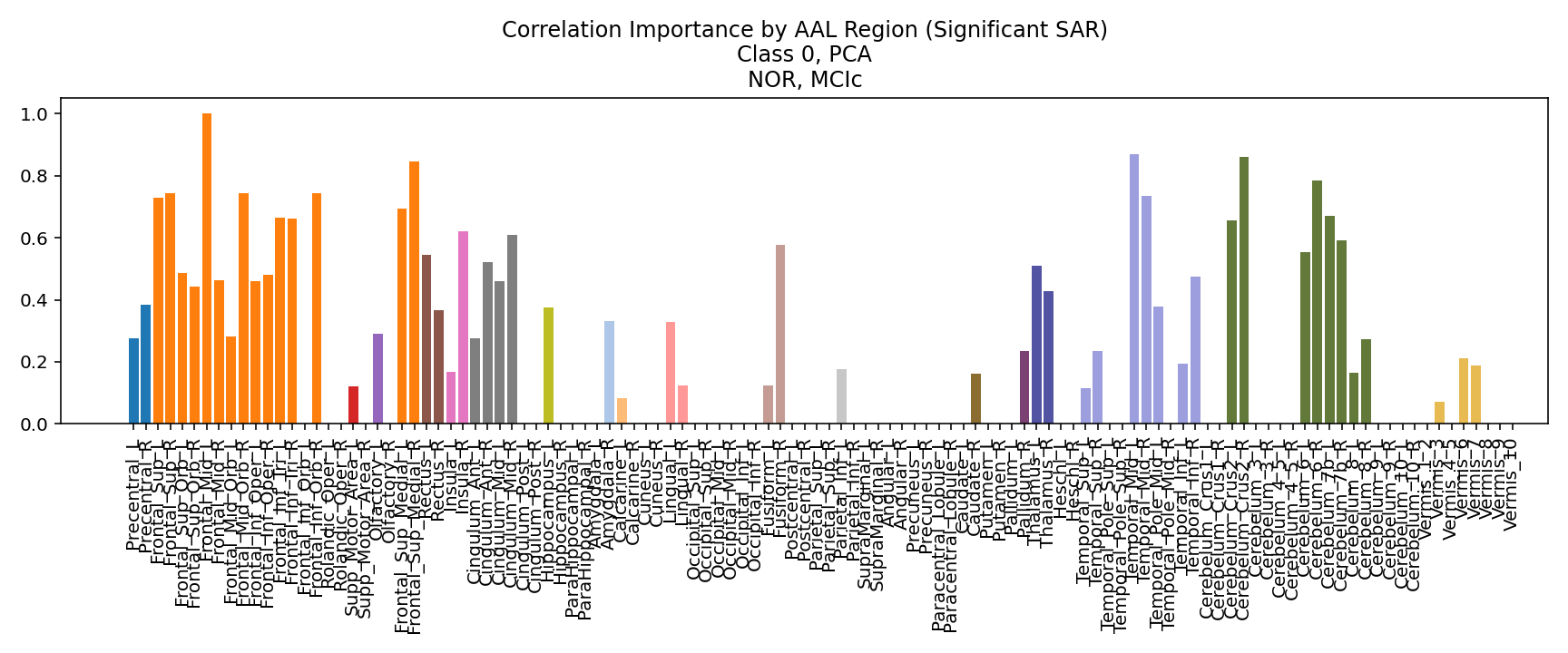}
\caption{Correlation importance for the NOR–MCIc comparison, showing (top) raw correlation values between latent-space features and AAL regions, (middle) results after applying significance testing (p-values), and (bottom) results after applying Statistical Agnostic Regression (SAR) for further bias correction. The progressive refinement highlights how uncorrected correlations can overestimate regional importance, while significance filtering and SAR lead to more robust and interpretable patterns. Colors indicate the anatomical brain region associated with each AAL region.}
\label{fig:corranal_NORMCIc}
\end{figure}

An analysis was performed using the correlation-based methodology described in previous sections. In that case, we focused on how the use of correlations in neuroimaging—or any statistical measure derived from high-dimensional manifolds—should be treated with caution. Raw correlation values between latent features and anatomical structures may appear high, but without statistical control such results can be misleading due to noise, multiple comparisons, or spurious associations. To illustrate this, we present three figures for the NOR–MCIc comparison in figure \ref{fig:corranal_NORMCIc}. The first shows the uncorrected mean correlation values, where inflated magnitudes are evident. The second applied significance testing using p-values, reducing many of the apparent associations. The third further corrected these results using SAR\cite{Gorriz2025} which accounted for spatial dependencies across brain regions and provided a robust validation framework. We have observed similar behavior across all other group comparisons indicating that the inflation of raw correlations and their progressive correction was a consistent phenomenon in our dataset. This highlights the necessity of rigorous statistical control when interpreting correlation-based results in high-dimensional neuroimaging data.

\subsection{Supervised correlation analysis of latent features}

To investigate the relationship between latent representations and regional brain patterns, we performed a supervised correlation analysis using the LRPC framework. A region was considered significant when the bound-corrected error was below 0.5, ensuring reliable predictive power under controlled error rates.

\subsubsection{LRPC analysis framework}

\begin{figure}[t]
\centering
\includegraphics[width=0.75\textwidth]{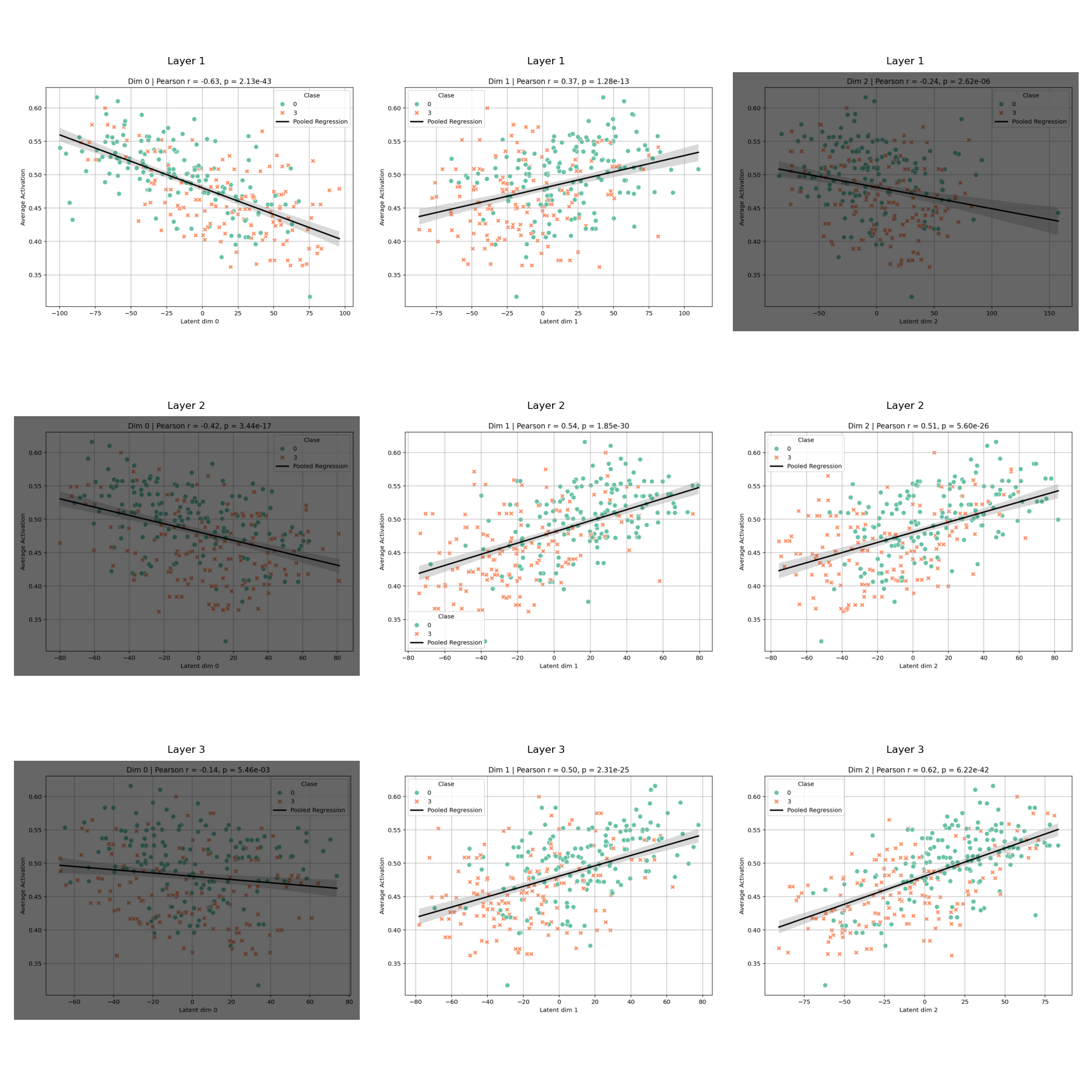}
\caption{LRCP analysis for region number 34, 'Cingulum\_Mid\_R,' across different AE layers and components in t-SNE projections. Significant results are highlighted.}
\label{fig:LRPC1}
\end{figure}

In figure \ref{fig:LRPC1}, we show the LRPC analysis of one of the regions highlighted in table \ref{tab:region_overlap}. We clearly observe how patterns were progressively extracted from Layer 1 to Layer 3. Regressions were statistically significant, but only those highlighted were clinically significant. We also notice how the correlation increased throughout the AE. A summary of the methods and the rest of the parameters and the LRPC analysis reveals regions established as part of the pattern of neurodegeneration, especially in the NOR vs MCI comparison, tables \ref{tab:PCA_PLS_summary} and \ref{tab:tSNE_UMAP_summary}.

\subsubsection{PCA and PLS-based correlation analysis}

Table~\ref{tab:PCA_PLS_summary} summarizes the number of significant and non-significant brain regions identified using PCA and PLS across different clinical groups and latent/dimension combinations. For PCA, the results show a strong pattern of significance in the NOR–AD group, especially in all the layers at dimensions 0 and 1, where all regions (116 out of 116) were found to be statistically significant in their association with the latent-space features under the adopted validation criterion. Notably, significance dropped dramatically in dimension 2 for several latent layers, with most regions classified as non-significant. This suggests that dimensions 0 and 1 captured the most clinically relevant information for PCA in the NOR\_AD group, while dimension 2 contained less informative or noisier features. In the NOR\_MCI group, PCA showed no significant regions across all layers and dimensions, indicating limited sensitivity in differentiating this clinical stage. The NOR\_MCIc group exhibited a more mixed pattern, with strong significance in some layer/dimension combinations (e.g., 100 regions in L1/D2 and 107 in L3/D1 showing significant associations with latent representations), but also non-significance in others. This variability might reflect the heterogeneity of this clinical group or differences in how PCA captures meaningful variation at specific latent/dimension pairs.

The PLS method had robust detection of significant regions across all groups, with all regions significant in all tested layers and dimensions. However, this apparent superiority was likely due to the use of clinical labels during the PLS fitting procedure which caused the extraction to be overfitted to the clinical condition. As a result, PLS detected significance more broadly, but with reduced generalizability.

It is important to note that this analysis was based on a classification framework in which a region was considered significant if the bound-corrected error was less than 0.5. This criterion ensures that significance reflects reliable predictive power with controlled error rates. Overall, PCA appeared sensitive to specific latent layers and dimensions, with dimension 2 often being less informative, while PLS consistently detected significance across groups and dimensions, likely due to overfitting. These findings highlight the importance of careful method selection and validation to avoid overfitting and to ensure clinically meaningful signal extraction.

\begin{table*}[]
\centering
\renewcommand{\arraystretch}{1.2}
\begin{tabular}{lllccl}
\toprule
\textbf{Method} & \textbf{Group} & \textbf{Latent/Dim} & \textbf{Significant} & \textbf{Non-significant} \\
\midrule
PCA & NOR\_AD   & L1/D0   & 116 & 0   \\
 &    & L1/D1   & 116 & 0   \\
 &    & L1/D2   & 3   & 113 \\
 &    & L2/D0   & 116 & 0   \\
 &    & L2/D1   & 116 & 0   \\
 &    & L2/D2   & 0   & 116 \\
 &    & L3/D0   & 116 & 0   \\
 &    & L3/D1   & 116 & 0   \\
 &    & L3/D2   & 0   & 116 \\
 & NOR\_MCI  & All     & 0   & 116 \\
 & NOR\_MCIc & L1/D2   & 100 & 16  \\
 &  & L2/D1   & 42  & 74  \\
 &  & L3/D1   & 107 & 9   \\
\midrule
PLS & NOR\_AD   & All     & 116 & 0   \\
 & NOR\_MCI  & All     & 116 & 0   \\
 & NOR\_MCIc & All     & 116 & 0  \\
\bottomrule
\end{tabular}
\caption{Summary of significant and non-significant regions for PCA and PLS by method, group, and latent/dimension.}
\label{tab:PCA_PLS_summary}
\end{table*}

\subsubsection{t-SNE and UMAP-based analysis}

Table \ref{tab:tSNE_UMAP_summary} presents the counts of significant and non-significant regions identified by t-SNE and UMAP across different clinical groups and latent/dimension pairs. For t-SNE, in the NOR\_AD group, the number of significant regions varied notably by latent layer and dimension, ranging roughly from 53 to 74 significant regions out of 116. Unlike PCA, which showed near-complete significance in several latent/dimension pairs for NOR\_AD, t-SNE presented a more moderate and variable significance pattern. This may reflect t-SNE’s nonlinear embedding characteristics which capture more complex relationships but with less uniform significance. In the NOR\_MCI group (see table \ref{tab:NORMCI_tsne_umap}), t-SNE showed very few significant regions (mostly under 11) with most regions  non-significant. This aligns with PCA’s limited sensitivity in this group but shows an even stronger contrast in significance. Similarly, the NOR\_MCIc group showed intermediate significance levels across layers and dimensions suggesting partial sensitivity of t-SNE to  neurodegenerative progression.

UMAP results for NOR\_AD generally exhibited higher counts of significant regions compared to t-SNE, with many latent/dimension pairs showing 70 or more significant regions. This suggests that UMAP better captured clinically relevant features in this group compared to t-SNE, while still showing some variability across dimensions. For the NOR\_MCI group (see table \ref{tab:NORMCI_tsne_umap}), UMAP’s detection of significant regions was limited (mostly 1 to 3 significant regions), again consistent with the trend observed in t-SNE and PCA, indicating difficulty in distinguishing this clinical stage. The NOR\_MCIc group showed moderate numbers of significant regions with UMAP, often higher than t-SNE, particularly in latent/dimension pairs such as L1/D0 and L2/D2. This indicates that UMAP might have better captured subtle clinical differences in this intermediate group.

\begin{table*}[]
\centering
\renewcommand{\arraystretch}{1.2}
\begin{tabular}{llll}
\toprule
\textbf{Method} & \textbf{Latent/Dim} & \textbf{Significant Regions} \\
\midrule
t-SNE & L1/D0 & \textbf{Caudate\_L}, Cerebelum\_10\_L, Cerebelum\_4\_5\_R, Cerebelum\_8\_R \\
      &       & \textbf{Cerebelum\_9\_L}, Cerebelum\_Crus2\_L, Frontal\_Inf\_Oper\_R, Vermis\_10 \\
 & L1/D1 & Cerebelum\_3\_L, \textbf{Heschl\_L}, Putamen\_R \\
 & L1/D2 & Cerebelum\_4\_5\_L, \textbf{Cerebelum\_8\_L}, Olfactory\_R, Pallidum\_L \\
      &       & \textbf{Postcentral\_L}, \textbf{SupraMarginal\_R}, \textbf{Temporal\_Pole\_Mid\_R} \\
 & L2/D0 & Calcarine\_R, \textbf{Cerebelum\_8\_L}, Frontal\_Mid\_Orb\_L, \textbf{Heschl\_L} \\
      &       & Lingual\_L, Olfactory\_L, \textbf{SupraMarginal\_R} \\
 & L2/D1 & \textbf{Caudate\_L}, Cingulum\_Post\_L, Frontal\_Mid\_R, Frontal\_Sup\_R \\
      &       & ParaHippocampal\_L, Parietal\_Inf\_L, \textbf{SupraMarginal\_R}, Temporal\_Inf\_R \\
 & L2/D2 & \textbf{Cerebelum\_9\_L} \\
 & L3/D0 & Cerebelum\_9\_R, Cerebelum\_Crus1\_R, Frontal\_Mid\_Orb\_R, Frontal\_Sup\_Orb\_R \\
      &       & Occipital\_Sup\_R, Precentral\_L, Temporal\_Inf\_L \\
 & L3/D1 & Cerebelum\_7b\_L, Cerebelum\_7b\_R, Cingulum\_Ant\_R, Cuneus\_R \\
      &       & Frontal\_Inf\_Tri\_L, Frontal\_Mid\_Orb\_L, Lingual\_L, Postcentral\_R \\
      &       & Rectus\_R, \textbf{Temporal\_Pole\_Mid\_R}, Vermis\_3 \\
 & L3/D2 & Cerebelum\_6\_R, Cerebelum\_Crus1\_L, Frontal\_Mid\_Orb\_L, Olfactory\_L \\
      &       & \textbf{Postcentral\_L}, Precuneus\_R, Putamen\_L, Rolandic\_Oper\_L \\
      &       & Temporal\_Pole\_Mid\_L, Temporal\_Sup\_L, Temporal\_Sup\_R \\
UMAP  & L2/D0 & Supp\_Motor\_Area\_L \\
  & L2/D2 & Cingulum\_Ant\_L, \textbf{SupraMarginal\_R} \\
  & L3/D0 & \textbf{Caudate\_L}, Cerebelum\_10\_R, Fusiform\_L \\
  & L3/D2 & ParaHippocampal\_L \\
\bottomrule
\end{tabular}
\caption{Comparison of significant regions for NOR\_MCI using t-SNE and UMAP. \\
Note: Regions highlighted in \textbf{bold} appear repeatedly across different latent/dimension projections.}
\label{tab:NORMCI_tsne_umap}
\end{table*}

\begin{table*}[]
\centering
\footnotesize
\renewcommand{\arraystretch}{1.2}
\begin{tabular}{lllccl}
\toprule
\textbf{Method} & \textbf{Group} & \textbf{Latent/Dim} & \textbf{Significant} & \textbf{Non-significant} \\
\midrule
t-SNE & NOR\_AD   & L1/D0   & 53  & 63  \\
 &    & L1/D1   & 74  & 42  \\
 &    & L1/D2   & 71  & 45  \\
 &    & L2/D0   & 66  & 50  \\
 &    & L2/D1   & 64  & 52  \\
 &    & L2/D2   & 70  & 46  \\
 &    & L3/D0   & 65  & 51  \\
 &    & L3/D1   & 70  & 46  \\
 &    & L3/D2   & 73  & 43  \\
 & NOR\_MCI  & L1/D0   & 8   & 108 \\
 &   & L1/D1   & 3   & 113 \\
 &   & L1/D2   & 7   & 109 \\
 &   & L2/D0   & 7   & 109 \\
 &   & L2/D1   & 8   & 108 \\
 &   & L2/D2   & 1   & 115 \\
 &   & L3/D0   & 7   & 109 \\
 &   & L3/D1   & 11  & 105 \\
 &   & L3/D2   & 11  & 105 \\
 & NOR\_MCIc & L1/D0   & 24  & 92  \\
 &  & L1/D1   & 19  & 97  \\
 &  & L1/D2   & 18  & 98  \\
 &  & L2/D0   & 23  & 93  \\
 &  & L2/D1   & 33  & 83  \\
 &  & L2/D2   & 19  & 97  \\
 &  & L3/D0   & 25  & 91  \\
 &  & L3/D1   & 28  & 88  \\
 &  & L3/D2   & 22  & 94  \\
\midrule
UMAP  & NOR\_AD   & L1/D0   & 83  & 33  \\
  &    & L1/D1   & 73  & 43  \\
  &    & L1/D2   & 76  & 40  \\
  &    & L2/D0   & 76  & 40  \\
  &    & L2/D1   & 81  & 35  \\
  &    & L2/D2   & 71  & 45  \\
 &    & L3/D0   & 68  & 48  \\
  &   & L3/D1   & 82  & 34  \\
  &    & L3/D2   & 76  & 40  \\
  & NOR\_MCI  & L2/D0   & 1   & 115 \\
  &   & L2/D2   & 2   & 114 \\
  &  & L3/D0   & 3   & 113 \\
  &   & L3/D2   & 1   & 115 \\
  & NOR\_MCIc & L1/D0   & 55  & 61  \\
  &  & L1/D1   & 35  & 81  \\
 &  & L1/D2   & 25  & 91  \\
  &  & L2/D0   & 45  & 71  \\
 &  & L2/D1   & 48  & 68  \\
  &  & L2/D2   & 56  & 60  \\
  &  & L3/D0   & 13  & 103 \\
  &  & L3/D1   & 12  & 104 \\
  &  & L3/D2   & 17  & 99  \\
\bottomrule
\end{tabular}
\caption{Summary of significant and non-significant regions for t-SNE and UMAP by method, group, and latent/dimension.}
\label{tab:tSNE_UMAP_summary}
\end{table*}

\subsubsection{Global trends across latent dimensions}

\begin{figure}[t]
\centering
\includegraphics[width=0.75\textwidth]{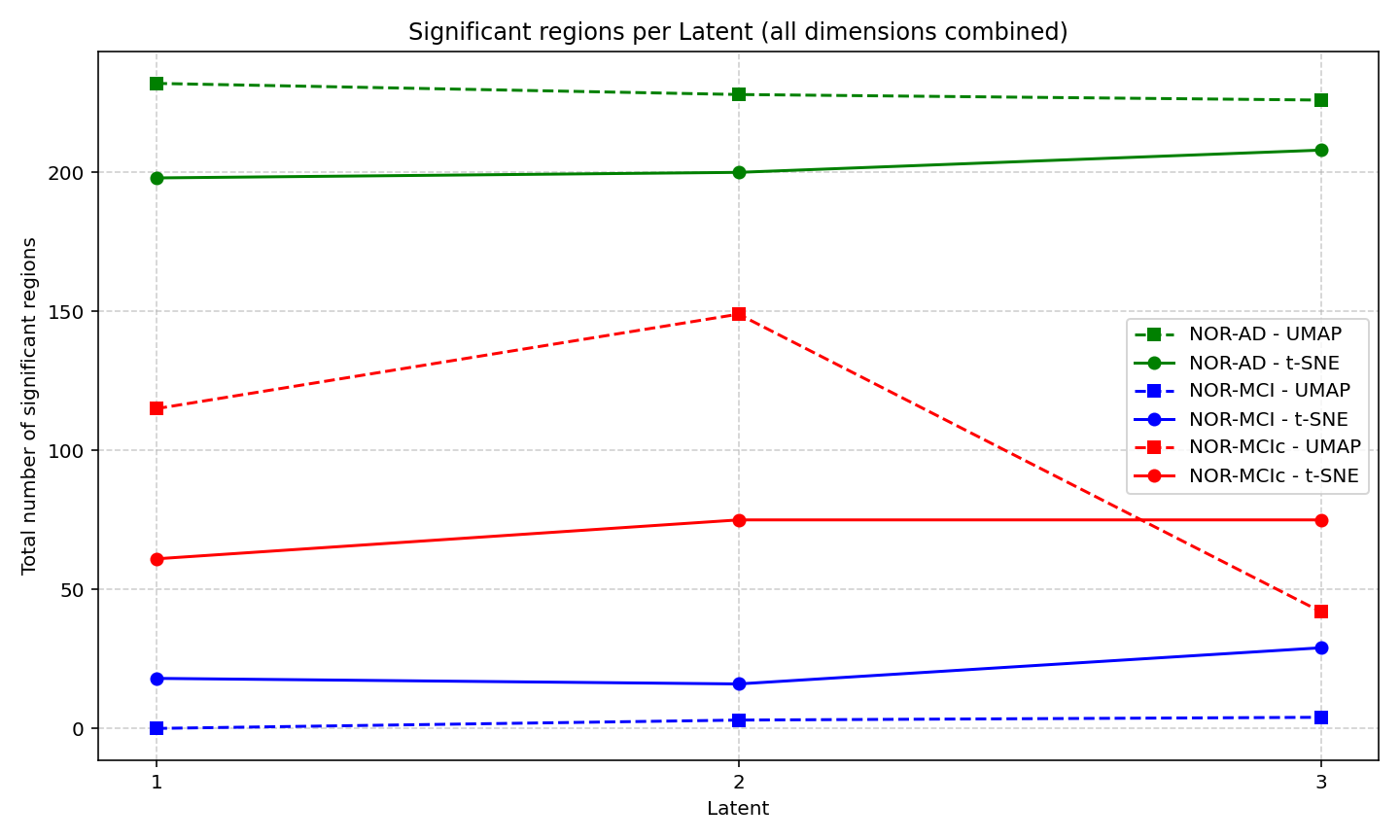}
\caption{Summary of significant and non-significant regions for t-SNE and UMAP by group and latent (adding dimensions)}
\label{fig:summarytabtSNE_UMAP_summary}
\end{figure}

Figure \ref{fig:summarytabtSNE_UMAP_summary} shows the number of significant regions as a function of latent dimension for different groups and methods. It can be observed that the NOR-AD group exhibited the highest number of significant regions while the NOR-MCIc and NOR-MCI groups showed fewer regions in comparison. Interestingly, for UMAP in the NOR-MCIc group, there was a decrease in the number of significant regions at latent 3, unlike the general trend.
Additionally, there was a slight overall increase in the number of significant regions with higher latent layers across all other groups, indicating a trend of increased significance at deeper latent components. Overall, the comparison highlights a clear group effect, with AD vs NOR showing the most pronounced difference, and a moderate latent effect across most conditions.

Several brain regions identified as significant in our NOR-MCI analysis—including the caudate nucleus, parahippocampal gyrus, cingulum, frontal operculum, and cerebellum—have been previously implicated in AD pathology compared to normal controls. The parahippocampal area and cingulum are well-known sites of early atrophy linked to memory impairment \cite{braak1991neuropathological, tondelli2012structural}, while alterations in the caudate and putamen relate to cognitive and motor dysfunction \cite{Seunghee, dejong2008putamen}. Cerebellar involvement, increasingly recognized in AD, may contribute to both cognitive and motor symptoms \cite{jacobs2018cerebellum, Cui2024}. Frontal and temporal regions, including the temporal pole and fusiform gyrus, also show structural decline correlating with executive and visual processing deficits \cite{risacher2010longitudinal}\cite{whitwell2007fusiform}. Overall, our findings align with established neuroanatomical changes in AD supporting the relevance of these significant regions as biomarkers distinguishing normal aging from AD even when individuals have MCI status.

\subsubsection{Spatial mapping of latent–region associations}

\begin{figure}[t]
\centering
\includegraphics[width=0.49\textwidth]{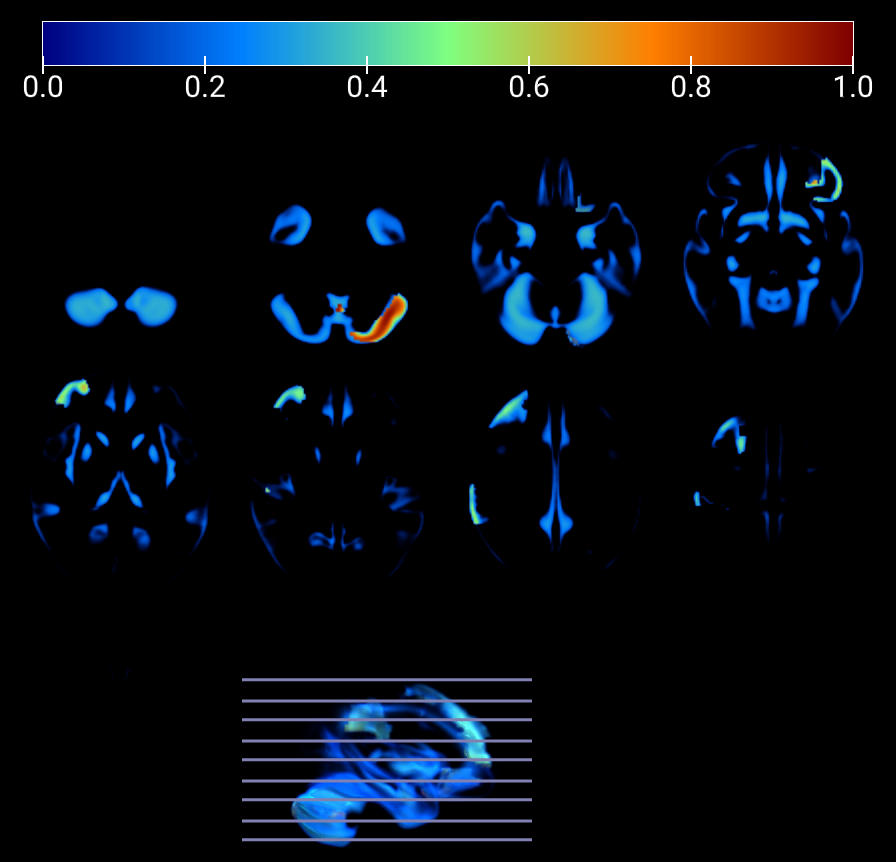}
\includegraphics[width=0.49\textwidth]{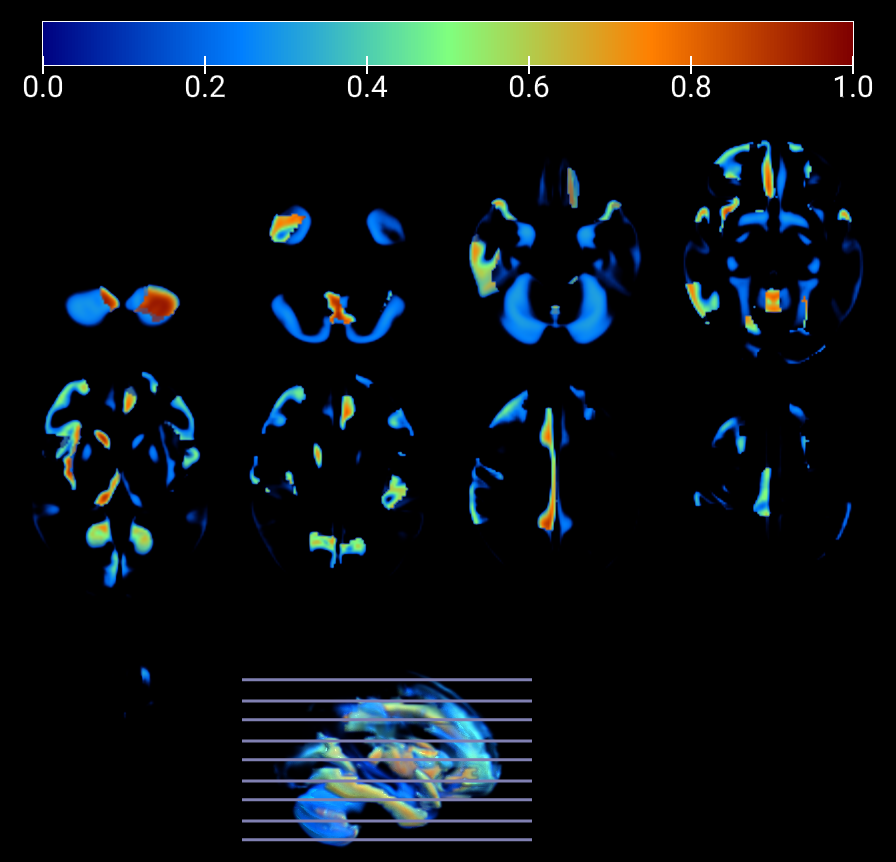}
\includegraphics[width=0.49\textwidth]{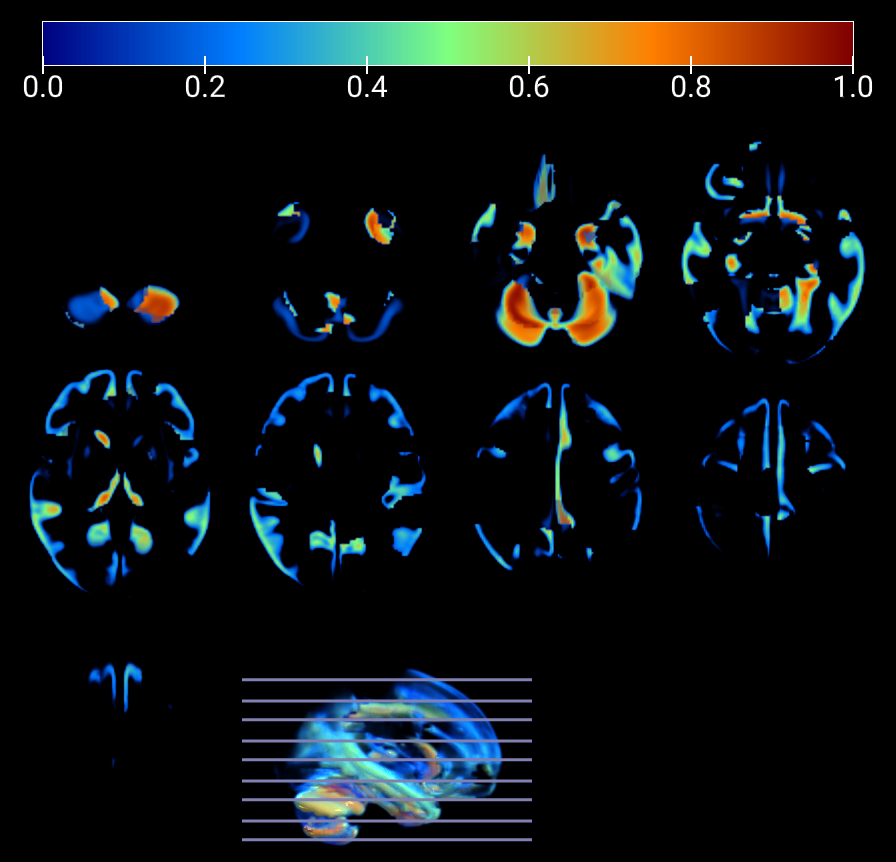}
\caption{LRCP analysis for the three binary groups showing the evolution of disease from MCI to AD. Results were obtained using t-SNE for DR, displaying the regional accuracy maps derived from the first latent component, which highlight spatial patterns of discriminative power across diagnostic groups. From top to bottom and left to right, rows correspond to MCI, MCIc, and AD.}
\label{fig:LRPC2}
\end{figure}

Finally, we applied the LRCP framework to generate spatial maps that highlighted how latent components related to regional brain variation across different diagnostic comparisons (see figure \ref{fig:LRPC2}). Specifically, we focused on the three binary groups and, for each case, projected the regional accuracy of the latent–region associations onto an anatomical atlas. By computing corrected significance rates for each latent–region pair and averaging them across participants, we obtained accuracy maps that indicated which brain regions consistently encoded discriminative information for each comparison. These maps provide an interpretable visualization of the spatial distribution of diagnostic relevance, facilitating a region-wise comparison of how latent dimensions captured biologically meaningful variation across the different binary groups.

\section{Discussion}
Our model successfully encoded MRI brain volumes into a compact latent representation that aligned with known anatomical and clinical patterns. The combination of autoencoding and DR allowed us to explore and interpret learned features. Results suggest potential applications in early diagnosis, subgroup identification, and biomarker discovery.

\subsection{Our approach to neuroimaging model validation}

The \textit{Latent–Regional Correlation Profiling} analysis is introduced as a supervised framework to evaluate the relationship between latent features extracted from the model and anatomically defined brain regions, while simultaneously assessing their discriminative power with respect to clinical labels. Unlike conventional correlation analysis, which only quantifies the linear association between a latent representation and a given region, LRCP incorporates a dual regression-classification evaluation. This enables the identification of regions that are not only statistically associated with the latent structure but also capable of distinguishing between clinical groups.  

In practice, Pearson correlation coefficients were first computed between each latent component and the regional measures, yielding a statistical assessment of association strength and significance. This regression-oriented step ensured that selected regions were relevant from a statistical standpoint. However, statistical significance alone does not guarantee practical discriminability in the context of exploratory neuroimaging analysis. For this reason, the second step of LRCP involved a supervised classification task using the same latent-regional mapping in which classification performance (e.g. empirical error correction) was used to determine whether the identified associations translated into effective group separation.  

This dual approach categorizes regions into four distinct cases:
\begin{enumerate}
    \item Both significant in correlation and classification -- regions that are statistically and practically relevant, representing the most robust biomarkers.
    \item Significant in correlation but not in classification -- regions that exhibit statistical associations but limited discriminative value.
    \item Significant in classification but not in correlation -- regions whose discriminative power arises from nonlinear or higher-order interactions not captured by Pearson correlation.
    \item Not significant in either test -- regions with no apparent statistical or practical relevance.
\end{enumerate}

For each group comparison and latent layer, we report the number of regions falling into each category, providing a global view of how statistical association and practical discriminability interact across the network hierarchy. Experimental results revealed distinct patterns of agreement and divergence between correlation-based and classification-based significance, underscoring the importance of combining both perspectives when performing latent feature interpretation in neuroimaging studies.

\subsection{Non-linear DR reveals differences in challenging groups}

Compared to PCA, which detected very few or no significant regions in key clinical groups such as NOR\_MCI and NOR\_MCIc, both t-SNE and UMAP had higher sensitivity when identifying significant regions, especially in these transitional groups. This difference likely stems from t-SNE and UMAP being nonlinear embedding methods that better capture complex, local data structures relevant to clinical progression, whereas PCA, as a linear method, fails to extract these subtle patterns. In the NOR\_AD group, t-SNE and UMAP revealed a variable number of significant regions across latent layers and dimensions, reflecting their ability to detect diverse nonlinear patterns. PCA, in contrast, had limited significant detection in these groups, underscoring its reduced sensitivity to the nonlinear relationships present in the data. UMAP tended to be more sensitive than t-SNE in detecting significant regions, particularly within the NOR\_MCIc group indicating a capacity to capture transitional clinical states or subtle changes.

Overall, these results suggest that nonlinear methods such as t-SNE and UMAP provide complementary and more effective approaches than PCA for identifying clinically relevant features, particularly in groups with subtle or early disease progression. This highlights the importance of choosing appropriate extraction methods tailored to the clinical context and data complexity.

\subsection{Limitations}

A key limitation of our method, and the majority of the approaches for image-to-image translation \cite{42} and feature attribution methods \cite{40,Bass2023}, is its reliance on the quality and representativeness of the dataset which may restrict the generalizability of findings to other populations or clinical settings. The use of AEs, complex generative models and, to a lesser extent DR techniques, such as PCA, t-SNE, and UMAP involves choices of hyperparameters and model architecture that influences the interpretation of latent patterns. Although statistical validation was strengthened by approaches like CUBV and SAR, results remained sensitive to sample size and potential biases in the data. Furthermore, correlations between latent components and anatomical regions do not imply causality or direct clinical relevance, and may have been affected by spatial autocorrelation and overfitting, especially when few significant regions were detected. Therefore, it is essential to complement these methods with external validation and longitudinal analyses to ensure the robustness and clinical utility of the findings.

A further limitation is that our analysis was performed at the level of AAL regions of interest rather than individual voxels, owing to the substantial computational demands of voxel-wise processing in large 3D MRI datasets. While this regional approach was motivated by practical constraints, it may have reduced spatial specificity and overlooked subtle local effects. Likewise, we focused on gray matter segmentations instead of raw MRI data to mitigate the impact of limited sample size and to enhance anatomical interpretability. However, this choice may have excluded potentially relevant information present in other tissue types or in the original images. These methodological decisions—although justified by computational and statistical considerations—are acknowledged as factors that may have influenced both the sensitivity and the generalizability of our results.

The proposed framework is specifically designed for statistically grounded population-level inference in high-dimensional neuroimaging settings where sample sizes are limited and data acquisition is costly, as commonly occurs in clinical research. Unlike predictive benchmarking approaches that primarily emphasize cross-cohort performance, our objective is to provide conservative inferential guarantees under explicit control of false positives and generalization risk through the CUBV/SAR framework. In this context, the validity of the analysis does not rely exclusively on the availability of multiple independent datasets, but rather on the robustness of the statistical learning framework used to assess whether the observed latent–regional associations reflect genuine population-level effects rather than spurious correlations.

\subsection{Remarks and future work}

Our findings echo a recurring theme in modern neuroimaging with deep learning:  seduction by convincing results without the necessary quantitative scrutiny. In AD MRI analysis, it is tempting to assume that ``we trained a model, therefore it works'', or that a visually appealing heatmap is sufficient validation. Yet, as our experiments with SHAP, correlation profiling, and SAR revealed, apparent structure in latent spaces may arise from noise, spurious associations, or methodological shortcuts.
The correlation between latent features and anatomical regions, while statistically significant in some cases, often fails to translate into practical discriminability—reminding us that ``correlation is not causation'', no matter how good it looks in a colormap. Interpretability tools, when unvalidated, risk becoming an exercise in ``science as seen through a colormap'', where bright colors mask weak evidence.
We have shown that without rigorous statistical control and robust evaluation—including checks across group comparisons and layers—latent space patterns may look meaningful on t-SNE plots yet fail to hold up under SAR or supervised discriminative testing. This work reinforces the idea that statistical significance is not a substitute for scientific validation, and that the true challenge lies not in finding patterns but in ensuring they reflect genuine neurobiological signals rather than artefacts of modeling or preprocessing. Ultimately, the aim is not to have ``one model to fool them all'', but rather a pipeline that earns trust through transparency, robustness, and reproducibility.

Future work may integrate GANs, as explored in previous studies \cite{Bass2022,42}; incorporate larger datasets with robust ground truth data (similar to the ADNI initiative with clinical follow-up); or apply alternative visualization and interpretability techniques such as SHAP or Grad-CAM, which have been used as baselines in the literature discussed in the introduction but with lower performance.

\section{Conclusion}

This study demonstrates the feasibility of using simple 3D convolutional autoencoders to extract clinically and anatomically meaningful features from brain MRI data. The autoencoder achieved consistently low reconstruction errors, with MSE values below 0.01 across all clinical groups, indicating high fidelity in preserving gray matter structure. Dimensionality reduction techniques revealed clear class separation, particularly with PLS, which consistently outperformed unsupervised linear methods such as PCA. This advantage is expected, as PLS explicitly incorporates diagnostic labels during projection, thereby maximizing covariance between latent features and group membership and increasing statistical power for group-level inference. In the NOR–AD comparison, PLS identified 100$\%$ of AAL regions ($116/116$) as statistically significant across multiple latent layers and dimensions, whereas PCA showed reduced sensitivity, especially in intermediate contrasts such as NOR–MCIc, where only a small subset of regions reached significance. While PLS offers increased inferential sensitivity, its supervised nature also entails a higher risk of overfitting in an inferential sense, potentially inflating significance if not properly controlled. These observations highlight the importance of careful method selection and rigorous validation to ensure that detected effects reflect clinically meaningful signals rather than label-driven artifacts.

Non-linear methods such as t-SNE and UMAP provided complementary and more nuanced insights into latent structure. UMAP identified up to 83 significant regions in the NOR–AD group, while t-SNE detected up to 74 regions, highlighting their ability to capture subtle anatomical differences, particularly in early disease stages (MCI and MCIc comparisons). SHAP analysis further confirmed the relevance of specific brain regions. In the NOR–AD comparison, regions such as the Right Supplementary Motor Area, Lingual Gyrus, and Middle Cingulum consistently showed high SHAP values, indicating their strong contribution to reconstruction error and their potential as biomarkers.

Finally, the SAR method corrected inflated correlation values, revealing that even moderate correlations (e.g., $r > 0.11$) can be misleading without proper statistical control. LRCP helped recover biologically meaningful patterns that were otherwise obscured. Overall, these findings underscore the importance of combining unsupervised representation learning with supervised and statistically rigorous downstream analyses. The proposed pipeline offers a transparent and reproducible framework for exploring latent neuroimaging features, with potential applications in early diagnosis, subgroup identification, and biomarker discovery in AD and related conditions.

\section*{Acknowledgements}

Data used in preparation of this article were obtained from the Alzheimer’s Disease Neuroimaging Initiative (ADNI) database (adni.loni.usc.edu). As such, the investigators within the ADNI contributed to the design and implementation of ADNI and/or provided data but did not participate in analysis or writing of this report. A complete listing of ADNI investigators can be found at: \href{https://adni.loni.usc.edu/wp-content/uploads/how_to_apply/ADNI_Acknowledgement_List.pdf}. We thank the ADNI data contributors and the institutions involved in data collection and curation.

This research is part of the PID2022-137451OB-I00 and PID2022-137629OA-I00 projects, funded by MICIU/AEI /10.13039/501100011033 and by ERDF, EU. We would also like to thank the reviewers for their contributions to improving this manuscript.

\newpage
\section{Appendices}
\subsection{Training Strategy}
For training the AE, balanced subsets of participants were created to form the following groupings: NOR–AD, NOR–MCI, NOR–MCIc, and NOR–MCI–MCIc–AD. These groupings were designed to allow for meaningful comparisons and robust representational learning across different stages of cognitive decline. To assess potential age-related confounding effects across diagnostic groups, we conducted a statistical comparison of subject ages using a two-sample Welch’s t-test. The results showed no statistically significant differences in age across any of the comparisons (all $p > 0.05$), indicating that the groups were well matched in terms of age and that age is unlikely to act as a confounding factor in the subsequent analyses. In addition, for each experimental setting, the number of images used in the analysis was explicitly controlled to ensure balanced comparisons. Specifically, the number of subjects selected from the NOR group was matched to the number of subjects in the corresponding comparison group, so that all analyses were performed on datasets with equal sample sizes. This strategy prevents biases arising from unequal group sizes and ensures that the results are not driven by differences in the number of samples.

The AE was trained using a mini-batch gradient descent strategy (stochastic regularization effect) with the Adam optimizer set with a learning rate of 0.001 over a maximum of 10 epochs. To avoid overfitting and promote generalization, an early stopping criterion was employed, with a patience threshold of 5 epochs based on the average reconstruction loss per epoch. In the experimental setup, three loss functions were evaluated: MSE, structural similarity index measure (SSIM), and a combined loss incorporating both MSE and SSIM, weighted by a parameter $\alpha = 0.5$. Only the reconstruction pathway of the AE was used for loss computation and gradient backpropagation. The training goal was twofold: first, to obtain a well-performing model capable of accurately reconstructing structural brain images from compressed representations; and second, to analyze how the encoder organizes and encodes relevant anatomical information in a low-dimensional latent space to support such reconstructions.

\begin{figure}[t]
    \centering
    \begin{subfigure}{\textwidth}
        \centering
        \includegraphics[width=\textwidth]{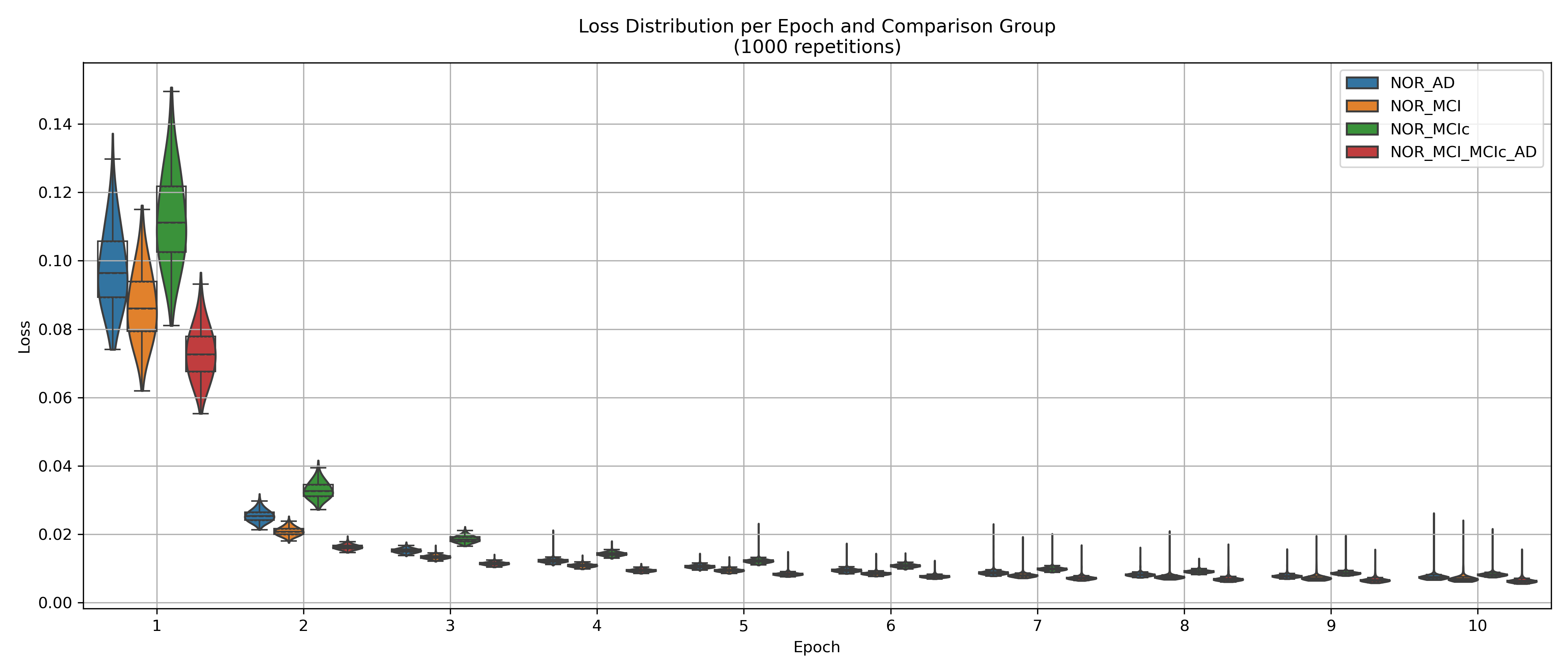}
        \label{fig:subfig3}
    \end{subfigure}
    \caption{Distribution of training loss across epochs (1 to 10) for each comparison group including NOR, AD, MCI, MCIc subjects. Violin plots illustrate the variability and central tendency (quartiles) of loss values over multiple experimental repetitions (1000), highlighting the convergence behavior and differences in training stability among groups.}
    \label{fig:learning}
\end{figure}

This procedure was repeated independently for each of the groupings: NOR–AD, NOR–MCI, NOR–MCIc, and NOR–MCI–MCIc–AD, allowing a comparative analysis of the latent space organization across different clinical conditions. In figure \ref{fig:learning}, we show the distribution of training loss for different reconstruction tasks, each reflecting problems of varying complexity depending on the heterogeneity of the  groups. Comparisons involving putatively dissimilar groups (e.g., NOR vs. AD or NOR vs. MCIc) correspond to more challenging reconstruction tasks for the AE resulting in slightly higher and more heterogeneous training losses. In contrast, tasks involving more similar groups (e.g., NOR vs. MCI) yielded lower and more consistent losses as the anatomical variability between classes is presumed to be more subtle. Although the differences in loss distributions were relatively modest—partly due to averaging across all voxels and epochs—they indicate that group dissimilarity increased the difficulty of the reconstruction task. These observations suggest that the AE’s final configuration reflects group-level differences, making it a potentially informative tool for further analysis by class and group comparison.

\subsection{Standard latent space analysis and visualization}\label{app:latent}

Additionally, we present a simplified yet interpretable baseline method for exploring latent representations in AEs trained on neuroimaging data. Unlike more complex approaches that rely on generative adversarial networks (GANs) to synthesize and classify images from latent features, our method leveraged direct analysis of the latent space without the need for image generation or additional classifier training. Standard validation strategies in the literature often rely on classification accuracy of synthetic images without reporting confidence intervals, and on the computation of Pearson correlations between attribution maps and group-level significance maps derived from statistical parametric mapping (SPM) analyses. These approaches may overestimate alignment due to spatial autocorrelation and lack robust individual-level validation.

\begin{figure}[t]
  \centering
  \includegraphics[width=0.49\textwidth]{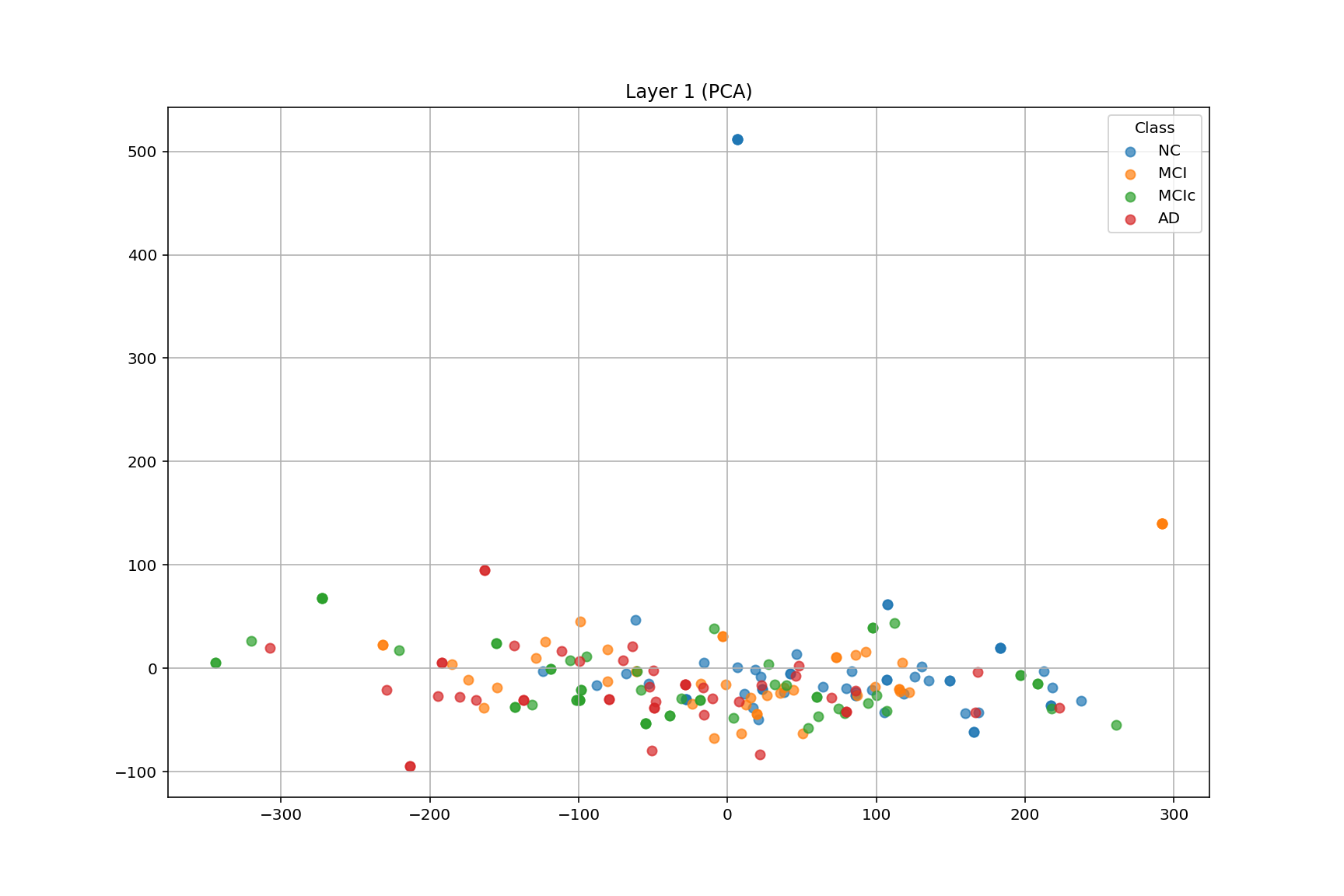}
  \includegraphics[width=0.49\textwidth]{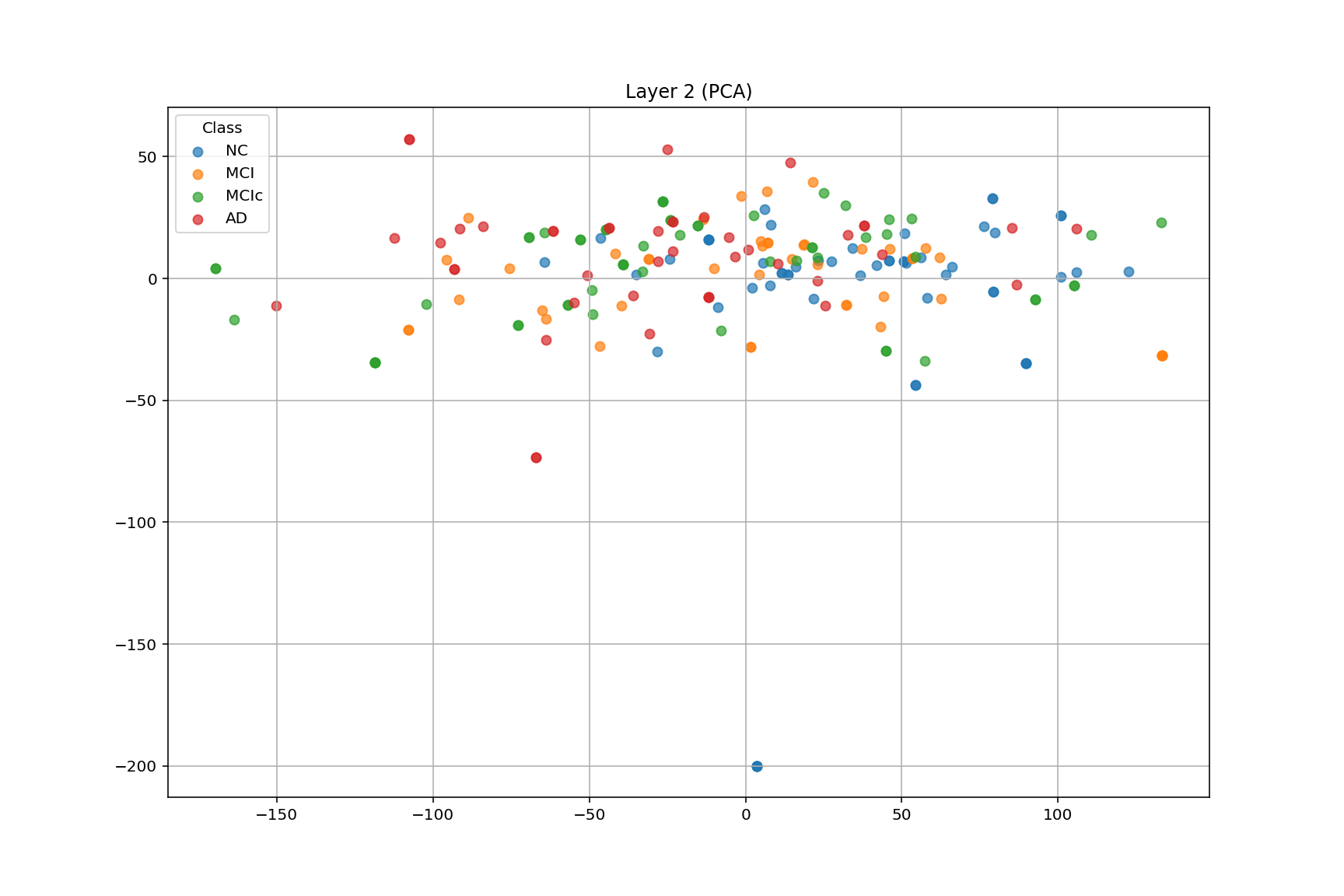}
  \includegraphics[width=0.49\textwidth]{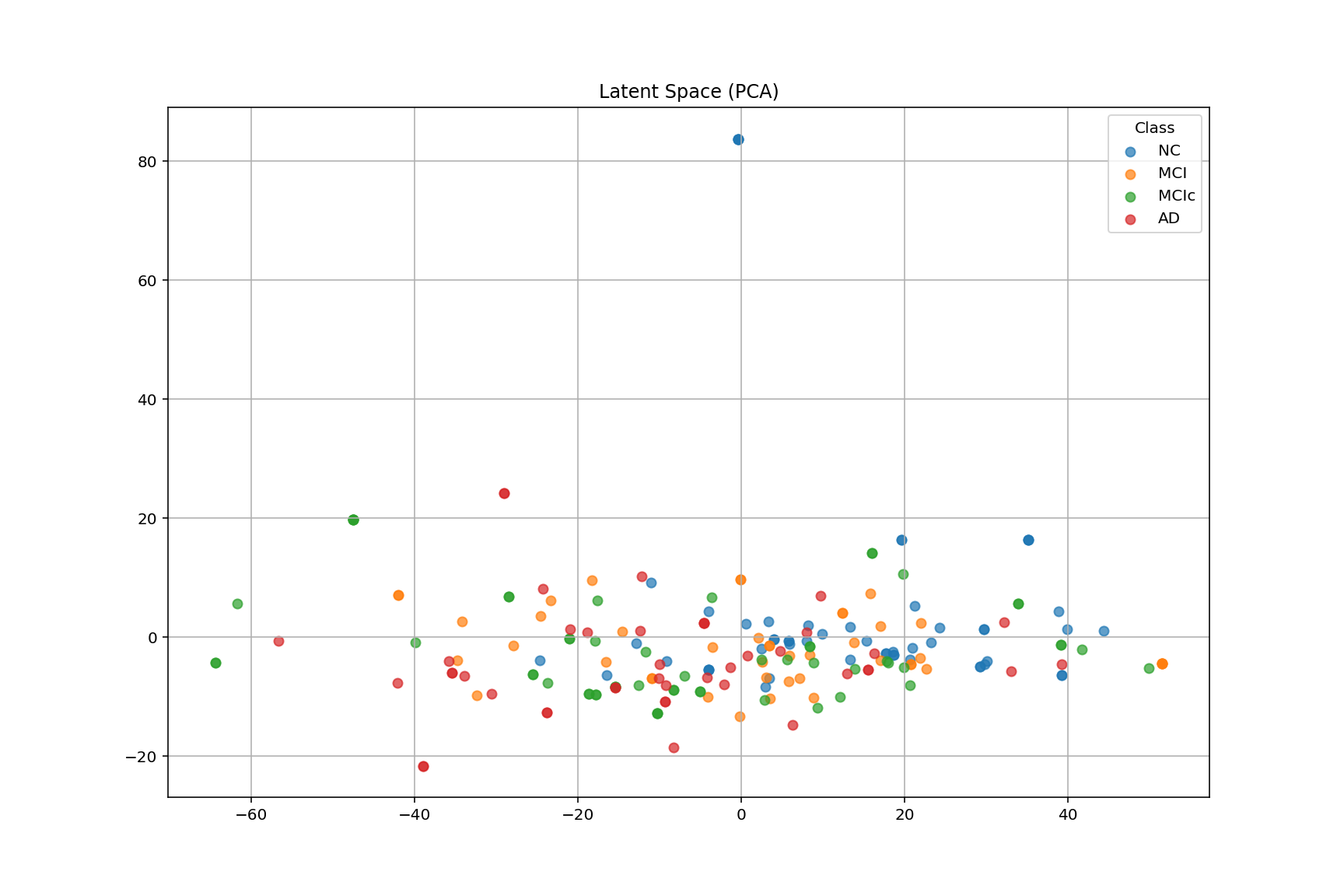}
  \caption{PCA projections of feature representations extracted from Layer 1 (top left), Layer 2 (top right), and the latent space (bottom) for the NOR–MCI–MCIc–AD grouping. Points correspond to individual subjects and are colored according to clinical labels. These projections provide a qualitative visualization of how the autoencoder organizes subjects in low-dimensional space across different levels of abstraction. While some degree of grouping can be visually observed, these patterns are exploratory and do not constitute statistical evidence of separability.}
  \label{fig:PCA}
\end{figure}

\begin{figure}[t]
  \centering
  \includegraphics[width=0.49\textwidth]{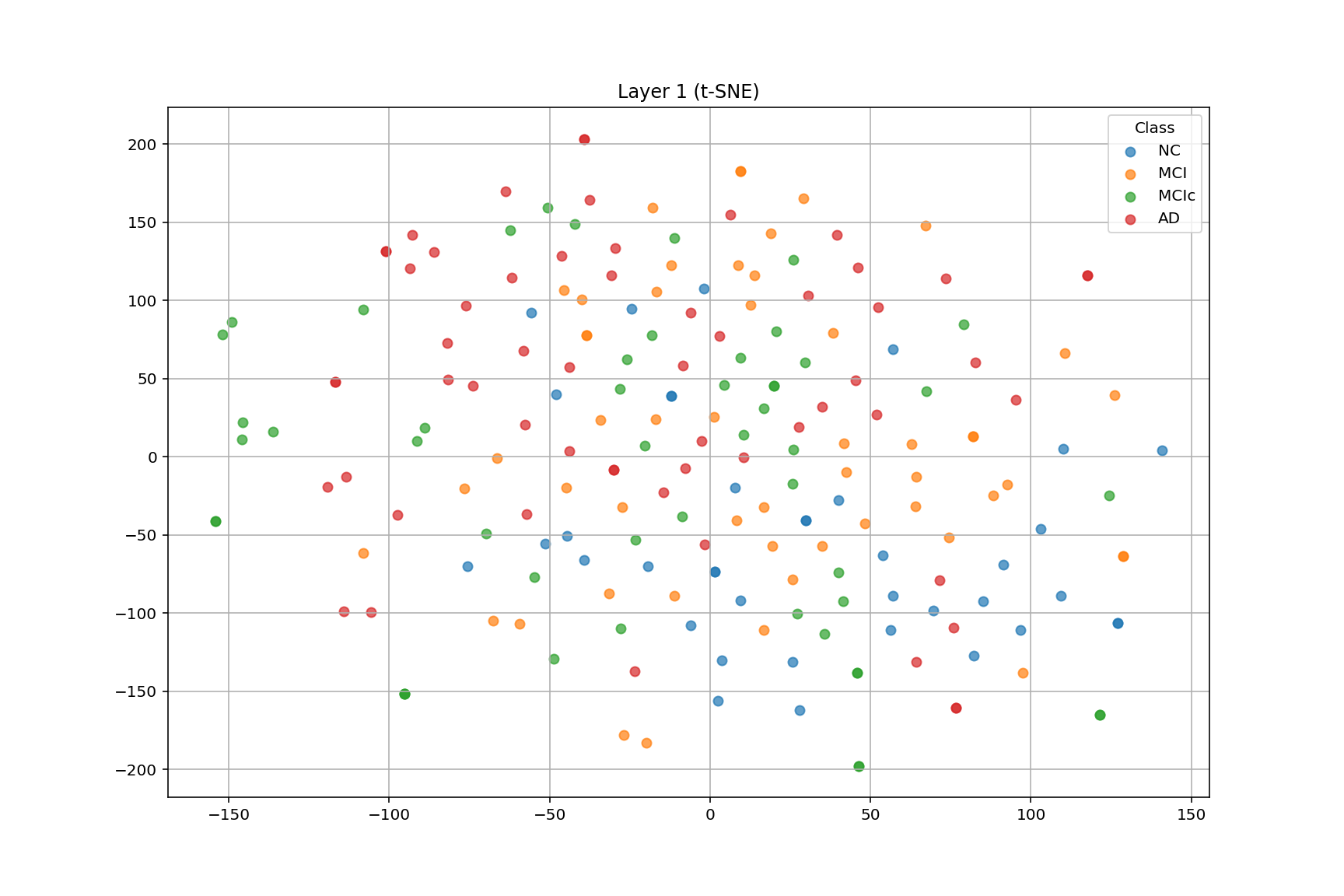}
  \includegraphics[width=0.49\textwidth]{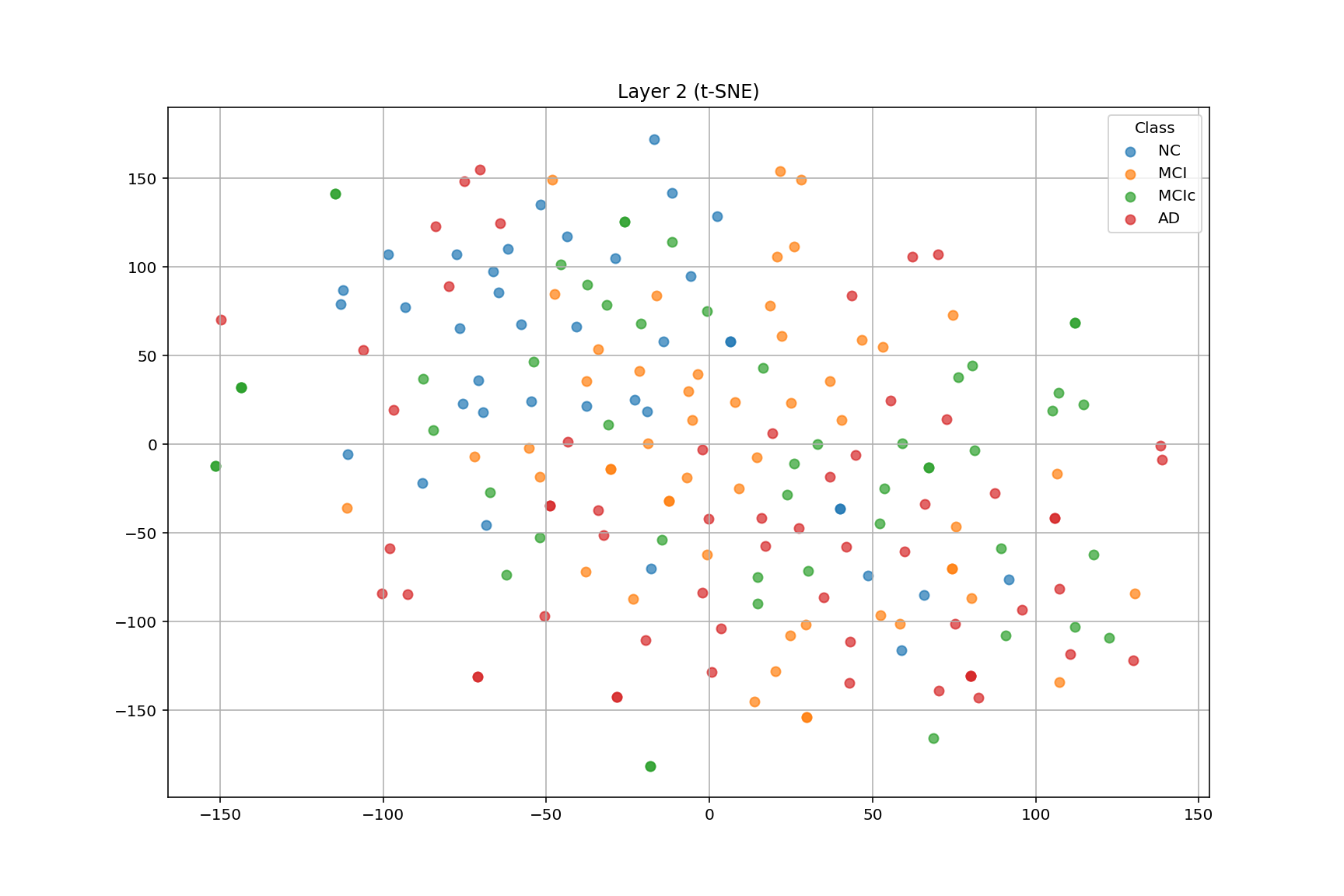}
  \includegraphics[width=0.49\textwidth]{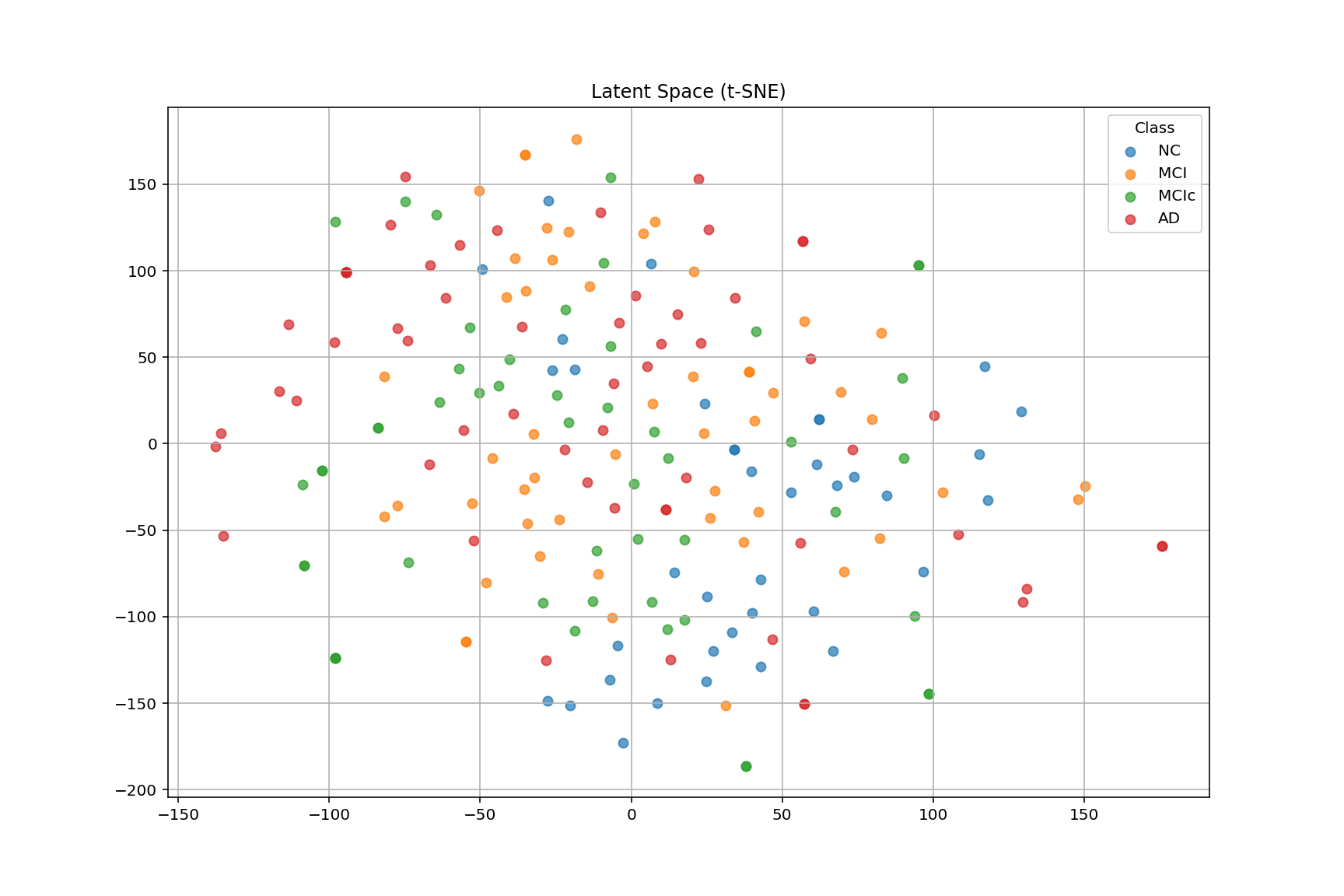}
  \caption{t-SNE projections of feature representations from Layer 1 (top left), Layer 2 (top right), and the latent space (bottom) for the NOR–MCI–MCIc–AD grouping. Each point represents a subject and is colored by clinical group. These nonlinear embeddings highlight potential local structure and clustering tendencies in the data; however, as with PCA, such visualizations are inherently qualitative and may be misleading without proper statistical validation, motivating the subsequent inferential analyses.}
  \label{fig:tsne}
\end{figure}

\begin{figure}[t]
  \centering
  \includegraphics[width=0.49\textwidth]{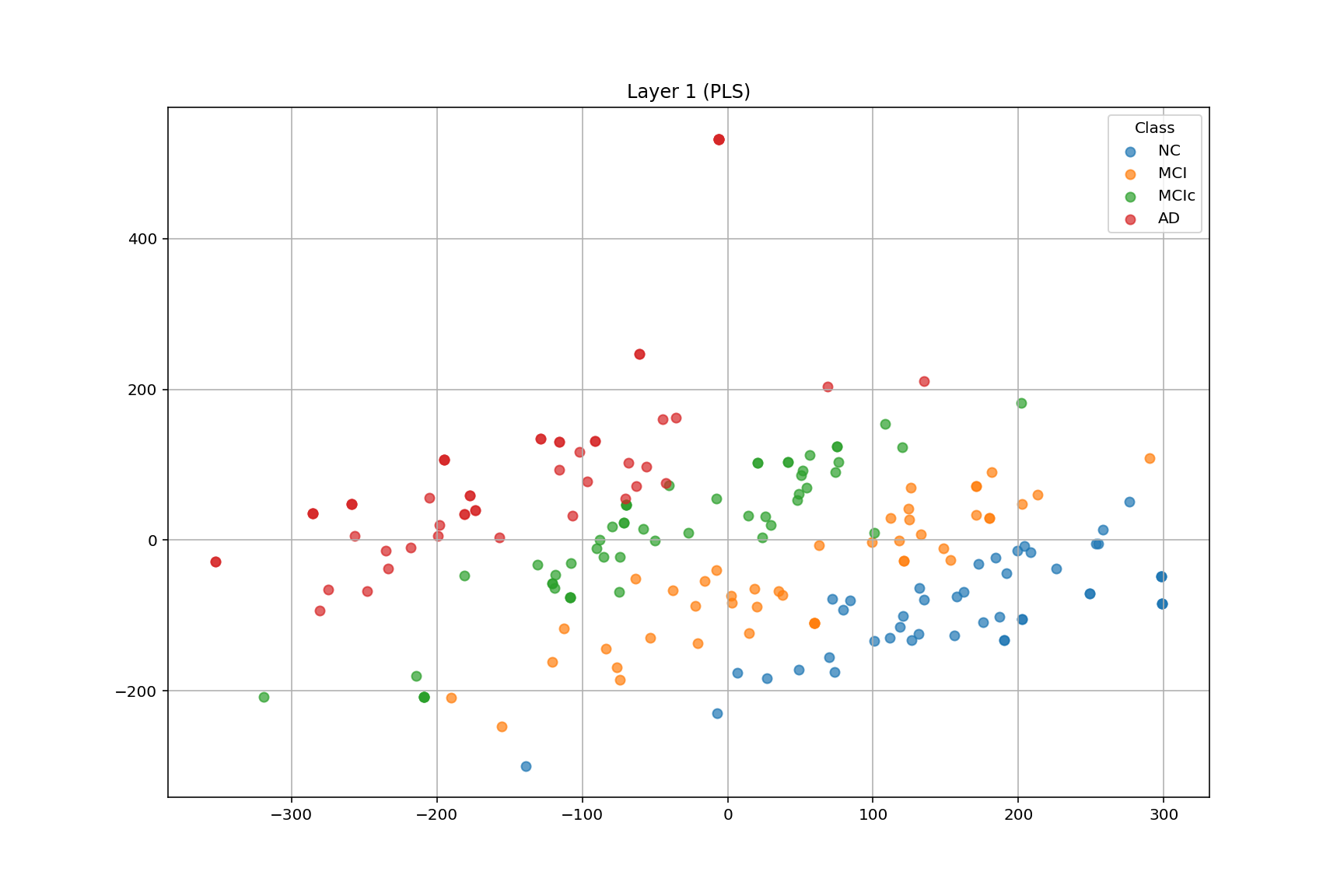}
  \includegraphics[width=0.49\textwidth]{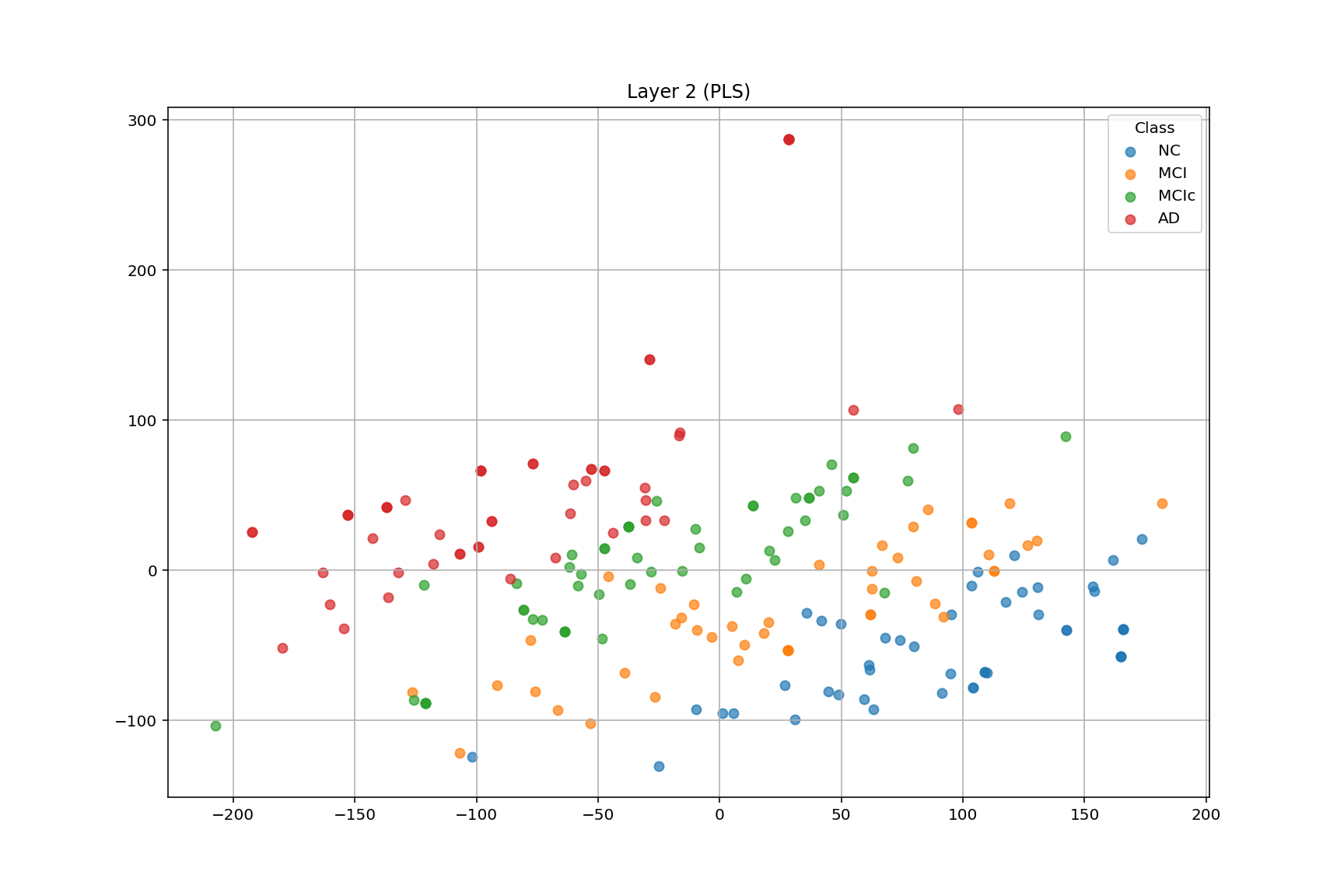}
  \includegraphics[width=0.49\textwidth]{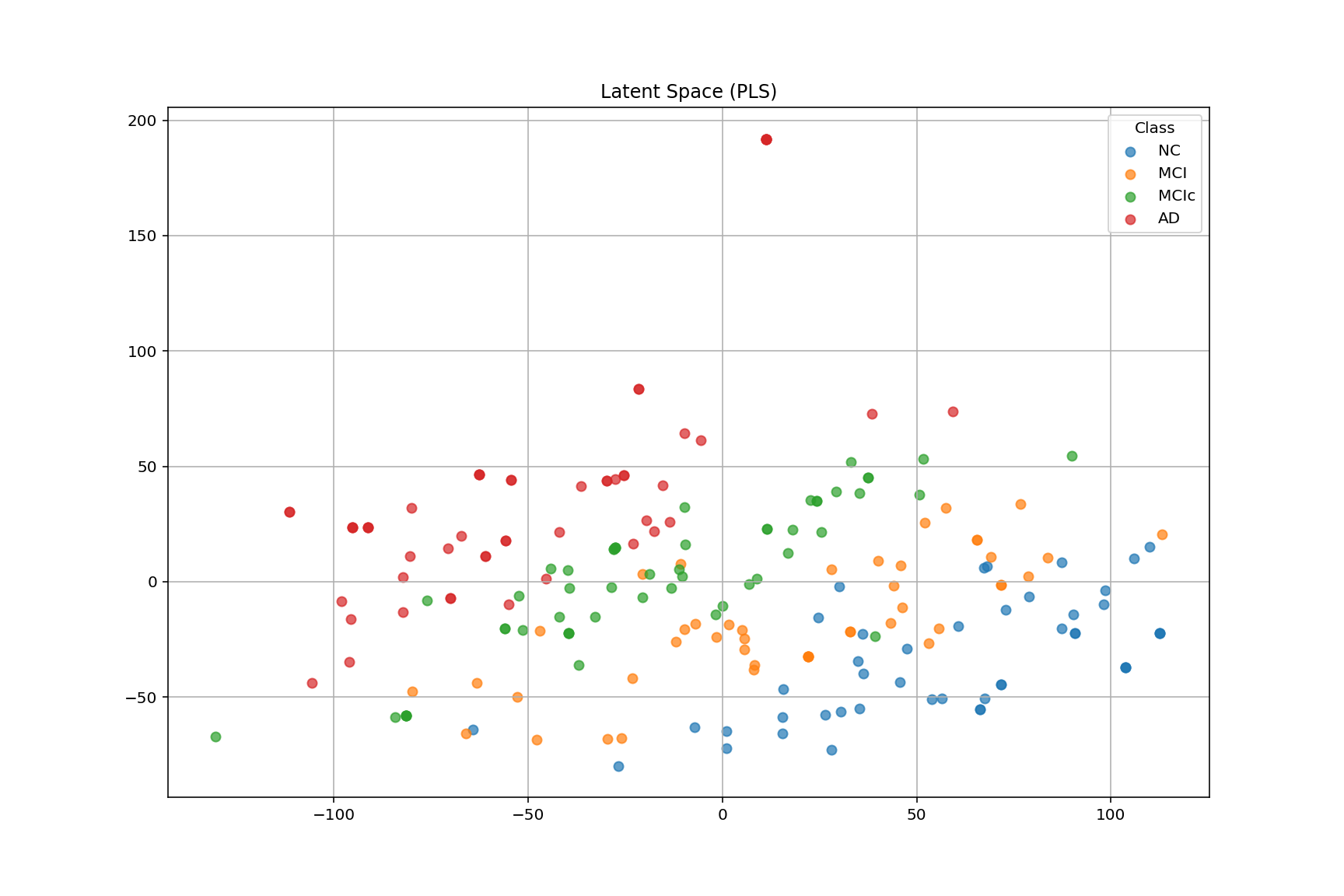}
  \caption{PLS projection of Layer 1, 2 and latent activations}
  \label{fig:PLS}
\end{figure}

\begin{figure}[t]
  \centering
  \includegraphics[width=0.49\textwidth]{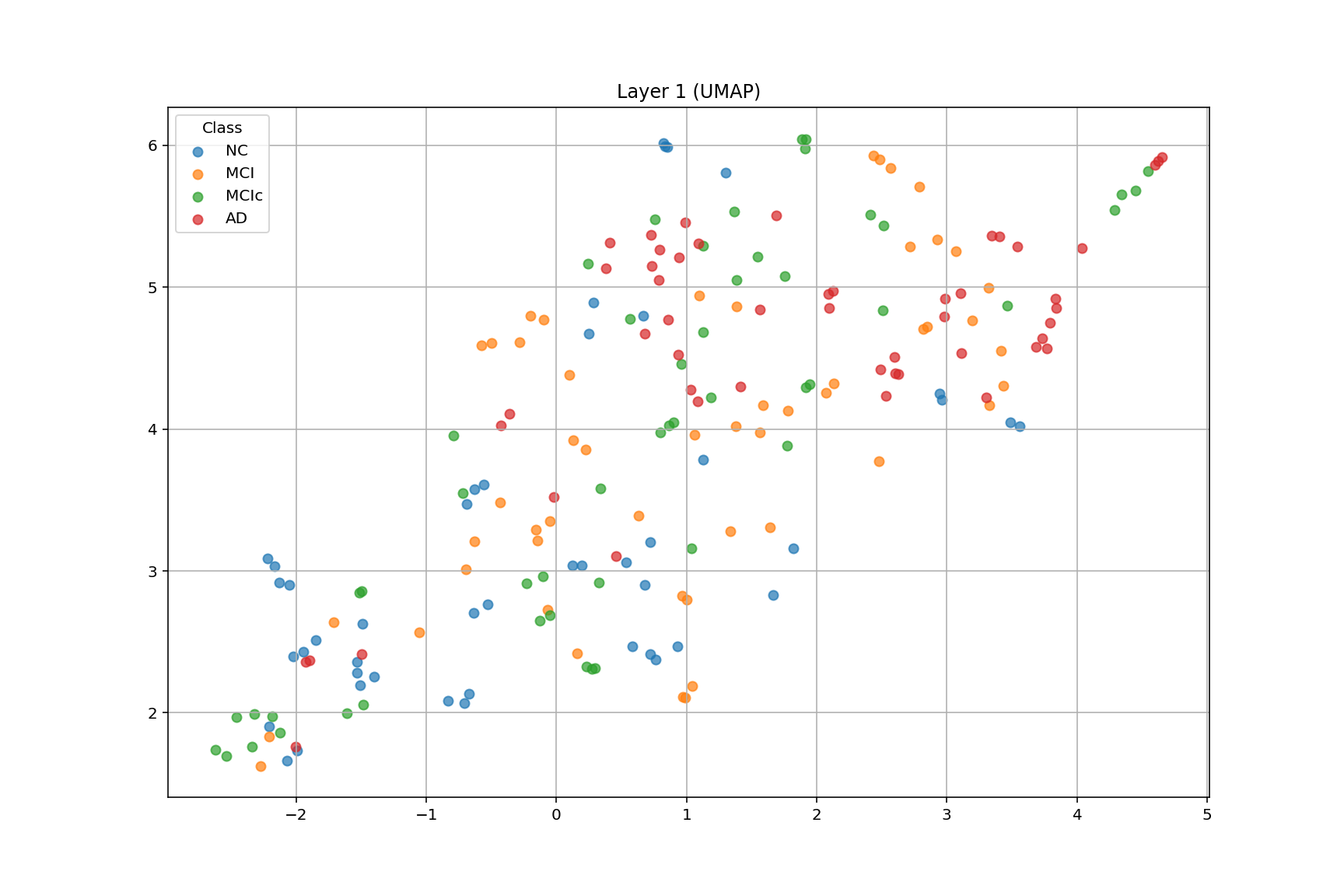}
  \includegraphics[width=0.49\textwidth]{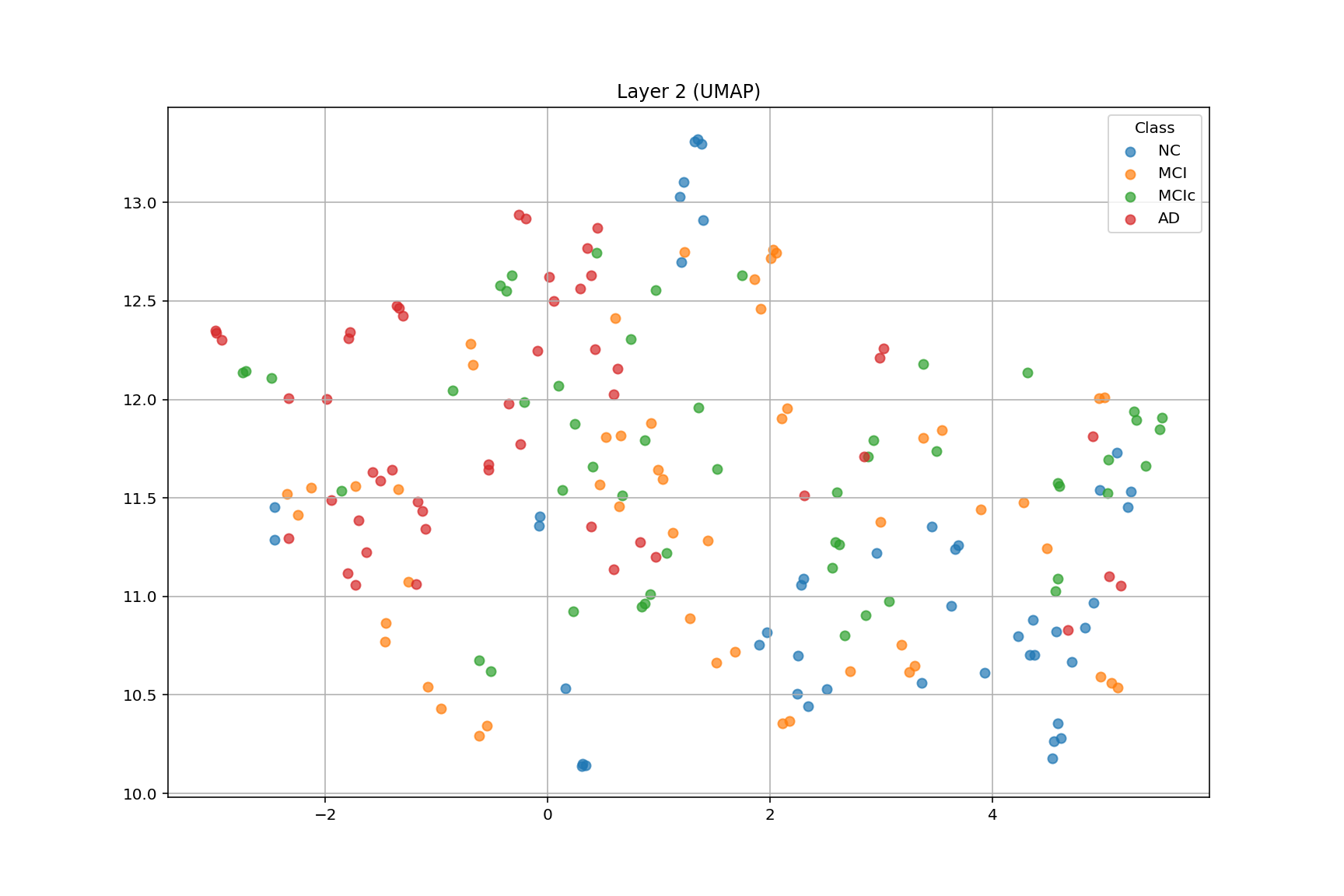}
  \includegraphics[width=0.49\textwidth]{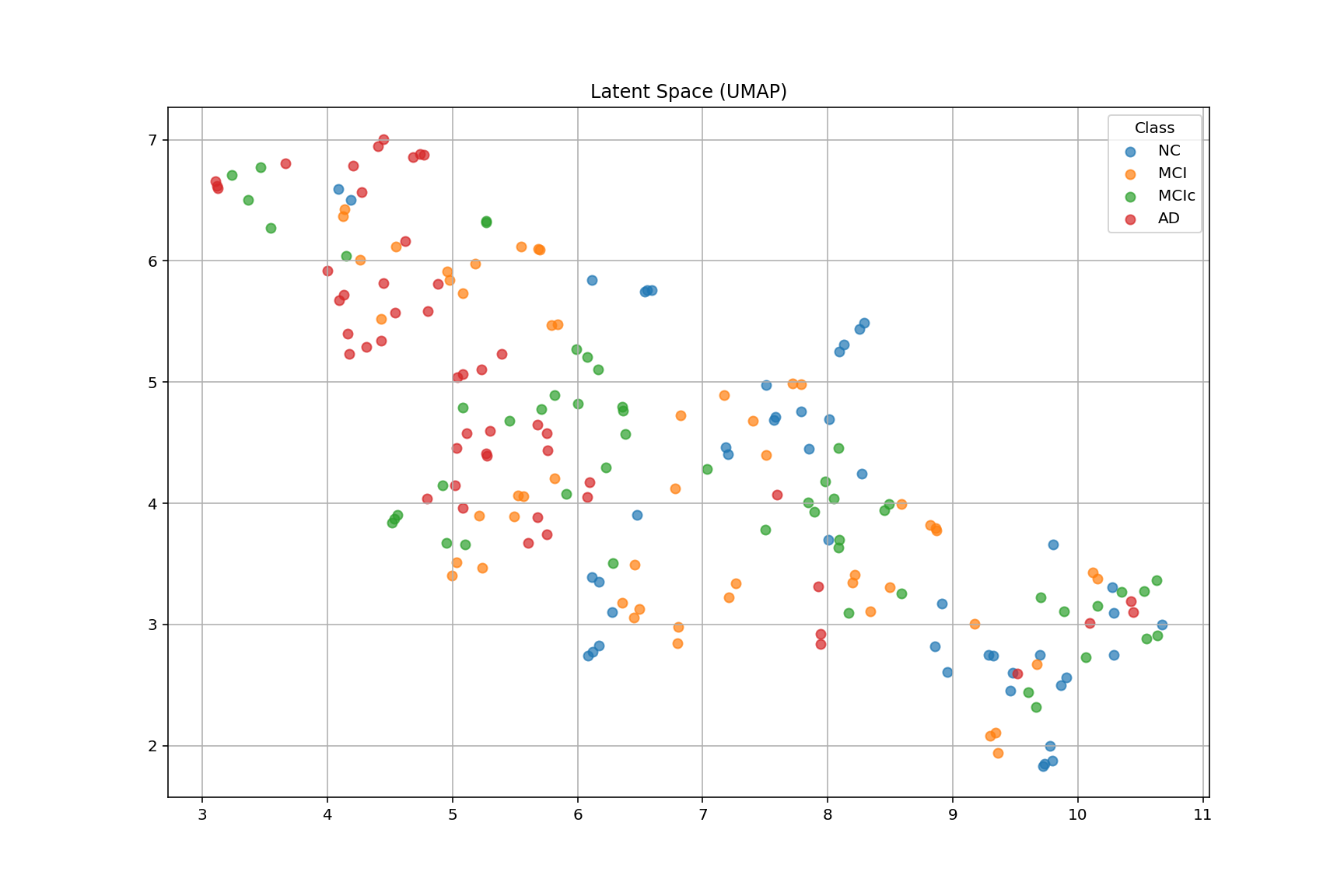}
  \caption{UMAP projection of Layer 1, 2 and latent activations}
  \label{fig:UMAP}
\end{figure}

After training the model independently on the previously defined clinical groupings, activations from intermediate layers and the latent space were extracted. PCA, t-SNE, UMAP and PLS were applied for DR. The resulting low-dimensional projections were visualized to assess whether class separation was preserved. Bootstrapping was performed with 200 samples. Figures \ref{fig:PCA} and \ref{fig:tsne} show 2D projections of intermediate layers and the latent space using PCA, and t-SNE. In figures \ref{fig:PLS} and \ref{fig:UMAP}, we show the global analysis of latent activation projection with UMAP and PLS. It is important to note the near-perfect separability of features in PLS, which is likely due to the use of clinical labels during feature extraction, enabling perfect overfitting to the clinical groups. Such feature representations—obtained either from simpler architectures \cite{Martinez} or from more complex ones \cite{Bass2022,42}, including generative models, have become a major focus in recent years for mapping neurological phenotypes.

\begin{figure}[t]
  \centering
  \includegraphics[width=0.49\textwidth]{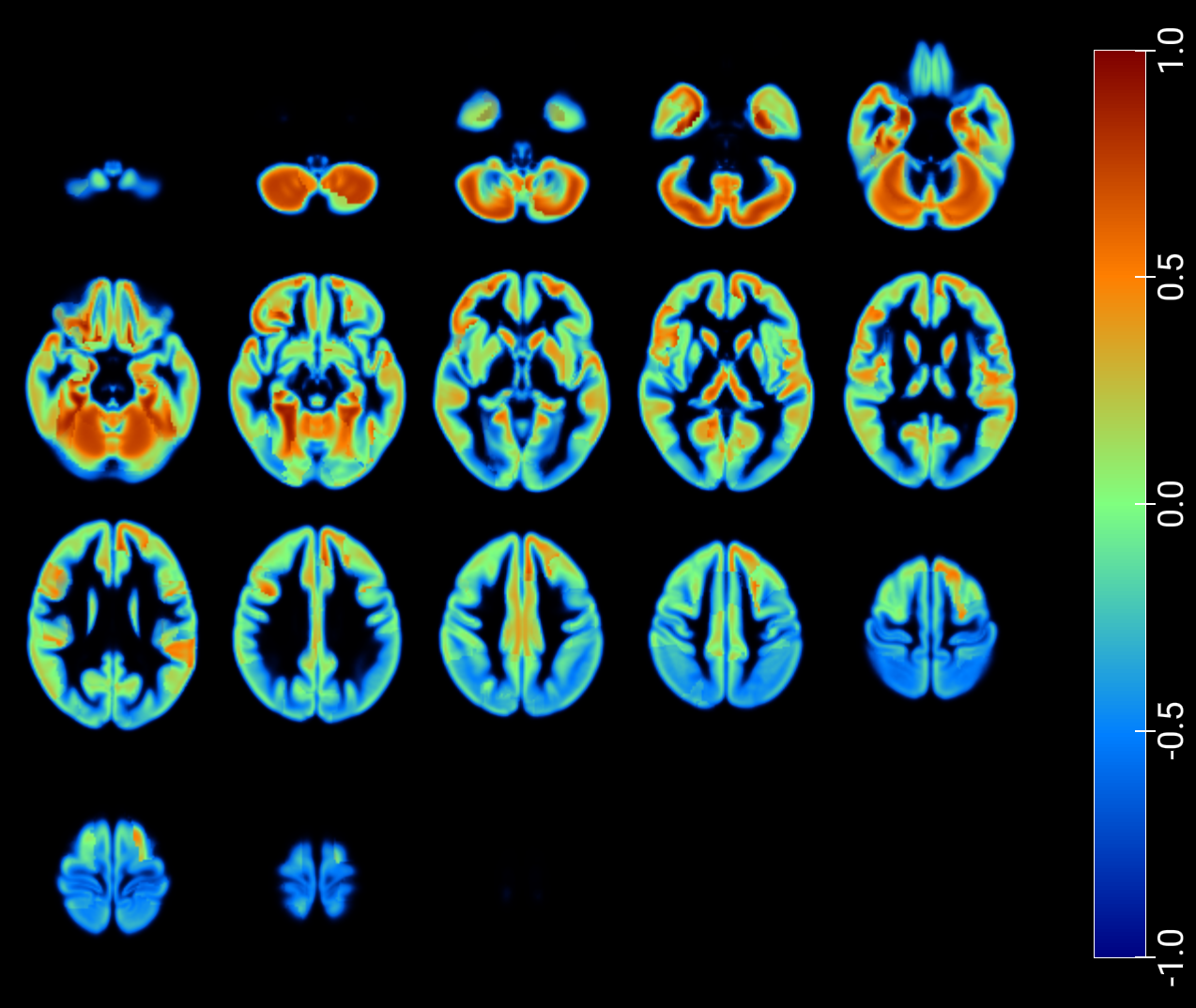}
  \includegraphics[width=0.49\textwidth]{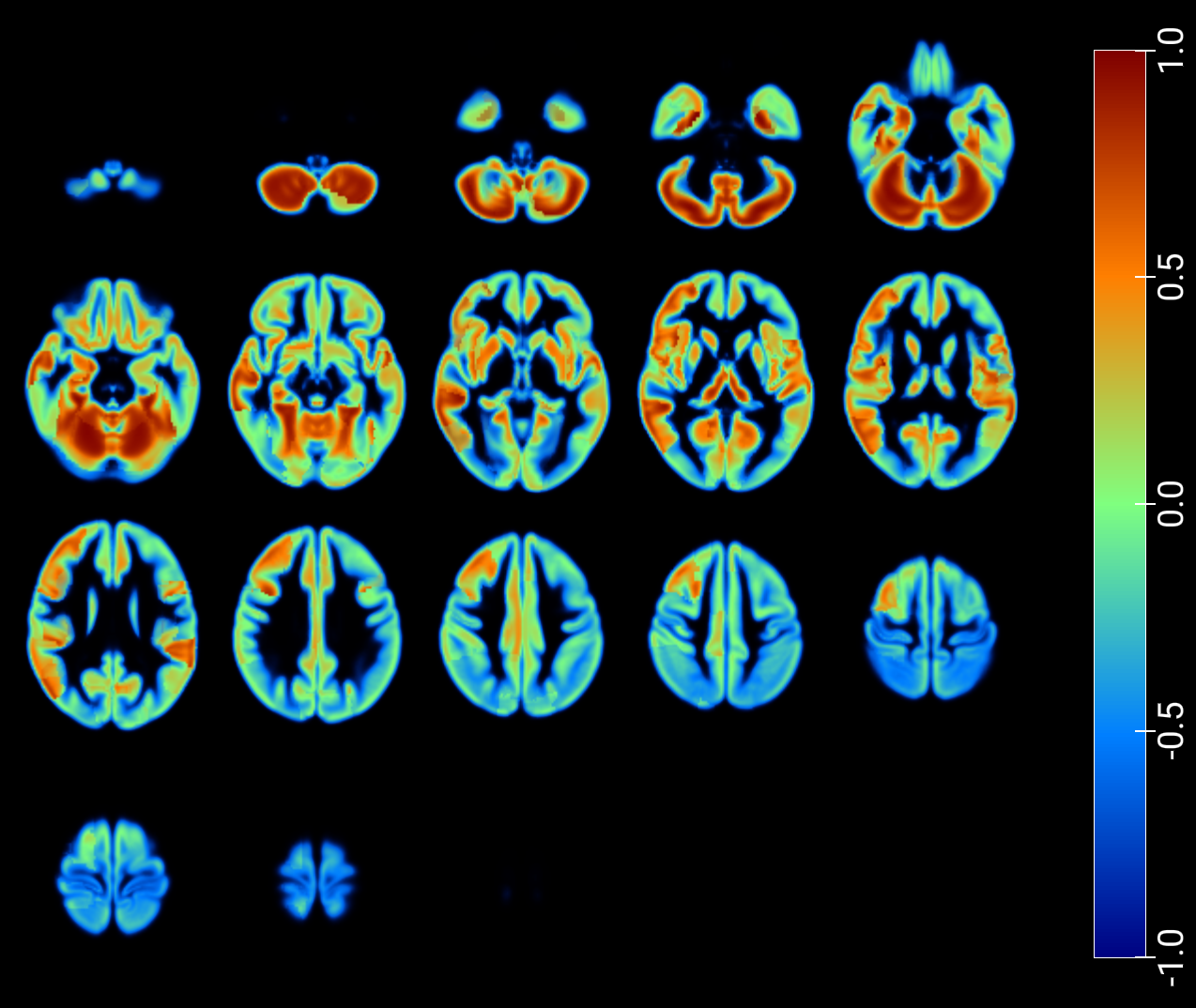}
  \includegraphics[width=0.49\textwidth]{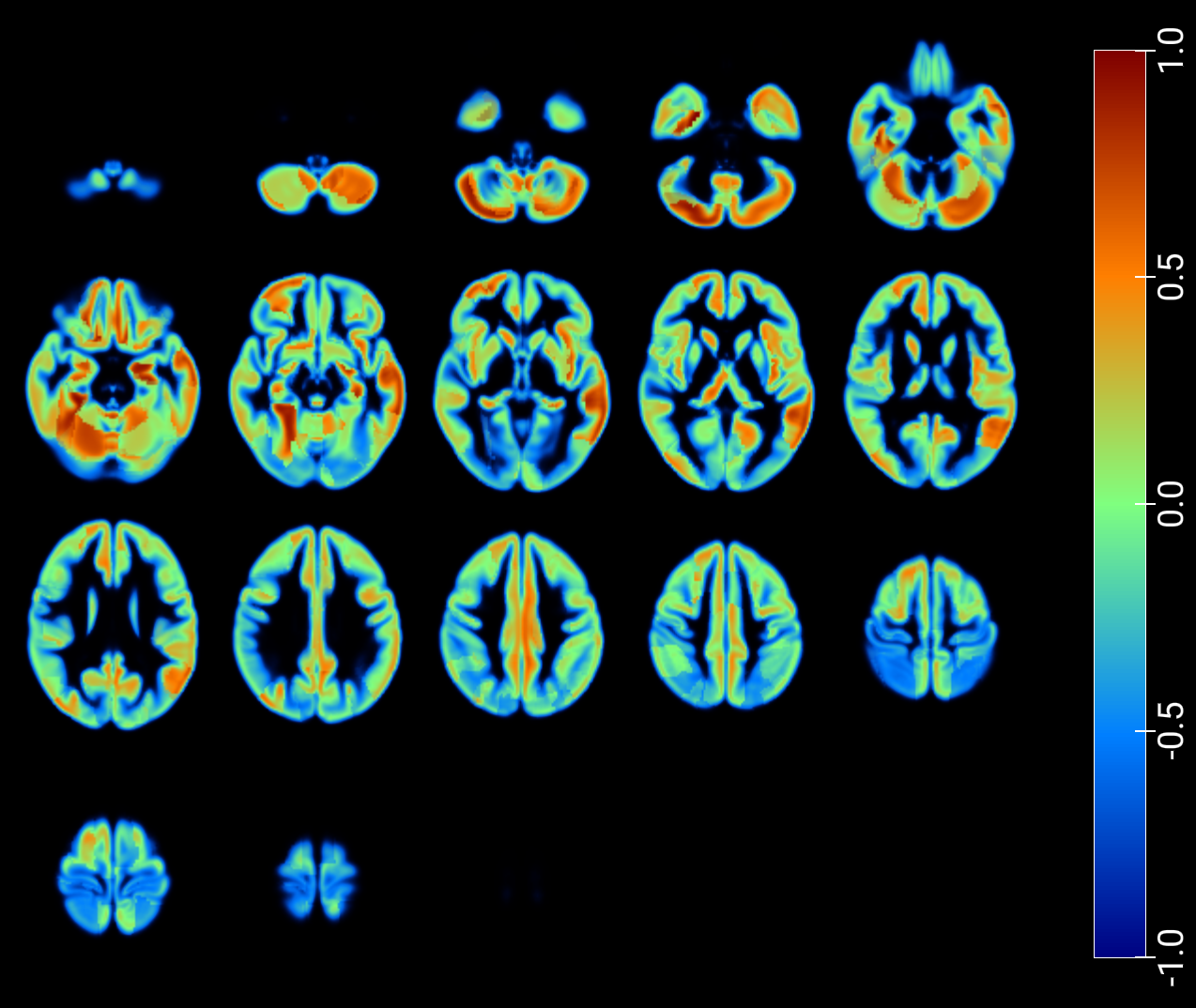}
  \includegraphics[width=0.49\textwidth]{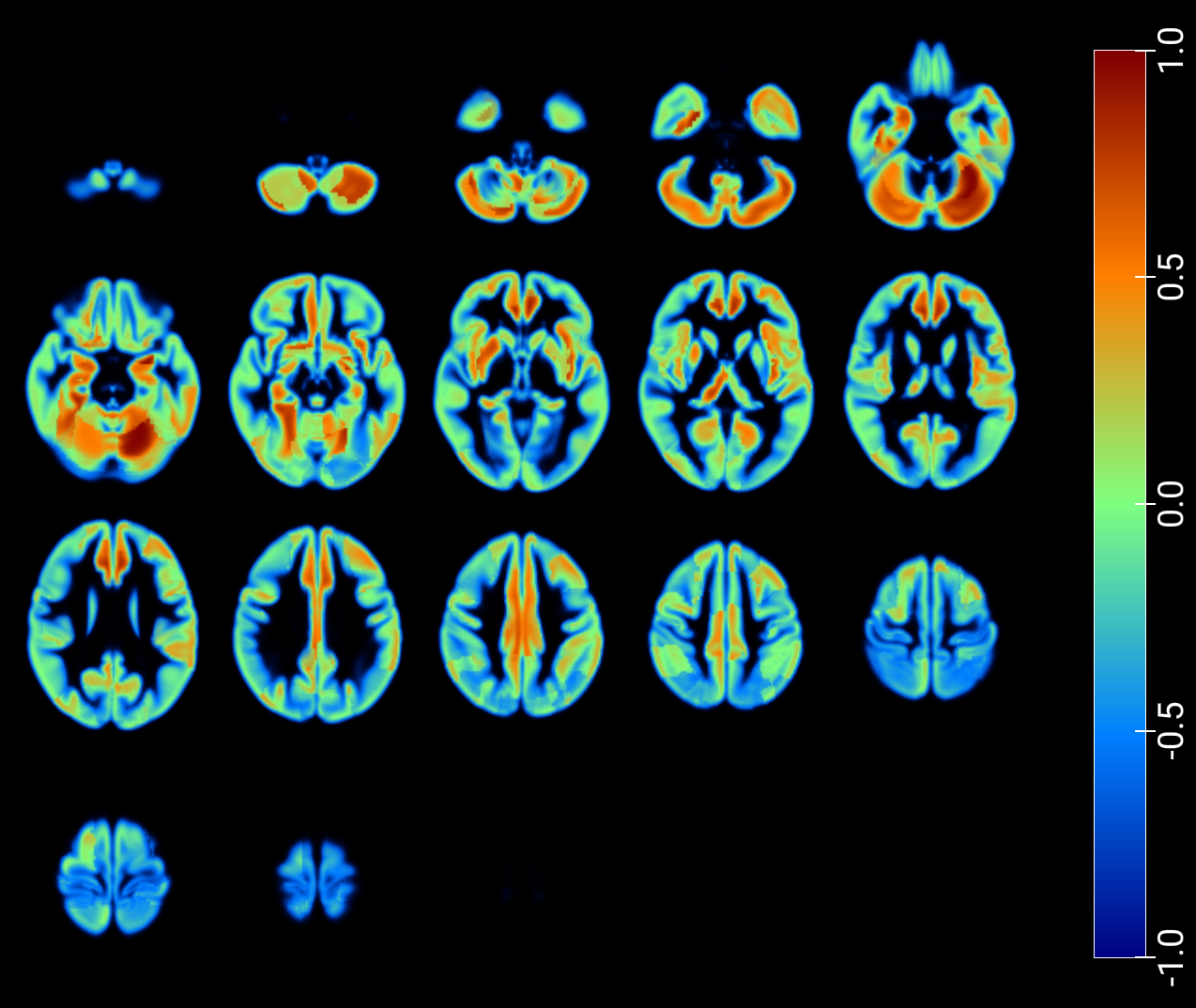}
  \caption{Fused neuroanatomical visualization of significant latent-to-anatomy correlations (PCA and t-SNE methods, \textbf{component 3} for NOR and AD classes at the latent layer.)}
  \label{fig:pcatsne3}
\end{figure}

\subsection{SHAP-based image analysis}

To assess the contribution of individual brain regions to the global reconstruction error across different diagnostic groups, we computed region-wise SHAP values based on the AE’s performance. 

For each image, mean intensity values were extracted within predefined anatomical regions using the AAL atlas, resulting in a feature matrix where each row corresponded to a participant and each column to a brain region. The total reconstruction error was computed globally for each subject. For each diagnostic class (i.e., NOR, MCI, MCIc, AD), a separate random forest regressor was trained to predict subject-wise reconstruction error from regional intensity profiles.

\subsubsection{SHAP computation and spatial mapping}\label{app:shap}

SHAP values were computed to estimate the contribution of each brain region to the predicted error. Region importance was quantified using the mean absolute SHAP value per region within each class. We define an aggregated SHAP importance map \( \mathbf{S} \in \mathbb{R}^{V} \) over the brain volume \( V \), where the value at voxel \( v_i \in V \) is given by:
\[
S(v_i) = 
\begin{cases}
\tilde{s}_r, & \text{if } v_i \in \text{region } r, \\
0, & \text{otherwise},
\end{cases}
\]
where \( \tilde{s}_r \) is the normalized mean SHAP value for region \( r \), computed as:
\[
\tilde{s}_r = \frac{s_r - \min(s)}{\max(s) - \min(s) + \epsilon},
\]
with \( s_r \) being the mean SHAP value for region \( r \), \( s = \{s_r\} \) for all \( r \in \{1, \dots, R\} \), and \( \epsilon \) a small constant to prevent division by zero.

\begin{figure}[t]
  \centering
  \includegraphics[width=0.49\textwidth]{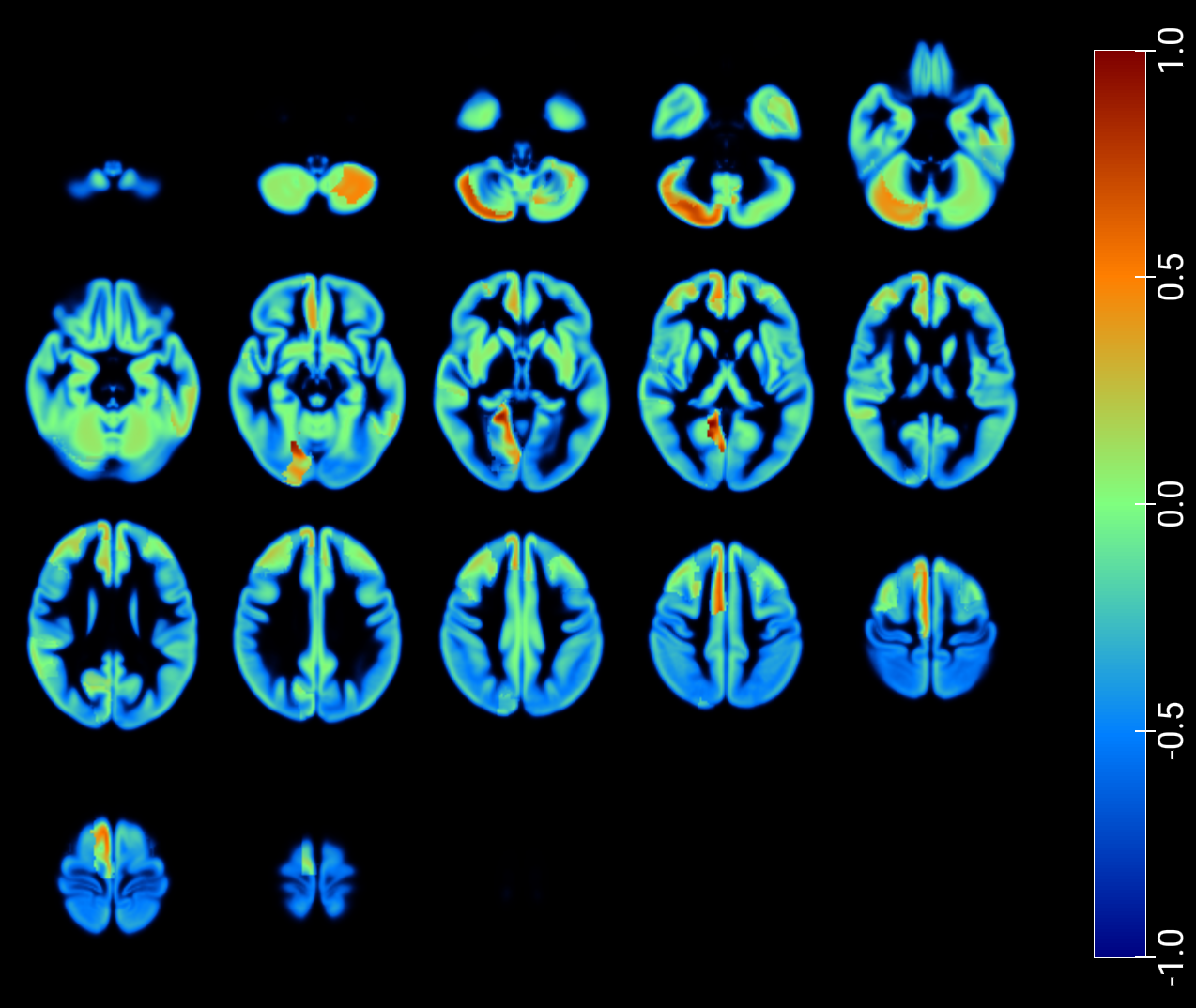}
  \includegraphics[width=0.49\textwidth]{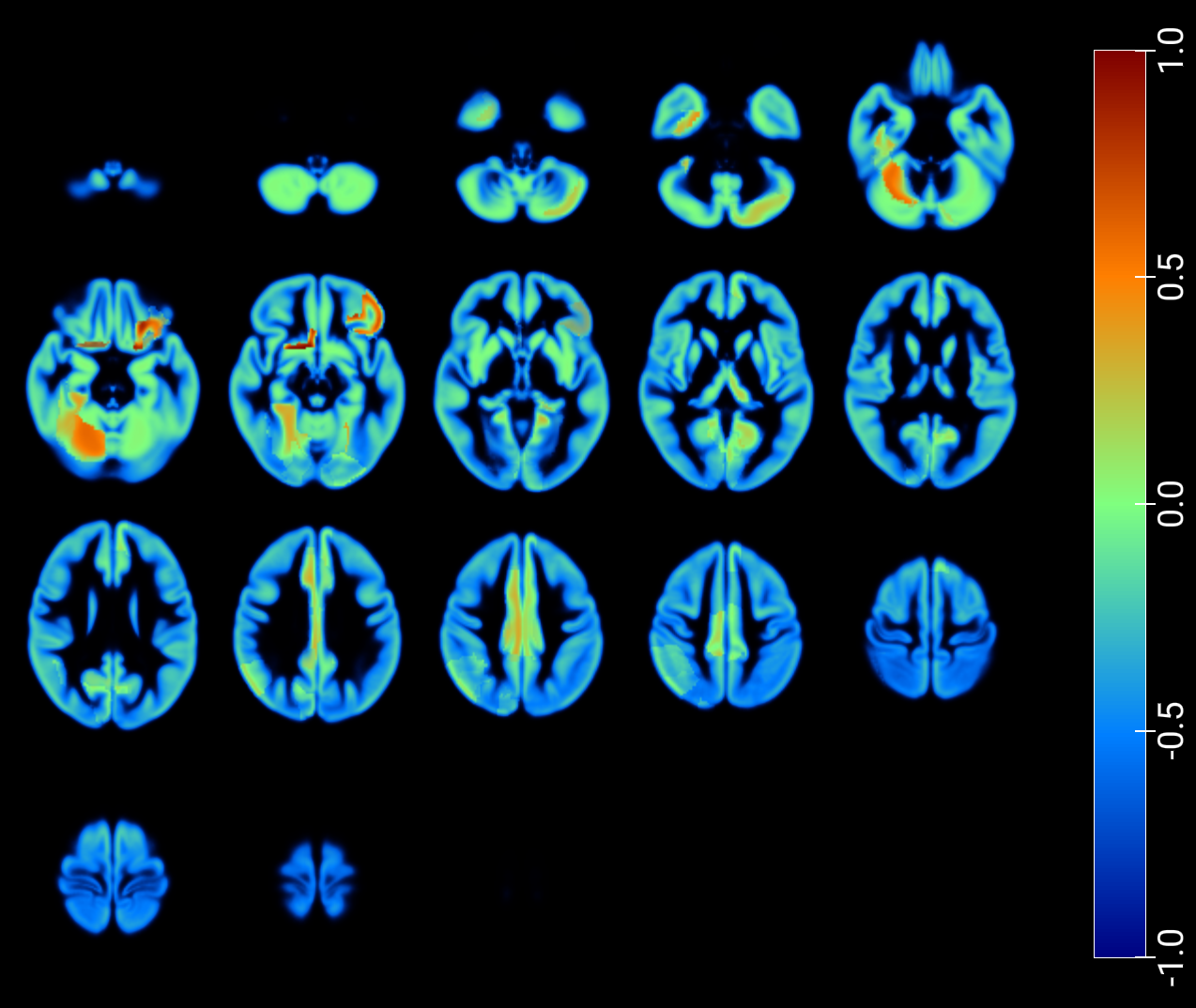}
  \includegraphics[width=0.49\textwidth]{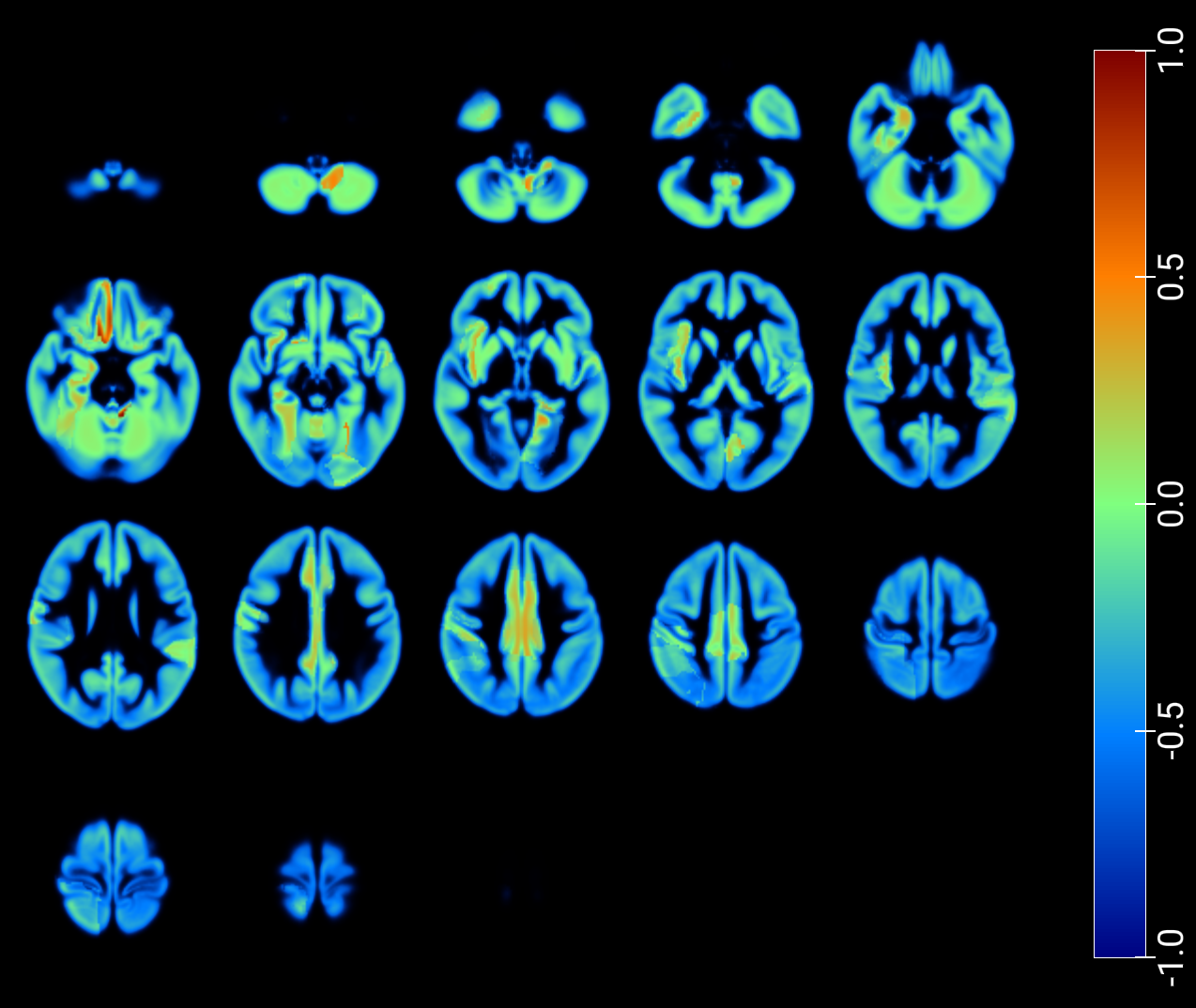}
  \includegraphics[width=0.49\textwidth]{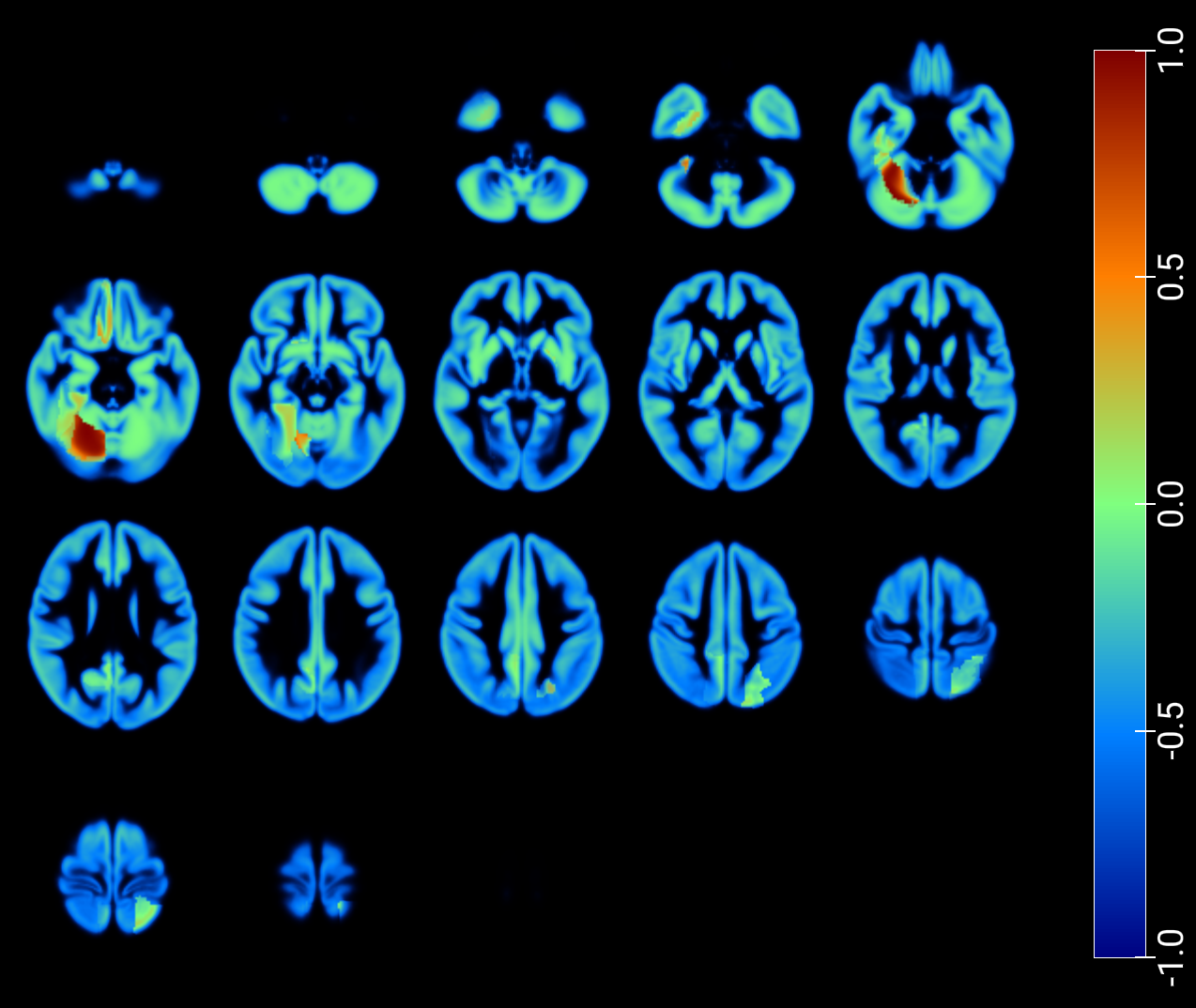}
  \caption{Fused neuroanatomical visualization of SHAP values mapped to anatomy: top row shows NOR (left) and AD (right) classes; bottom row shows NOR (left) and MCIc (right) classes.}
  \label{fig:fus_shap}
\end{figure}

SHAP scores were mapped onto a high-resolution anatomical MRI using the corresponding AAL label indices. To improve anatomical specificity and reduce noise, the SHAP maps were further masked using a gray matter probability map. For visualization, fused overlays of SHAP importance maps on anatomical MRI slices were generated to enhance interpretability (see examples in figure~\ref{fig:fus_shap}).

\subsubsection{Region-wise SHAP importance}

Table~\ref{tab:shap_regions} summarizes the four most important AAL regions by SHAP value for each diagnostic class and comparison group. Certain regions, such as Rectus\_R, Insula\_R, Lingual\_L, and Parietal\_Sup\_L, consistently appeared across comparisons within the same class, indicating robust relevance for group characterization. Additionally, overlaps across classes suggest shared neuroanatomical patterns, while regions appearing exclusively in a single class may act as discriminative markers. In the context of AD, regions such as the superior parietal lobule, fusiform gyrus, and Heschl's gyrus have been associated with disease progression and cognitive decline. For example, atrophy in parietal and temporal cortices has been linked to impaired memory and visuospatial processing \cite{frisoni2010clinical, whitwell2007patterns}, while alterations in Heschl's gyrus may reflect disruptions in auditory processing and broader cortical networks \cite{fitzhugh2019heschl}. These findings support the relevance of the regions identified by SHAP.

\begin{table*}[]
\centering
\renewcommand{\arraystretch}{1.2}
\begin{tabular}{ccl}
\toprule
\textbf{Class} & \textbf{Group} & \textbf{Four Most Important Regions (AAL)} \\
\midrule
NOR   & NOR - MCI           & \textbf{Supp\_Motor\_Area\_R}, \underline{\textbf{Supp\_Motor\_Area\_L}}, Cerebellum\_8\_L, \\
      &                     & \textbf{Cingulum\_Post\_L}\\
      & NOR - MCIc          & Frontal\_Mid\_L, \underline{\textbf{Supp\_Motor\_Area\_L}}, \underline{\textbf{Cerebellum\_Crus1\_R}}, \\
      &                     & \textbf{Cingulum\_Post\_L} \\
      & NOR - AD            & Lingual\_R, \textbf{Supp\_Motor\_Area\_R}, \underline{Frontal\_Sup\_Medial\_R}, \\
      &                     & Cerebellum\_Crus2\_R \\
      & NOR - MCI - MCIc - AD & \textbf{Supp\_Motor\_Area\_L}, Frontal\_Mid\_R, \underline{\textbf{Cerebellum\_Crus1\_R}}, \\
      &                     & \textbf{Supp\_Motor\_Area\_R} \\
\midrule
MCI   & NOR - MCI           & Temporal\_Sup\_R, Cingulum\_Ant\_R, \underline{Frontal\_Sup\_Medial\_R}, \\
      &                     & \textbf{Frontal\_Mid\_Orb\_L} \\
      & NOR - MCI - MCIc - AD & \textbf{Frontal\_Mid\_Orb\_L}, Angular\_L, Temporal\_Inf\_L, \\
      &                     & Frontal\_Inf\_Orb\_L \\
\midrule
MCIc  & NOR - MCIc          & \textbf{Temporal\_Pole\_Sup\_R}, \underline{\textbf{Lingual\_L}}, \underline{\textbf{Supp\_Motor\_Area\_L}}, \\
      &                     & \underline{Calcarine\_R} \\
      & NOR - MCI - MCIc - AD & \textbf{Temporal\_Pole\_Sup\_R}, \underline{\textbf{Lingual\_L}}, Parietal\_Inf\_L, \\
      &                     & \underline{\textbf{Supp\_Motor\_Area\_L}} \\
\midrule
AD    & NOR - AD            & \underline{Frontal\_Inf\_Orb\_L}, Olfactory\_R, Cerebellum\_6\_R, \\
      &                     & \underline{Lingual\_L} \\
      & NOR - MCI - MCIc - AD & Angular\_R, \underline{Calcarine\_R}, \underline{Cerebellum\_Crus1\_R}, Caudate\_L \\
\bottomrule
\end{tabular}
\caption{Summary of the four most important AAL regions identified by SHAP for each class and group. Regions in \textbf{bold} appear repeatedly across comparisons within a class, while underlined regions appear across different comparisons.}
\label{tab:shap_regions}
\end{table*}

\subsubsection{Class-specific SHAP patterns}

\begin{figure}[t]
\centering
\includegraphics[width=0.75\textwidth]{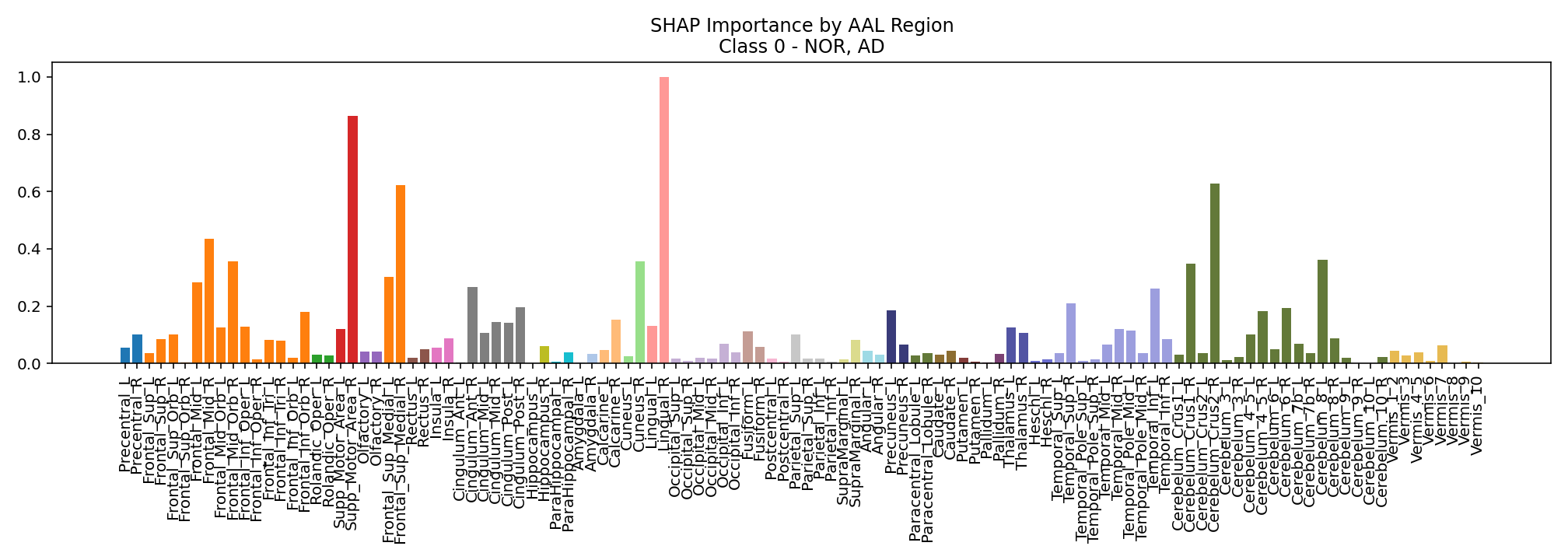}
\caption{AAL regions ranked by SHAP importance for class 0 (NOR) when compared with AD.}
\label{fig:shap_bar_class0}
\end{figure}

\begin{figure}[t]
\centering
\includegraphics[width=0.75\textwidth]{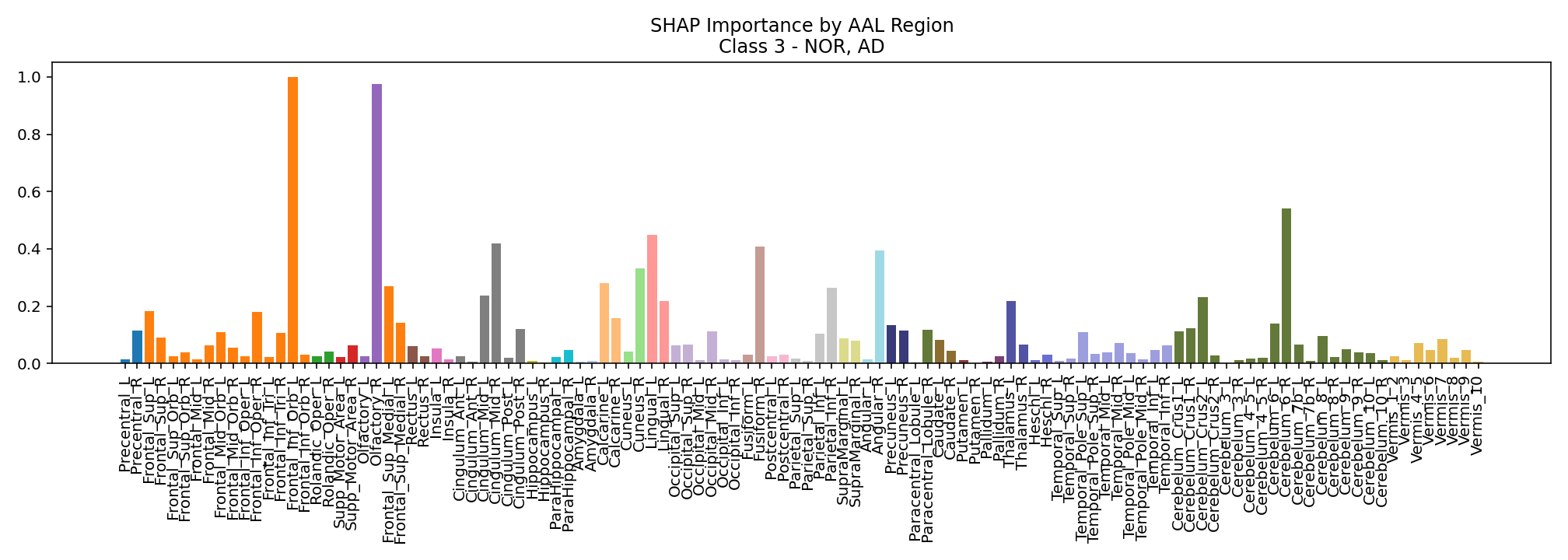}
\caption{SHAP regions for class 3 (AD), reflecting contributions aligned with known atrophy patterns.}
\label{fig:shap_bar_class3}
\end{figure}

Bar plots (figures~\ref{fig:shap_bar_class0} and \ref{fig:shap_bar_class3}) show normalized regional SHAP importance, while violin plots (figure~\ref{fig:shap_violin_class03}) illustrate inter-subject variability. For the NOR class, regions such as the Supp\_Motor\_Area, Cuneus, and Lingual gyrus were among the most influential contributors when contrasted against AD. These regions were consistently identified across comparisons, reinforcing their interpretability. For the AD class, temporal and parietal structures—including the Fusiform gyrus, Frontal\_Inf\_Orb, and Cingulum—showed strong contributions, aligning with known neurodegenerative patterns. These results provide anatomical grounding to the model’s reconstruction behavior.

\subsubsection{Inter-subject variability and feature interpretation}

\begin{figure}[t]
\centering
\includegraphics[width=0.75\textwidth]{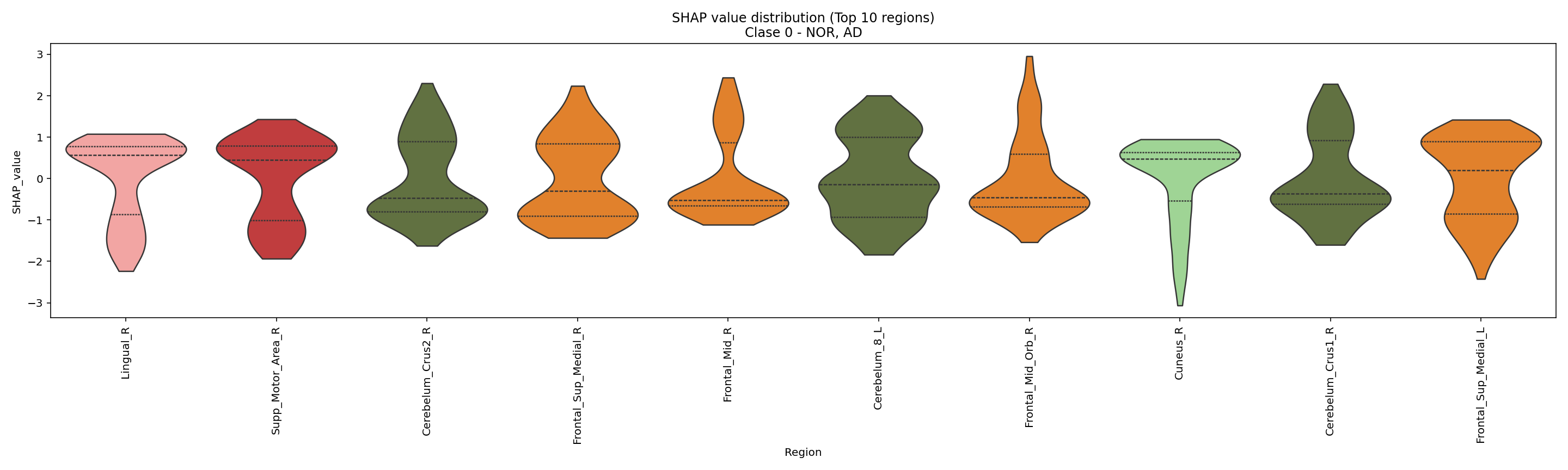}
\includegraphics[width=0.75\textwidth]{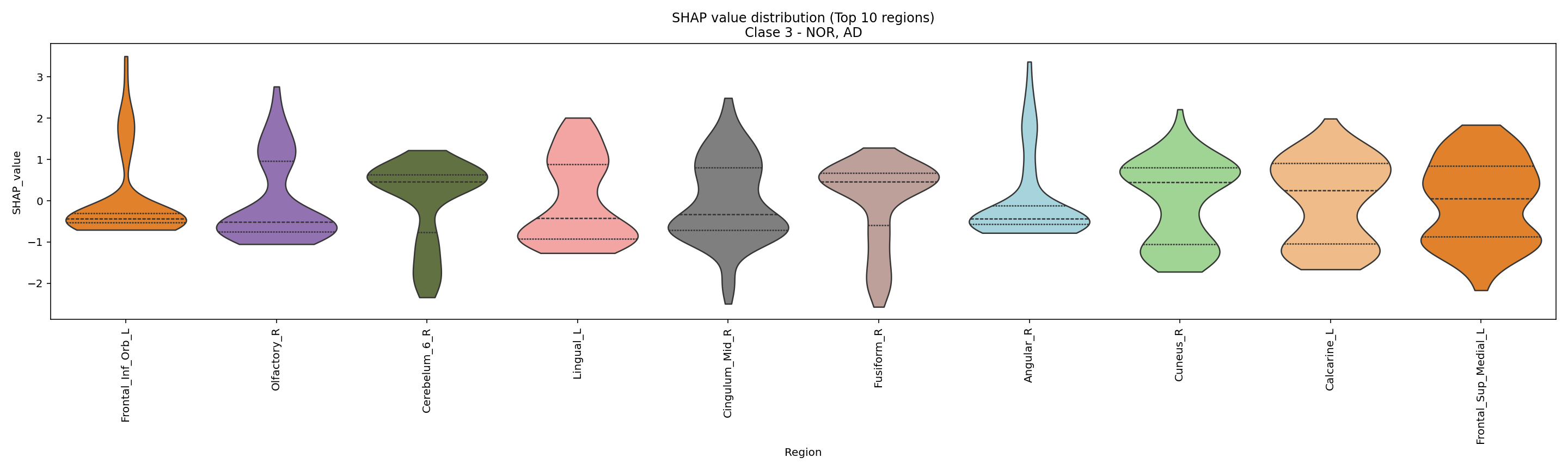}
\caption{Violin plots of SHAP values for NOR and AD classes.}
\label{fig:shap_violin_class03}
\end{figure}

\begin{figure}[t]
\centering
\includegraphics[width=\textwidth]{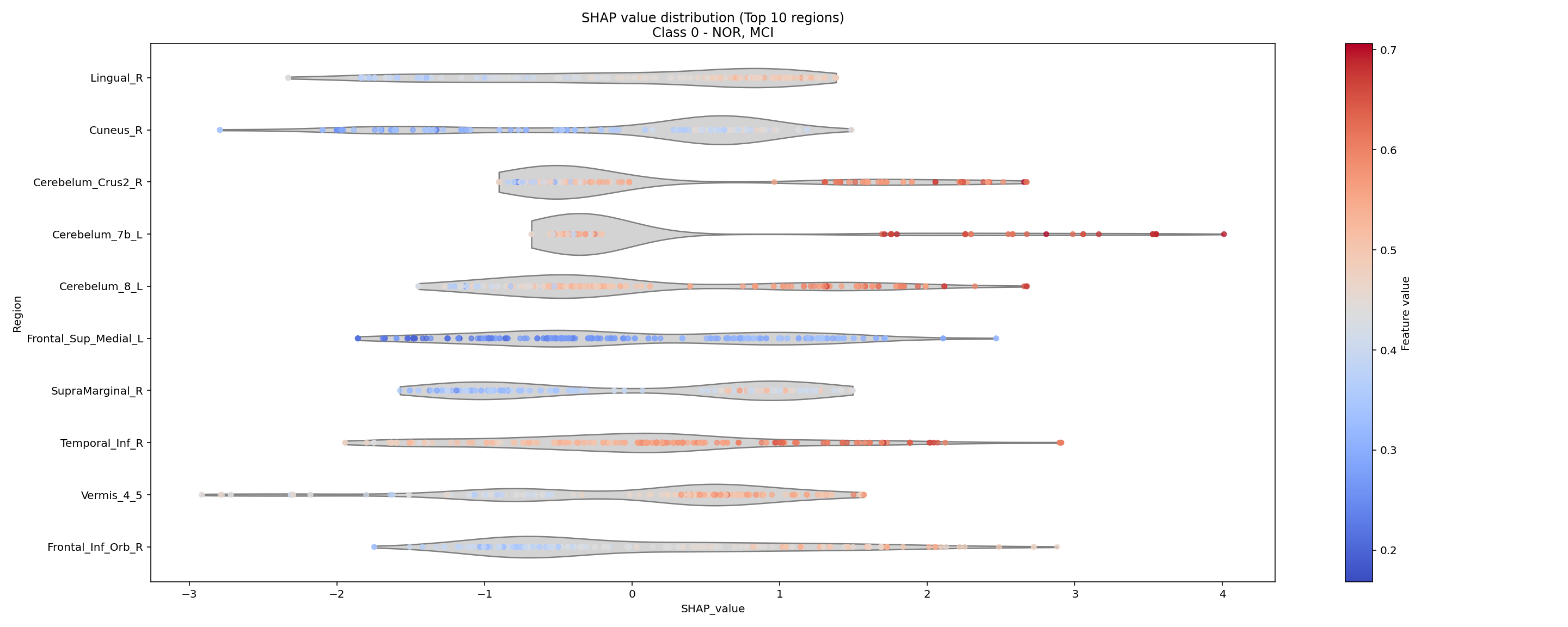}
\includegraphics[width=\textwidth]{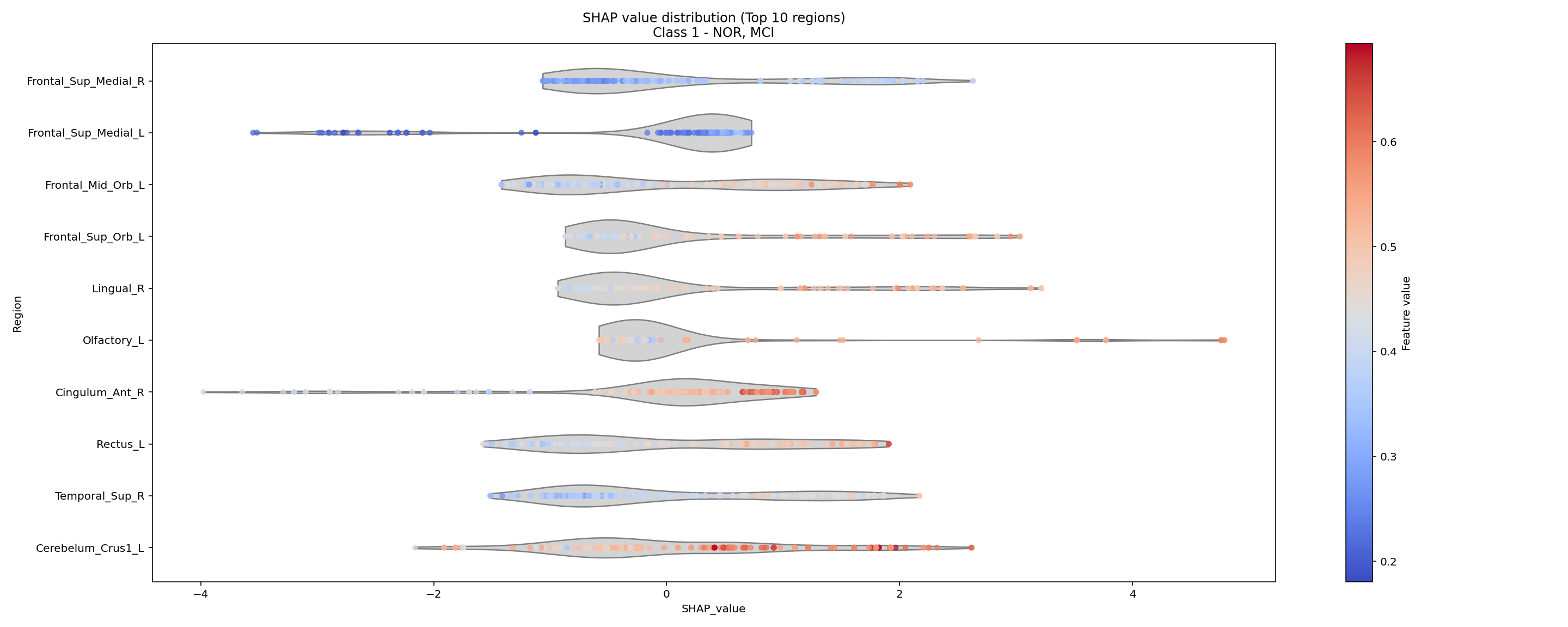}
\caption{Distribution of SHAP values and corresponding feature values for the NOR–MCI comparison. Each violin represents the distribution of SHAP values associated with a specific anatomical region, while the color encoding reflects the underlying feature value (regional gray matter intensity). This joint visualization allows the relationship between feature magnitude and model attribution to be explored. Regions exhibiting broader or shifted distributions may indicate differential relevance between classes; however, these patterns are exploratory and primarily serve to motivate subsequent statistically grounded analyses.}
\label{fig:shap_violin_feature01}
\end{figure}

Violin plots of SHAP values across subjects (figure~\ref{fig:shap_violin_class03}) reveal that regions with high mean importance may also exhibit substantial dispersion, suggesting heterogeneous structure–function relationships within diagnostic groups. Additional plots combining SHAP values with original feature values (figure~\ref{fig:shap_violin_feature01}) further highlight this variability. Feature contributions can be interpreted directionally, indicating consistent regional effects across subjects. Overall, the SHAP analysis supports the identification of meaningful, class-specific neuroanatomical markers. Several regions contribute both to reconstruction performance and to discrimination across clinical stages, supporting their relevance for downstream tasks such as dimensionality reduction and feature selection.

\subsection{Comparison of SHAP and correlation analyses in challenging classes}\label{app:shap-corr}

\begin{figure}[t]
  \centering
  \includegraphics[width=0.49\textwidth]{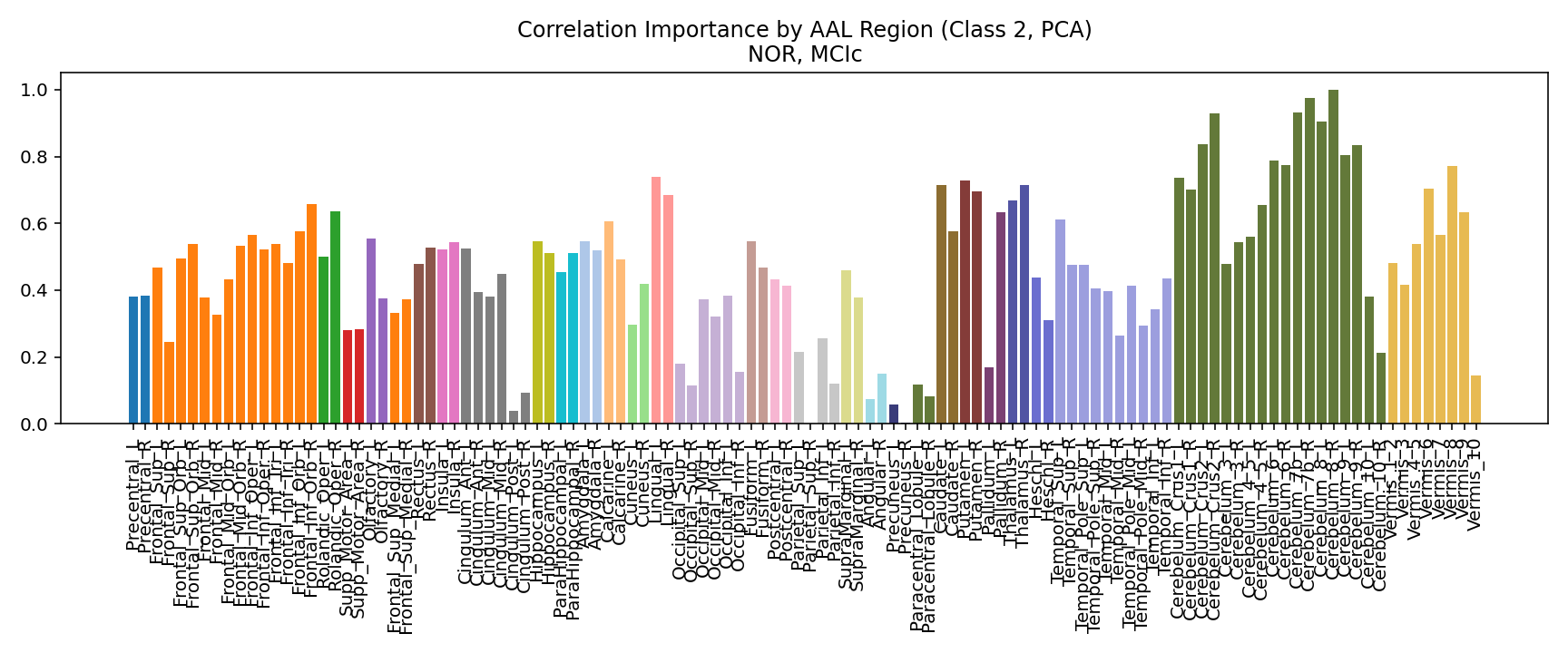}
  \includegraphics[width=0.49\textwidth]{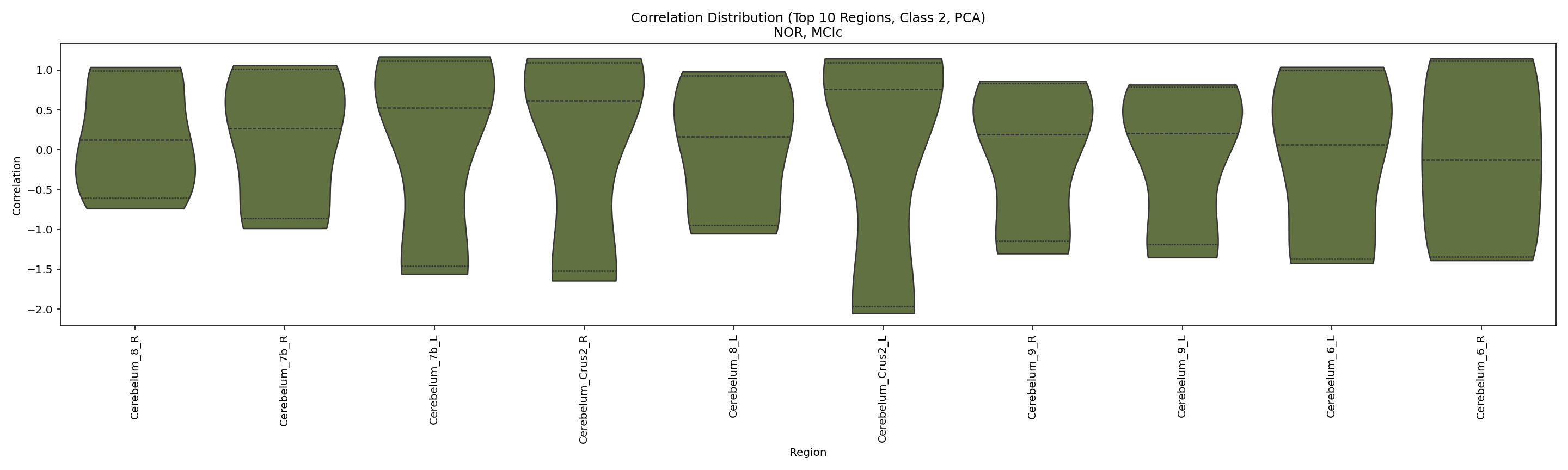}
  \includegraphics[width=0.49\textwidth]{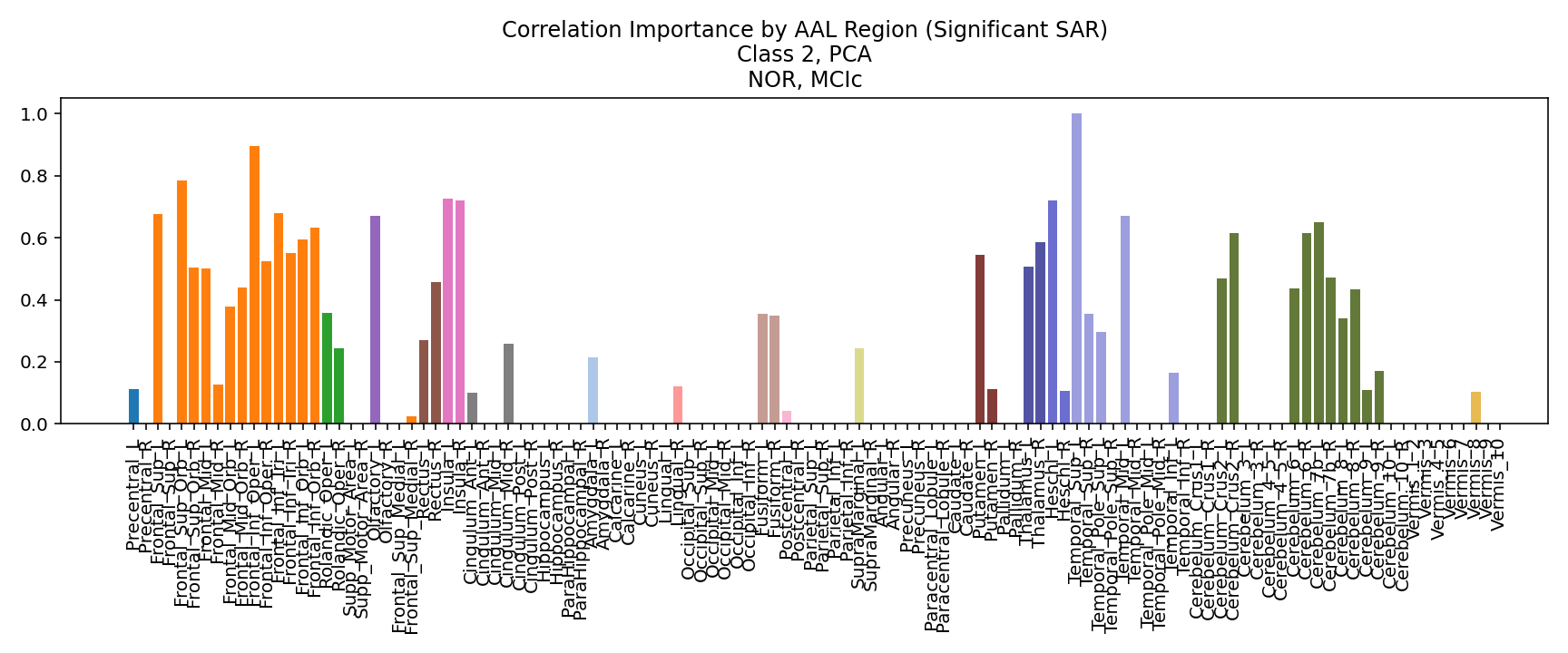}
  \includegraphics[width=0.49\textwidth]{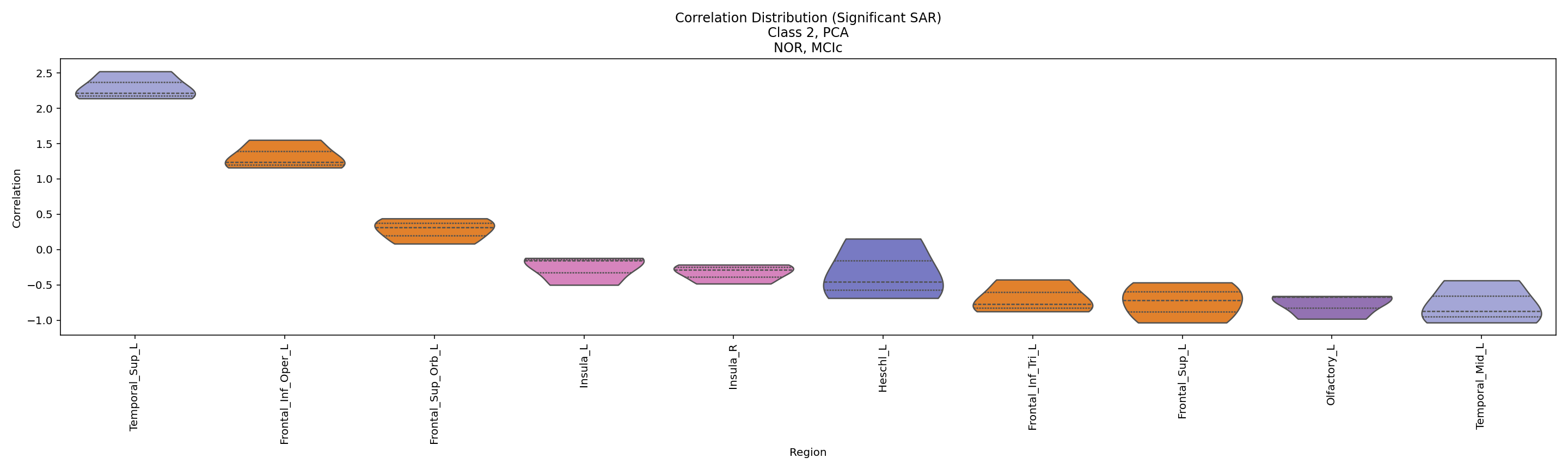}
  \caption{Distribution of the ten most relevant AAL regions for the MCIc group obtained using correlation analysis (\textbf{z-scored}) with and without SAR correction. The SAR-corrected results (bottom) highlight several regions also identified as highly relevant in the AD group, which are not present or are less prominent in the uncorrected analysis (top). This demonstrates how SAR can recover consistent neuroanatomical patterns that may be obscured when statistical corrections are omitted.}
  \label{fig:comparison_MCIc}
\end{figure}

As an illustrative example, we present in this appendix the distribution of the most relevant AAL regions for the MCIc group obtained through the SAR-corrected correlation analysis (figure \ref{fig:comparison_MCIc}), alongside the uncorrected results. This comparison revealed that several brain regions—previously identified as highly relevant in the AD group—also emerged as top-ranked for MCIc when applying SAR. These regions, however, were absent or appeared substantially less important in the uncorrected analysis suggesting that the lack of statistical control may have obscured biologically meaningful patterns. This example highlights how SAR can recover and validate subtle but consistent neuroanatomical signals that would otherwise be lost when relying solely on raw correlation values.

\begin{figure}[t]
  \centering
  \includegraphics[width=0.49\textwidth]{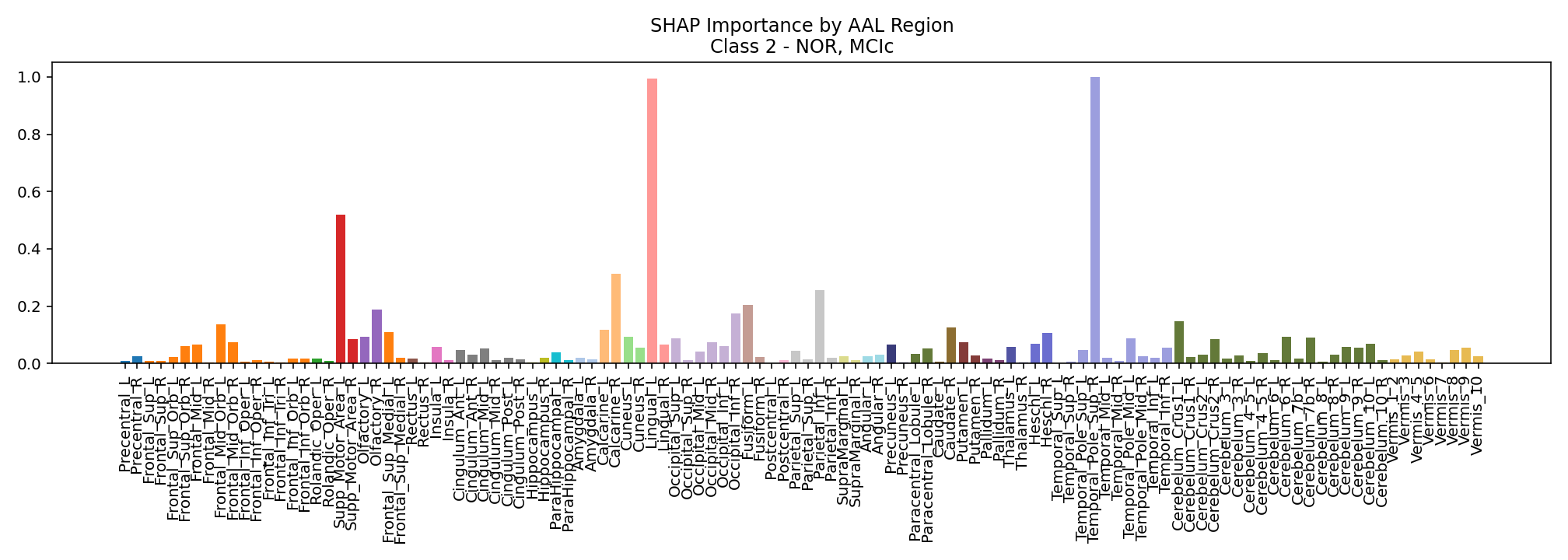}
  \includegraphics[width=0.49\textwidth]{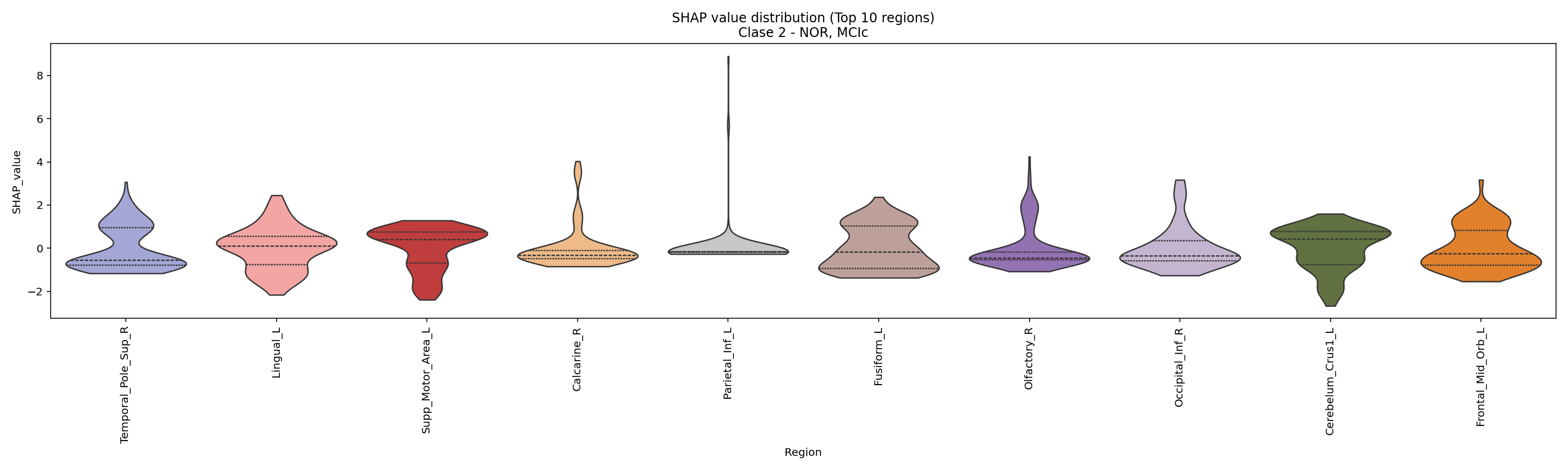}
  \caption{Distribution of anatomical region importance (AAL) according to SHAP values for class 2 (NOR, MCIc). Only a few regions show prominent importance in the model, while most display low values. Distribution of SHAP values for the top 10 most important regions in the classification of class 2 (NOR, MCIc). Each violin plot illustrates the variability and magnitude of each region’s contribution to the model output.}
  \label{fig:SHAPMCIc}
\end{figure}

When we ran the same analysis based on SHAP features, we obtained similar results with a clear reduction in SHAP importance for the main regions involved in neurodegeneration (figure \ref{fig:SHAPMCIc}). While SHAP provided a robust framework for identifying the most influential regions in a model’s output, it is important to acknowledge its limitations. SHAP’s attribution mechanism adjusts for feature interactions and redundancy, which can result in the exclusion of regions that may still exert meaningful influence, especially if their effects are correlated with other features. Furthermore, although SHAP refines the selection beyond simple correlation, it remained sensitive to the underlying statistical relationships and did not establish causality. Consequently, there is a risk that SHAP may discard regions whose importance was masked by complex dependencies or shared variance, potentially overlooking anatomically relevant areas. This highlights the need for complementary analyses and careful interpretation when using SHAP to prioritize features in neuroimaging studies.

\subsection{SLT-Based inferential Framework, a CUBV/PAC-Bayes Approach}

Let \( R(f) \) denote the true risk of a model \( f \), and \( R_{\text{emp}}(f) \) the empirical risk estimated from the data (e.g., via K-fold CV). The CUBV framework proposes the following inequality:

\begin{equation}
R(f) \leq R_{\text{CV}}(f) + \Psi(n, \delta)
\end{equation}

\noindent where \( R_{\text{CV}}(f) \) is the empirical risk estimated via CV -in this exploratory analysis, using a resubstitution approach- \( \Psi(n, \delta) \) is a confidence-based concentration bound \cite{Boucheron13}, \( n \) is the sample size, and \( \delta \in (0,1) \) is the confidence level. The concentration term \( \Psi \) typically takes the form:
\begin{equation}
\Psi(n, \delta) = \sqrt{\frac{C \log(1/\delta)}{2n}}
\end{equation}

\noindent where \( C \) is a constant that depends on the complexity of the hypothesis space (e.g., the VC-dimension or Rademacher complexity). This formulation ensures that, with probability at least \( 1 - \delta \), the true risk does not exceed the estimated CV risk plus the uncertainty margin.

We implemented a PAC-Bayes-based upper bound analysis to evaluate the classification error rate of linear models. This bound was used to assess whether the classification performance was significantly better than chance incorporating a theoretical correction based on model complexity and generalization capacity. Specifically, the empirical accuracy was adjusted using a PAC-Bayes bound derived from the model parameters with a dropout rate $\eta$ yielding a corrected rate (please see \cite{Gorrizarxiv}). If this corrected rate exceeded 0.5, the classification was considered statistically significant. This approach provided a model-agnostic and theoretically grounded validation of latent space–region associations, enhancing the interpretability and reliability of neuroimaging-based ML models.

\subsection{Projection methods}\label{app:projection}

The DR methods employed default hyperparameters commonly recommended in the literature and standard implementations. For t-SNE, a perplexity of $30$ was used to balance local and global data structure, with a learning rate of $200$ controlling the optimization step size, and $1000$ iterations to ensure algorithm convergence. UMAP was configured with 15 neighbors, determining the local context for reduction, and a minimum distance ($0.1$) influencing the compactness of points in the embedded space. Both methods utilized the Euclidean metric for sample distance calculations. For PCA and PLS, default parameters were maintained, such as the automatic solver selection and internal data normalization, respectively, ensuring reproducible results and comparability with previous studies.

\subsubsection{PCA}
PCA seeks an orthogonal linear transformation that maps the data to a new coordinate system where the greatest variance lies along the first axis (principal component), the second greatest variance along the second axis, and so forth. Let \( \mathbf{X} \in \mathbb{R}^{n \times p} \) be the data matrix of latent features. PCA solves the eigenvalue problem:

\begin{equation}
\mathbf{S} \mathbf{w}_i = \lambda_i \mathbf{w}_i,
\end{equation}

\noindent where \( \mathbf{S} = \frac{1}{n-1} \mathbf{X}^\top \mathbf{X} \) is the empirical covariance matrix, and \( \mathbf{w}_i \) is the \( i \)-th principal axis.

\subsubsection{PLS}
PLS finds components that maximize the covariance between predictors \( \mathbf{X} \) (e.g., latent representations) and responses \( \mathbf{Y} \) (e.g., clinical or anatomical labels). The first latent variable \( \mathbf{t}_1 \) is obtained by:

\begin{equation}
\mathbf{t}_1 = \mathbf{X} \mathbf{w}_1, \quad \text{where } \mathbf{w}_1 = \arg\max_{\|\mathbf{w}\|=1} \text{Cov}^2(\mathbf{X} \mathbf{w}, \mathbf{Y}),
\end{equation}

\noindent and subsequent components are computed on deflated versions of \( \mathbf{X} \) and \( \mathbf{Y} \).

\subsubsection{t-SNE}
t-SNE is a non-linear method that maps high-dimensional data to a low-dimensional space by minimizing the Kullback–Leibler divergence between two distributions: one representing pairwise similarities in the high-dimensional space and another in the low-dimensional space. The objective is:

\begin{equation}
\text{KL}(P \| Q) = \sum_{i \neq j} p_{ij} \log \frac{p_{ij}}{q_{ij}},
\end{equation}

\noindent where \( p_{ij} \) and \( q_{ij} \) denote the joint probabilities of similarity in the high- and low-dimensional spaces, respectively.

\subsubsection{UMAP}
UMAP is a manifold learning technique grounded in topological data analysis. It constructs a weighted k-nearest neighbor graph in the high-dimensional space and optimizes a cross-entropy loss to embed the data in a lower dimension. Formally, the optimization minimizes:

\begin{equation}
C = \sum_{(i,j)} w_{ij} \log \left( \frac{w_{ij}}{\hat{w}_{ij}} \right) + (1 - w_{ij}) \log \left( \frac{1 - w_{ij}}{1 - \hat{w}_{ij}} \right),
\end{equation}

\noindent where \( w_{ij} \) are the edge weights in the high-dimensional space and \( \hat{w}_{ij} \) their corresponding weights in the low-dimensional embedding.

\end{document}